\documentclass[a4paper,USenglish,cleveref]{lipics-v2021-fork}

\usepackage[vlined,boxed,linesnumbered]{algorithm2e}
\usepackage{amsthm}
\usepackage{multirow}
\usepackage[normalem]{ulem}
\usepackage{amsthm}
\usepackage{xspace}
\usepackage{tabularx}
\usepackage{makecell}
\usepackage{dsfont} 

\usepackage[backend=biber,giveninits=true,maxbibnames=99,citetracker=true,url=false,doi=false,giveninits=true,sortcites]{biblatex}
\defbibheading{bibliography}[\refname]{\section*{#1}} 

\makeatletter
\@ifpackageloaded{algorithm2e}{%
\PackageInfo{cleveref}{`algorithm2e' support loaded}%
\crefalias{algocf}{algorithm}%
\crefalias{algocfline}{line}%
\crefalias{AlgoLine}{line}%
\let\cref@old@algocf@nl@sethref\algocf@nl@sethref%
\renewcommand{\algocf@nl@sethref}[1]{%
    \cref@old@algocf@nl@sethref{#1}%
    \cref@constructprefix{AlgoLine}{\cref@result}%
    \@ifundefined{cref@AlgoLine@alias}%
        {\def\@tempa{AlgoLine}}%
        {\def\@tempa{\csname cref@AlgoLine@alias\endcsname}}%
    \xdef\cref@currentlabel{%
        [\@tempa][\arabic{AlgoLine}][\cref@result]%
        \csname p@AlgoLine\endcsname\csname theAlgoLine\endcsname}}%
}{}
\makeatother

\renewcommand{\paragraph}[1]{\smallskip\noindent \textbf{#1.} \;}

\newcommand{\cac}{cac}
\newcommand{\CAC}{CAC\xspace}
\newcommand{\RC}{RC\xspace}
\newcommand{\GC}{GC\xspace}
\newcommand{\CC}{CC\xspace}

\newcommand{\beText}{be}
\newcommand{\be}{be\xspace}

\usepackage{tikz}
\usetikzlibrary{arrows.meta}
\usetikzlibrary{shapes}
\usetikzlibrary{decorations.pathreplacing}
\usetikzlibrary{math}
\usetikzlibrary{shapes.geometric}

\newcommand{\choice}{\ensuremath{\mathsf{choice}}\xspace}

\newcommand{\return}{\ensuremath{\mathsf{return}}\xspace}

\newcommand{\wait}{\ensuremath{\mathsf{wait}}\xspace}

\newcommand{\broadcast}{\ensuremath{\mathsf{be\_broadcast}}\xspace}
\newcommand{\broadcastText}{\beText-broadcast\xspace}

\newcommand{\receivedd}{\ensuremath{\mathsf{received}}\xspace}

\newcommand{\cacpropose}{\ensuremath{\mathsf{cac\_propose}}\xspace}
\newcommand{\cacaccept}{\ensuremath{\mathsf{cac\_accept}}\xspace}

\newcommand{\Verify}{\ensuremath{\mathsf{Verify}}\xspace}

\newcommand{\snaming}{\ensuremath{\mathsf{shortnaming}}\xspace}
\newcommand{\snamingclaim}{\ensuremath{\mathsf{\snaming\_Claim}}\xspace}

\newcommand{\mcprefix}{\ensuremath{\mathsf{Max\_Common\_Prefix}}\xspace}
\newcommand{\choosename}{\ensuremath{\mathsf{Choose\_Name}}\xspace}

\newcommand{\Names}{\ensuremath{\mathit{Names}}\xspace}
\newcommand{\claimlist}{\ensuremath{\mathit{Claim\_dict}}\xspace}
\newcommand{\commitlist}{\ensuremath{\mathit{Commit\_dict}}\xspace}
\newcommand{\prop}{\ensuremath{\mathit{prop}}\xspace}
\newcommand{\currname}{\ensuremath{\mathit{curr\_name}}\xspace}

\newcommand{\pk}{\ensuremath{\mathit{pk}}\xspace}

\newcommand{\timerstart}{\ensuremath{\mathsf{start}}\xspace}
\newcommand{\timerend}{\ensuremath{\mathsf{end}}\xspace}
\newcommand{\timerstop}{\ensuremath{\mathsf{stop}}\xspace}

\newcommand{\CCons}{Cascading Consensus\xspace}
\newcommand{\ccons}{ccons\xspace}
\newcommand{\cconspropose}{\ensuremath{\mathsf{\ccons\_propose}}\xspace}
\newcommand{\cconsdecide}{\ensuremath{\mathsf{\ccons\_decide}}\xspace}

\newcommand{\rcons}{rcons\xspace}
\newcommand{\rconspropose}{\ensuremath{\mathsf{\rcons\_propose}}\xspace}
\newcommand{\rconsdecide}{\ensuremath{\mathsf{\rcons\_select}}\xspace}
\newcommand{\rconserror}{\ensuremath{\mathsf{\rcons\_no\_selection}}\xspace}

\newcommand{\CACValidity}{\textsc{CAC-Validity}\xspace}
\newcommand{\CACPrediction}{\textsc{CAC-Prediction}\xspace}
\newcommand{\CACNonTriviality}{\textsc{CAC-Non-triviality}\xspace}
\newcommand{\CACLocalTermination}{\textsc{CAC-Local-termination}\xspace}
\newcommand{\CACGlobalTermination}{\textsc{CAC-Global-termination}\xspace}

\newcommand{\CValidity}{\textsc{C-Validity}\xspace}
\newcommand{\CAgreement}{\textsc{C-Agreement}\xspace}
\newcommand{\CIntegrity}{\textsc{C-Integrity}\xspace}
\newcommand{\CTermination}{\textsc{C-Termination}\xspace}

\newcommand{\RCValidity}{\textsc{RC-Weak-validity-2}\xspace}
\newcommand{\RCValidityExtra}{\textsc{RC-Weak-validity-1}\xspace}
\newcommand{\RCAgreement}{\textsc{RC-Weak-agreement}\xspace}
\newcommand{\RCIntegrity}{\textsc{RC-Integrity}\xspace}
\newcommand{\RCTermination}{\textsc{RC-Termination}\xspace}

\newcommand{\SNUnicity}{\textsc{SN-Unicity}\xspace}
\newcommand{\SNAgreement}{\textsc{SN-Agreement}\xspace}
\newcommand{\SNShortNames}{\textsc{SN-Short-names}\xspace}
\newcommand{\SNTermination}{\textsc{SN-Termination}\xspace}

\newcommand{\sigendorse}{\ensuremath{\mathit{endorse\_sigs}}\xspace}
\newcommand{\sigretract}{\ensuremath{\mathit{retract\_sigs}}\xspace}
\newcommand{\sigretractsingular}{\ensuremath{\mathit{retract\_sig}}\xspace}

\newcommand{\proofset}{\boldsymbol{\pi}\hspace{-.1em}}

\newcommand{\tcons}{gcons\xspace}
\newcommand{\tconspropose}{\ensuremath{\mathsf{\tcons\_propose}}\xspace}
\newcommand{\tconsdecide}{\ensuremath{\mathsf{\tcons\_decide}}\xspace}

\newcommand{\timercc}{\ensuremath{T_{\mathit{CC}}}\xspace}
\newcommand{\timerrc}{\ensuremath{T_{\mathit{RC}}}\xspace}

\newcommand{\retract}{\ensuremath{\mathit{retract}}\xspace}

\newcommand{\initialproof}{\ensuremath{\pi}\xspace}

\newcommand{\checkdecision}{\ensuremath{\mathsf{check\_selection}}\xspace}

\newcommand{\maxfn}{\ensuremath{\mathsf{max}}\xspace}

\SetInd{0.3em}{0.4em}

\newcommand{\rememberlines}[1]{\expandafter\xdef\csname #1\endcsname{\number\value{AlgoLine}}}
\newcommand{\resumenumbering}[1]{\setcounter{AlgoLine}{\csname #1\endcsname}}

\newcommand{\OptionalElse}[1]{\Else{#1}}
\newcommand{\mylinelabel}[1]{\label{line:#1}}
\newcommand{\cdbalgoTitle}{}

\SetAlgoVlined
\SetArgSty{upshape}

\SetKwHangingKw{Init}{init:}{}{}
\SetKwProg{Operation}{operation}{ is}{}
\SetKwProg{WhenReceived}{when}{ is \receivedd do}{}
\SetKwProg{When}{when}{ do}{}
\SetKwProg{WhenReceivedFrom}{when}{ is \receivedd from $p_j$ do}{}
\SetKwProg{InternalOperation}{internal operation}{ is}{}
\SetKwProg{ForAll}{for all}{ do}{}
\SetKwIF{If}{ElseIf}{Else}{if}{then}{else if}{else}{}
\SetKwProg{Function}{function}{ is}{}
\SetKwProg{InternalOperation}{internal operation}{ is}{}

\newcommand{\cOr}{\textup{\textbf{or}}\xspace}
\newcommand{\cAnd}{\textup{\textbf{and}}\xspace}
\newcommand{\SuchThat}{\textup{\textbf{such that}}\xspace}

\SetKwComment{Comment}{$\rhd$}{}
\SetCommentSty{textit}

\DeclareMathOperator{\Forall}{\forall\,}
\DeclareMathOperator{\Exists}{\exists\,}
\DeclareMathOperator{\Nexists}{\nexists\,}

\newcounter{sbcbassumption}

\newcommand{\valid}{\ensuremath{\mathit{valid}}\xspace}

\newcommand{\sigcount}{\ensuremath{\mathit{sigcount}}\xspace}
\newcommand{\blacklist}{\ensuremath{\mathit{blacklist}}\xspace}
\newcommand{\sigs}{\ensuremath{\mathit{sigs}}\xspace}
\newcommand{\sigsI}[1]{\ensuremath{\mathit{sigs_{#1}}}\xspace}

\newcommand{\ttrue}{\ensuremath{\mathtt{true}}\xspace}
\newcommand{\ffalse}{\ensuremath{\mathtt{false}}\xspace}

\newcommand{\accepted}{\ensuremath{\mathit{accepted}}\xspace}
\newcommand{\candidates}{\ensuremath{\mathit{candidates}}\xspace}

\newcommand{\acceptedI}[1]{\ensuremath{\mathit{accepted_{#1}}}\xspace}

\newcommand{\mytextsc}[1]{{\normalfont\textsc{#1}}}

\newcommand{\witness}{\mytextsc{witness}\xspace}
\newcommand{\ready}{\mytextsc{ready}\xspace}
\newcommand{\witnessm}{\mytextsc{witnessMsg}\xspace}
\newcommand{\readym}{\mytextsc{readyMsg}\xspace}
\newcommand{\witnessms}{\mytextsc{witnessMsg}s\xspace}
\newcommand{\readyms}{\mytextsc{readyMsg}s\xspace}
\newcommand{\witm}{\witnessm\xspace}
\newcommand{\rdm}{\rdm\xspace}

\newcommand{\rdms}{\rdms\xspace}

\newcommand{\witsig}{\mytextsc{witSig}\xspace}
\newcommand{\readysig}{\mytextsc{readySig}\xspace}
\newcommand{\witsigs}{\mytextsc{witSig}s\xspace}
\newcommand{\readysigs}{\mytextsc{readySig}s\xspace}

\newcommand{\bundlem}{\mytextsc{bundle}\xspace}

\newcommand{\witcount}{\ensuremath{\mathsf{wit\_count}}\xspace}

\newcommand{\ie}{\textit{i.e.},\xspace}
\newcommand{\eg}{\textit{e.g.},\xspace}
\newcommand{\etc}{\textit{etc.}\xspace}

\newcommand{\witfunc}{\ensuremath{\mathsf{witness}}\xspace}
\newcommand{\readyfunc}{\ensuremath{\mathsf{ready}}\xspace}

\newcommand{\rconsmess}{{\textsc{rcons-sig}}\xspace}
\newcommand{\rconsretractmess}{{\textsc{rcons-retract}}\xspace}

\makeatletter
\newtheorem*{rep@theorem}{\rep@title}
\newcommand{\newreptheorem}[2]{%
\newenvironment{rep#1}[1]{%
 \def\rep@title{#2 \ref{##1}}%
 \begin{rep@theorem}}%
 {\end{rep@theorem}}}
\makeatother

\newreptheorem{theorem}{Theorem}
\newreptheorem{lemma}{Lemma}

\newtheorem{subobservation}{Observation}[theorem]

\crefname{line}{line}{lines}

\newcommand{\ccomment}[1]{}

\newcommand{\tagbox}[2]{\colorbox{#1}{\bfseries\sffamily\tiny\textcolor{white}{#2}}}

\newcommand{\annote}[3]{{\color{#3}%
    \tagbox{#3}{#2}
    $\blacktriangleright$\em #1$\blacktriangleleft$}
}

\newcommand{\sep}{\newline@@@@@@@@@@@@@@@@@@@@@@@@@@@@@@@@@@@@@@@@@@@@@@@@@@\newline}

\newcommand{\df}[1]{\annote{#1}{DF}{green}}
\newcommand{\ft}[1]{\annote{#1}{FT}{magenta}}
\newcommand{\mg}[1]{\annote{#1}{MG}{orange}}

\newcommand{\TA}[1]{\annote{#1}{TA}{blue}}

\newcommand{\todo}[1]{\annote{#1}{Todo}{red}}

\makeatletter   
\newcommand\ifnotfloat[1]{\@ifundefined{@captype}{#1}{}}
\newcommand\ifnotfloatelse[2]{\@ifundefined{@captype}{#1}{#2}}
\makeatother

\newboolean{withinSout}
\setboolean{withinSout}{false}

\newcommand{\ftasout}[1]{%
  \ifthenelse{\boolean{withinSout}}{
    #1
  }
  {%
    \setboolean{withinSout}{true}\sout{#1}\setboolean{withinSout}{false}%
  }%
}%

\newcommand{\ftanote}[2]{%
  \ifthenelse{\boolean{withinSout}}{}{%
    \ifnotfloat{\protect\marginpar{%
      \tagbox{#1}{#2}}%
    }%
  }%
}

\newrobustcmd{\hsout}[1]{\texorpdfstring{\ftasout{#1}}{#1}}

\newrobustcmd{\change}[4]{{%
    {\color{#4}#1}\ifx\relax#2\relax\else{ \color{gray}{\hsout{#2}}}\fi%
    \texorpdfstring{\ftanote{#4}{#3}}{}%
  }%
}

\newcommand{\insertft}[1]{\change{#1}{}{FT}{magenta}}

\newcommand{\insertmg}[1]{\change{#1}{}{MG}{orange}}

\renewcommand{\annote}[3]{}
\renewcommand{\change}[4]{#1}
\nolinenumbers

\title{Contention-Aware Cooperation}
\titlerunning{Contention-Aware Cooperation}

\author{Timoth\'e Albouy}{IMDEA Software Institute, Madrid, Spain}{timothe.albouy@imdea.org}{https://orcid.org/0000-0001-9419-6646}{}

\author{Davide Frey}{Univ Rennes, Inria, CNRS, IRISA, 35042 Rennes-cedex, France}{davide.frey@inria.fr}{https://orcid.org/0000-0002-6730-5744}{}

\author{Mathieu Gestin}{Université Paris-Saclay, CEA, List, F-91120, Palaiseau, France}{mathieu.gestin@cea.fr}{https://orcid.org/0009-0004-4045-6560}{}

\author{Michel Raynal}{Univ Rennes, Inria, CNRS, IRISA, 35042 Rennes-cedex, France}{michel.raynal@irisa.fr}{https://orcid.org/0000-0002-3355-8719}{}

\author{Fran\c{c}ois Ta\"{i}ani}{Univ Rennes, Inria, CNRS, IRISA, 35042 Rennes-cedex, France}{francois.taiani@irisa.fr}{https://orcid.org/0000-0002-9692-5678}{}

\authorrunning{Albouy, Frey, Gestin, Raynal, and Taïani}

\ccsdesc[100]{Theory of computation 
$\rightarrow$ Distributed algorithms}

\Copyright{Timothé Albouy, Davide Frey, Mathieu Gestin, Michel Raynal, and Fran\c{c}ois Ta\"{i}ani}

\keywords{
Agreement, Asynchronous message-passing system, Byzantine processes, Conflict detection, Consensus, Cooperation abstraction, Distributed computing, Fault tolerance, Optimistically terminating consensus, Short-naming.
}

\acknowledgements{
This work was partially supported by the French ANR projects ByBloS (ANR-20-CE25-0002-01) and PriCLeSS (ANR-10-LABX-07-81) devoted to the modular design of building blocks for large-scale Byzantine-tolerant multi-users applications, and by the SOTERIA project.
SOTERIA has received funding from the European Union's Horizon 2020 research and innovation programme under grant agreement No 101018342.
This content reflects only the author's view.
The European Agency is not responsible for any use that may be made of the information it contains.
}

\EventEditors{Andrei Arusoaie, Emanuel Onica, Michael Spear, and Sara Tucci-Piergiovanni}
\EventNoEds{4}
\EventLongTitle{29th International Conference on Principles of Distributed Systems (OPODIS 2025)}
\EventShortTitle{OPODIS 2025}
\EventAcronym{OPODIS}
\EventYear{2025}
\EventDate{December 3--5, 2025}
\EventLocation{Ia\c{s}i, Romania}
\EventLogo{}
\SeriesVolume{361}
\ArticleNo{9}

\bibliography{bibliography}
\begin{document}

\maketitle
\begin{abstract}
As shown by Reliable Broadcast and Consensus, cooperation among a set of independent computing entities (sequential processes) is crucial in fault-tolerant distributed computing. Considering $n$-process asynchronous message-passing systems where some processes may be Byzantine, this paper introduces a novel cooperation abstraction, Contention-Aware Cooperation (CAC). While Reliable Broadcast is a one-to-$n$ cooperation abstraction and Consensus is an $n$-to-$n$ cooperation abstraction, CAC is a $d$-to-$n$ cooperation abstraction where $d$ ($1 \leq d \leq n$) varies with each run and remains unknown to the processes. Correct processes accept the same set of $\ell$ pairs $\langle v,i \rangle$ ($v$ is the value proposed by $p_i$) from the $d$ proposer processes, where $1 \leq  \ell \leq d$ and (as $d$) $\ell$ remains unknown to the processes (except in specific cases). Those $\ell$ values are accepted one at a time, potentially in different orders at each process. In addition, CAC provides each process with an imperfect oracle that provides insights into the values that they may accept in the future. Interestingly, the CAC abstraction is particularly efficient in favorable circumstances, when the oracle becomes accurate, which processes can detect. To illustrate its practical utility, the paper details two applications leveraging CAC: a fast consensus implementation optimized for low contention (named Cascading Consensus), and a novel naming problem that can be solved under full asynchrony. All algorithms presented require signatures.
\end{abstract}

\mg{Attention ! Version avec commentaires. NE PAS ENLEVER CE COMMENTAIRE !
}
\section{Introduction}

Distributed computing is the science of algorithm-based cooperation.
It consists in defining (using precise specifications) and implementing distributed abstractions (distributed objects) that allow a set of computing entities (denoted processes, nodes, peers, actors, \etc) to cooperate to reach a common goal.
In the following, we use the term {\it process} to denote a sequential computing entity.
Considering asynchronous $n$-process message-passing systems, this paper focuses on cooperation abstractions that have to cope with Byzantine processes (\ie processes that may behave arbitrarily, as defined in~\cite{LSP82,PSL80}).

The gold standard of cooperation abstractions is consensus. It allows a set of processes to propose values, and to eventually agree on one of those values.
This abstraction makes it possible to implement a deterministic state machine, \ie it makes it possible for the processes to run any deterministic algorithm.
However, consensus algorithms cannot be deterministically implemented in an asynchronous distributed system in the presence of faulty processes~\cite{FLP85}.
A way to circumvent this impossibility consists in weakening one of its assumptions: weakening the full asynchrony assumption~\cite{BMR15,DDS87,DLS88}, assuming scheduling constraints, \eg~\cite{BT85,CL02, L98}, weakening determinism by allowing the processes to use random numbers, \eg~\cite{B83,MXCSS16,DRZ18}, providing the processes with as-weak-as-possible information on failures~\cite{CHT96}, or using an appropriate  combination of some of the previous weakenings. 
An issue with these solutions lies in their high cost in terms of latency and in the number of messages exchanged.
A survey of these approaches can be found in~\cite{R18}.
Another interesting approach consists in the design of optimistic algorithms that terminate quickly in predefined favorable  circumstances (fast-path) and pay a degrading  additional price (layered fast-paths)
in the other cases, \eg~\cite{BGMR01,SVW22,SR08,KTZ21}. Interestingly, the fast-path part of those algorithms usually does not rely on synchrony assumptions.
However, the optimistically terminating consensus algorithms do not separate the fast-path from the rest of their consensus algorithm.
It is incorporated in the algorithm without considerations for its specificities.
In this paper, we propose to isolate the essence of fast-path mechanisms in a standalone abstraction: the \emph{Contention-Aware Cooperation}.

\subsection{Content of the paper} \label{sec:intro:content}

This paper focuses on optimistic termination under \emph{limited contention}, \ie the ability to exploit a fast-path strategy when no or little disagreement exists in the system. This occurs for instance when only a few actions may conflict, or only a few participants are active at any given time.
To address such cases, the paper introduces a harmoniously degrading ladder of 
fast-paths and integrates them  into a novel standalone communication abstraction called 
{\it Contention-Aware Cooperation} (\CAC). This abstraction provides many-to-many communication with weak agreement capabilities that informally work as follows.
\begin{itemize}
    \item Within a \CAC instance, some arbitrary 
    number, $d$ (where $1\leq d\leq n$), of processes propose values.
    \item During the subsequent \CAC execution, each process accepts pairs, $\langle v,i\rangle$ (where $v$ is the value proposed by $p_i$), so that, eventually, all processes accept the same set of $\ell$ pairs, where $1 \leq \ell \leq d$. A process accepts pairs one after the other in some arbitrary order, which may vary from one process to another.
\end{itemize}
Both $d$ (the number of actual proposers) and $\ell$ (the number of eventually accepted values) depend on the run of the \CAC instance (\eg the (unknown) number of proposers, asynchrony, and the behavior of Byzantine processes) and are unknown to participating processes.
In particular, a process cannot generally conclude it has accepted the $\ell$ pairs composing the final set of accepted pairs (except in some specific favorable cases that will be discussed later).

To help reach agreement in favorable cases,
the \CAC abstraction endows each process with \emph{an imperfect oracle} that offers information about the set of pairs that might get accepted in the future.
This oracle is imperfect in the sense that it may produce false positives (it might return pairs that never get accepted).
The oracle cannot, however, produce false negatives: any pair excluded by the oracle will never be accepted by any correct process.

In favorable cases, the oracle's predictions allow processes to detect they have converged to the final set of accepted pairs, and that they will not accept any other pair. 
As a result, \CAC provides a weak form of agreement, and falls into an intermediate class of cooperation abstractions that are less powerful than consensus (and thus feasible even in fully asynchronous environments), but that can achieve fast-path agreement under propitious conditions~\cite{D82,FM88,YNG98,AW23}. Specifically, \CAC can dynamically adjust to the number of proposers and the conflicts between proposed values, thereby expanding the design space of existing weak-agreement abstractions.

\begin{table}[tb]
\footnotesize
\caption{Comparison of \CAC (\Cref{sec:opti-algorithm}) with existing fast paths 
of optimistic cooperation algorithms}
\vspace{-1em}
\newcommand{\threeTPlusOne}{$\mathbf{3t+1}$}%
\resizebox{\textwidth}{!}{
\begin{tabular}{|c|c||c|c|}
\hline
\textbf{Implementation} & \textbf{Optimistic condition} & \makecell{\textbf{Best}\\\textbf{latency}\footnotemark} & \makecell{\textbf{Byzantine} \\ \textbf{resilience}} \\
\hline \hline
Zyzzyva \cite{KADCW07} & Synchronous period and correct leader & $3$ & \threeTPlusOne \\
Thunderella \cite{PSNR18} & Synchronous period and correct leader & 3 & $4t+1$ \\
Parsimonious BFT \cite{RC06} & Synchronous period and correct leader &  $3$ & \threeTPlusOne \\
Fast Byzantine consensus \cite{MA06} & Synchronous period and correct leader & \textbf{2} & $5t+1$ \\
Optimal Fast Byzantine Consensus \cite{KTZ21} & Synchronous period and correct leader & \textbf{2} & $5t-1$\\
PBFT \cite{CL02}& Synchronous period and unanimity & $3$ & \threeTPlusOne \\
Algorand \cite{GHMVZ17}& Synchronous period and unanimity  &  $3$ & \threeTPlusOne \\
Optimistic Byzantine agreement \cite{P06} & \makecell{No Byzantine behaviour \\and synchronous period} & $3$ & \threeTPlusOne \\

Fault scalable BFT services \cite{AGGRW05} & Failure-free and unanimity  & 3 & $5t+1$ \\
HQ Replication \cite{CMLRS06} & Unanimity  & $4$ & \threeTPlusOne \\
Bosco \cite{SR08} & Unanimity  & \textbf{2} & $5t+1$ \\
Consensus on demand \cite{SVW22} & Unanimity  & \textbf{2} & $5t+1$ \\
Optimistically terminating consensus \cite{P06} & Correct leader and unanimity  & \textbf{2} & $5t+1$ \\
\hline
\hline
\CAC (\Cref{sec:opti-algorithm}) & Unanimity  & \textbf{2} & $5t+1$\\
\CAC (\Cref{sec:opti-algorithm})& \makecell{%
Unanimity
} & $3$ & \threeTPlusOne\\

\hline
\end{tabular}}
\label{tab:cac:parameters}
\end{table}
\footnotetext{Latency is measured in terms of
consecutive asynchronous rounds for all algorithms.}%

\subsection{The \CAC abstraction to address low contention problems}
\label{sec:intro:cac:to:implement:low:contention}
The \CAC abstraction is designed to solve distributed problems efficiently under low contention. 
This is particularly useful when implementing objects whose state is determined by the sequence of executed operations, but only a subset of these operations conflict with one-another.

When the probability of contention is low, processes can leverage \CAC{}'s imperfect oracle to detect contention and identify competing processes.
In the absence of contention or competitors, processes can terminate prematurely (fast path), and fall back on more advanced (and more costly) strategies in the remaining cases. 
This hybrid strategy is typical of fast/adaptive cooperation distributed algorithms, and ensures safety in all cases, while delivering high performance in favorable ones.
In the case of consensus, \CAC's focus on contention makes it possible to realize a consensus algorithm that can terminate optimistically even when multiple (different) values are proposed (\Cref{sec:cas-cons-tot}).
This ability contrasts with existing optimistically terminating consensus algorithms~\cite{MR00,SR08,Z06,SVW22,KKSD23}, which typically either require unanimity and/or synchrony to activate their fast-path mechanism.
To illustrate this point, \Cref{tab:cac:parameters} compares the fast-path conditions of existing optimistically terminating consensus to those of \CAC.
The numbers indicated for \CAC are those of the optimized algorithm we present in \Cref{sec:opti-algorithm}, which can terminate in three asynchronous rounds when $n>3t$ and finishes in two asynchronous rounds in the best cases if $n>5t$ (\Cref{sec:opti-algorithm}) \insertft{using signatures}.
Although \CAC does not implement consensus, its fast-path conditions directly transfer to the consensus algorithm based on \CAC that we present in \Cref{sec:cas-cons-tot}, which motivates this comparison. 
In \Cref{tab:cac:parameters}, \emph{unanimity} means that there is no contention on the proposed values. In other words, all proposing processes propose the same value.

The main results of this comparison are that our \CAC implementation is equivalent to the best \insertft{existing fast paths of} optimistically terminating consensus in favorable conditions, \ie an agreement can be reached in $2$ asynchronous rounds when $n\ge 5t+1$ and there is a unanimity of proposed values.
The capabilities of our \CAC implementation go, however, beyond this optimal best case. In particular,it can also be used to improve
the latency of intermediate cases, \ie when some contention exists but remains limited. 
For instance, the \CAC-based consensus algorithm we introduce in \Cref{sec:cas-cons-tot} (dubbed \emph{\CCons}) can still reach an agreement if less than $k$ processes endorsed the messages that do not have the majority of endorsements.\footnote{This condition is explained in \Cref{sec:opti-algorithm}, $k$ is a parameter of the algorithm.} As a result, it exhibits a ladder of escalating reconciliation strategies, with intermediate cases limiting reconciliation to the subset of conflicting processes, a capability that is of direct practical relevance when these processes happen to be geographically close to one another~\cite{BCRCVB24}. This contention management strategy is made possible thanks to the imperfect oracle of the \CAC abstraction.

\insertft{Note that \CAC's interest extends beyond consensus, in particular when there exists a deterministic back-off strategy that can be implemented under full asynchrony.} 
In that case, the \CAC abstraction can be used to construct fully asynchronous agreement algorithms, 
whereas other solutions would have required consensus.
\insertft{To illustrate this strategy, \Cref{sec:sname} introduces the \emph{short-naming problem}, a novel coordination task in which processes seek to adopt unique names with the lowest possible information-theoretic entropy. Using \CAC, we present a Byzantine-tolerant algorithm that solves that problem in a fully asynchronous network.}

\subsection{Benefits of the \CAC abstraction}

To summarize the benefits of \CAC, \insertft{this novel abstraction along with our proposed implementations} 
make it possible:
\begin{itemize}
    \item to implement algorithms with \insertft{expanded} ``graceful conditions,'' \insertft{enhancing the efficiency of fast-paths in optimistically terminating consensus algorithms;}
    \item to adjust precisely fast-path parameters \insertft{to optimally align with algorithmic requirements;}
    \item to isolate the fast-path components of consensus algorithms and implement them in isolation for enhanced modularity;
    \item to implement new and more efficient contention resolution methods when fast-path conditions are not met;
    \item to implement new asynchronous solutions to problems weaker than consensus (\eg such as the short naming problem).
\end{itemize}

\subsection{Related work}
The work described in this paper places itself in the context of {\it fast/adaptive cooperation distributed algorithms where an arbitrary, \textit{a priori} unknown subset of processes try to modify a shared object}.
These algorithms seek to terminate as rapidly as possible in favorable circumstances  (\eg no or few faults, no or little contention) and with as few as possible actions from non-participating processes while maintaining strong safety guarantees in the general case.
Such algorithms have been investigated in earlier works.

Considering shared memory systems, the reader will find more developments of this approach in~\cite{R13,T06}.

\paragraph{Fast/adaptive consensus in message-passing asynchronous crash-prone/Byzantine systems}
As stated in~\cite{L06} (which introduces the fast Paxos algorithm), the notion of fast consensus algorithm in crash-prone message-passing asynchronous system was introduced in~\cite{BGMR01}.
This algorithm was then extended to Byzantine asynchronous systems in~\cite{SR08}.
Many efficiency-oriented Byzantine consensus algorithms have since been designed (\eg~\cite{MR00,MA06,PS03,KKSD23,KTZ21} to cite a few).


\paragraph{Structuring the space of weak agreement problems}
The algorithms just discussed are specific to a single problem.
In~\cite{AW23}, Attiya and Welch go one step further and introduce a new problem termed {\it Multivalued Connected Consensus}, which \insertft{unifies} a range of weak agreement problems such as crusader agreement~\cite{D82}, graded broadcast~\cite{FM97} and adopt-commit agreement~\cite{G98}.
Differently from consensus, these agreement problems can be solved without requiring additional computational power such as synchrony constraints~\cite{BMR15}, randomization~\cite{B83}, or failure detectors~\cite{CHT96}. 

Interestingly, the decision space of these weak agreement problems can be represented as a spider graph.
Such a graph has a central clique (which can be a single vertex) and a set of $|V|$ paths (where $V$ is a finite set of totally ordered values) of length $R$.
Two asynchronous message-passing algorithms that solve Multivalued Connected Consensus are described in~\cite{AW23}.
Let $n$ be the number of processes and $t$ the maximal number of processes that can fail.
The first algorithm considers crash failures and assumes $t<n/2$, and the second considers Byzantine failures and assumes $t<n/5$.  
For both of them, the instance with $R=1$ solves crusader agreement, and the instance $R=2$ solves graded broadcast and adopt-commit.

Albeit bearing some resemblance to our CAC abstraction, these agreements are one-shot agreements with only one output, generally, either a value is decided, or the processes are informed that other processes may have decided a value, then they terminate. No two different values can be decided by a single process. Whereas the \CAC abstraction makes it possible to decide multiple values, and the oracle informs the processes about the values they may accept in the future. 

\subsection{Roadmap}
The article is structured into~\ref{sec:conclusion} sections.
First, \Cref{sec:model} introduces the computing model while \Cref{sec:CAC-definition}  provides a formal definition of the \CAC cooperation abstraction. 
Then, on the feasibility front, \Cref{sec:simple-algorithm-1} showcases a first implementation of the \CAC primitive that assumes $t<n/4$ 
\insertft{but is easy to explain and understand}.
Next, \Cref{sec:CAC-in-action} demonstrates how \CAC can be used to solve two synchronization problems of immediate practical relevance: \snaming (\Cref{sec:sname}), which provides processes with unique names while minimizing their information-theoretic entropy, and \emph{Consensus} with optimistic termination (\Cref{sec:cas-cons-tot}). The consensus implementation we present (termed \emph{\CCons}) exploits the \CAC abstraction to offer a ladder of harmoniously degrading fast paths that directly arise from the optimistic performance of our CAC implementation (\Cref{sec:opti-algorithm}) to consensus. 
Finally, \Cref{sec:conclusion} concludes the article.  
Due to page limitations, additional developments are provided in the appendices, including an implementation of \CAC which only requires $t<n/3$ 
and detailed proofs.

\section{Computing Model}
\label{sec:model}
\paragraph{Model}
The system is made up of a set $\Pi$ of $n$ processes, denoted $p_1, \cdots, p_n$, that communicate using message-passing over asynchronous channels.
``Asynchronous'' means that each message can be delayed an arbitrary finitely long time, and that processes can take an arbitrary but finitely long time to process an incoming message. However, channels are reliable, \ie no message is dropped.
Among the processes, at most $t$ are Byzantine.
A Byzantine process can arbitrarily deviate from its prescribed algorithm.
The other processes (that are at least $n-t$ and at most $n$) are called correct; they follow their prescribed algorithm and do not stop prematurely.
We assume an adversary that controls the scheduler and all Byzantine processes.
We further assume that cryptographic primitives cannot be forged, namely, we assume an unforgeable signature scheme resistant against chosen message attacks.\footnote{We conjecture that the \CAC abstraction cannot be implemented without cryptographic signatures.}

In this paper, the word {\it message} refers to messages sent by the algorithm at  the network level to implement an abstraction, they are sent and received.
The word {\it value}, on the other hand, refers to the payloads disseminated at the user level by the abstractions, they are proposed and accepted (or decided in the case of consensus).

\sloppy
Finally, the \CAC abstraction uses a best-effort (unreliable as a result of process failures) broadcast abstraction, noted \broadcast, as an underlying communication primitive.
An invocation of $\broadcast$ \textsc{msg} by a correct process $p_i$ sends the same message \textsc{msg} to all processes in $\Pi$ if not specified otherwise.\footnote{In \Cref{sec:rcons:impl}, processes $\broadcast$ messages to processes in a specific subset of $\Pi$}
We say that messages are ``\be-broadcast'' and ``received''.

\paragraph{Notations}
We denote by $\langle v_1,...,v_k \rangle$ the $k$-tuple containing the sequence of $k$ values $v_1$ to  $v_k$.
The $\star$ symbol is used as the wildcard symbol (any value can be matched).

\section{Contention-Aware Cooperation: Definition}
\label{sec:CAC-definition} 
\subsection{Definition}

The Contention-Aware Cooperation (\CAC) object provides each process $p_i$ with (1) an operation denoted $\cacpropose$ that allows it to propose a value and (2) two sets denoted $\acceptedI{i}$ and $\candidates_i$.\footnote{The $\candidates_i$ set is the imperfect oracle mentioned in this paper's introduction.\ft{CHECK: Can we move this remark directly into the main text?}\TA{I think so}}
When a process $p_i$ invokes $\cacpropose(v)$, we say that ``$p_i$ \cac-proposes (in short ``proposes'') the value $v$'' (for clarity sometimes we also say that ``$p_i$ \cac-proposes the pair $\langle v,i \rangle$'').
When a pair $\langle v,j \rangle$ is added to the set $\acceptedI{i}$ of a process $p_i$, we say that ``$p_i$ \cac-accepts (in short ``accepts'') $\langle v,j \rangle$''. For the sake of simplicity, when a pair $\langle v,j \rangle$ is \cac-accepted, a $\cacaccept(v,j)$ callback is triggered.

\begin{itemize}
    \item The set $\acceptedI{i}$ is initially empty.
    It then grows monotonically, progressively adding a pair $\langle v,j \rangle$ for each value $v$ that is \cac-accepted by $p_i$ from $p_j$.
    Eventually, $\acceptedI{i}$ contains all the pairs $\langle v,j \rangle$ accepted by the \CAC abstraction (and only them).
    
    \item The set $\candidates_i$ is initialized to $\top$, where $\top$ is defined as a symbolic value representing the identity element of the $\cap$ operation.\footnote{That is to say, for any set $S$, $S \cap \top = \top \cap S = S$, and the statement $S \subseteq \top$ is always true.}
    Then, $\candidates_i$ shrinks monotonically, and contains a dynamic estimation of the pairs $\langle v,j \rangle$ that have been or will be \cac-accepted by process $p_i$. Hence, $\acceptedI{i} \subseteq \candidates_i$ always holds.
    More concretely, $\candidates_i$ contains all the pairs $\langle v,j \rangle$ that have been already added to the $\acceptedI{i}$ set locally by $p_i$ along with some pairs $\langle v,k \rangle$ that may (or may not) be added to the set $\acceptedI{i}$ later on.
    Formally, if $\tau_1$ and $\tau_2$ are two arbitrary time points in the execution of $p_i$ (in no particular order, \ie with either $\tau_1\leq \tau_2$ or $\tau_1\geq \tau_2$) and $x_i^{\tau_k}$ represents the value of variable $x_i$ at time $\tau_k$, then $\candidates_i$ satisfies $\accepted^{\tau_2}_i \subseteq \candidates^{\tau_1}_i$.
    As a result, if a pair $\langle v,k\rangle$ is not in $\candidates_i$ at some point, $p_i$ will never \cac-accept this pair.
    Furthermore, if at some point $\tau$, $p_i$ observes $\accepted^{\tau}_i = \candidates^{\tau}_i$, then $p_i$ knows it has \cac-accepted all values for this \CAC instance.
    Let us notice that this case may never happen (see \Cref{sec:termination}).
    The behavior of both types of sets is summarized in \Cref{fig:sets-evolution,fig:sets-intersection}.
\end{itemize}
\begin{figure}[t]
\centering
\begin{minipage}{.47\textwidth}
    \centering
    \begin{tikzpicture}
\definecolor{NavyBlue}{RGB}{20, 20, 180}
\definecolor{ForestGreen}{RGB}{63, 150, 54}


\tikzmath{
\sizeCL = 2.5;
\sizeAL = .4;
\sizeCM = 1.7;
\sizeAM = .7;
\sizeCR = 1.3;
\sizeAR = 1;
\sizeLCinv = \sizeCL-.6;
\sizeLAinv = \sizeAL+.6;
\xGap = .5;
\xM = \sizeCL/2 + \xGap + \sizeCM/2;
\xR = \xM + \sizeCM/2 + \xGap + \sizeCR/2;
}


\node[
    ellipse, minimum width=\sizeLCinv cm, minimum height=\sizeLCinv cm
] (CLinvis) at (0, 0) {};
\node[
    ellipse, minimum width=\sizeLAinv cm, minimum height=\sizeLAinv cm
] (ALinvis) at (0, 0) {};

\node[
    ellipse, thick, draw=NavyBlue, fill=blue!15,
    minimum width=\sizeCL cm, minimum height=\sizeCL cm
] (CL) at (0, 0) {};

\draw[->, draw=NavyBlue] (CL.north) -- (CLinvis.north);
\draw[->, draw=NavyBlue] (CL.north east) -- (CLinvis.north east);
\draw[->, draw=NavyBlue] (CL.east) -- (CLinvis.east);
\draw[->, draw=NavyBlue] (CL.south east) -- (CLinvis.south east);
\draw[->, draw=NavyBlue] (CL.south) -- (CLinvis.south);
\draw[->, draw=NavyBlue] (CL.south west) -- (CLinvis.south west);
\draw[->, draw=NavyBlue] (CL.west) -- (CLinvis.west);
\draw[->, draw=NavyBlue] (CL.north west) -- (CLinvis.north west);

\node[
    ellipse, thick, draw=ForestGreen, fill=ForestGreen!20,
    minimum width=\sizeAL cm, minimum height=\sizeAL cm
] (AL) at (0, 0) {};

\draw[->, draw=ForestGreen] (AL.north) -- (ALinvis.north);
\draw[->, draw=ForestGreen] (AL.north east) -- (ALinvis.north east);
\draw[->, draw=ForestGreen] (AL.east) -- (ALinvis.east);
\draw[->, draw=ForestGreen] (AL.south east) -- (ALinvis.south east);
\draw[->, draw=ForestGreen] (AL.south) -- (ALinvis.south);
\draw[->, draw=ForestGreen] (AL.south west) -- (ALinvis.south west);
\draw[->, draw=ForestGreen] (AL.west) -- (ALinvis.west);
\draw[->, draw=ForestGreen] (AL.north west) -- (ALinvis.north west);

\node[text=NavyBlue] (Clabel) at (0, .625) {\scriptsize $\candidates_i$};

\node[text=ForestGreen] (Alabel) at (0, -.7) {\scriptsize $\accepted_i$};
\draw[dotted, draw=ForestGreen, thick] (Alabel) edge[bend right] (AL);


\node[
    ellipse, thick, draw=NavyBlue, fill=blue!15,
    minimum width=\sizeCM cm, minimum height=\sizeCM cm
] (CM) at (\xM, 0) {};
\node[
    ellipse, thick, draw=ForestGreen, fill=ForestGreen!20,
    minimum width=\sizeAM cm, minimum height=\sizeAM cm
] (AM) at (\xM, 0) {};


\node[
    ellipse, thick, draw=NavyBlue, fill=blue!15,
    minimum width=\sizeCR cm, minimum height=\sizeCR cm
] (CR) at (\xR, 0) {};
\node[
    ellipse, thick, draw=ForestGreen, fill=ForestGreen!20,
    minimum width=\sizeAR cm, minimum height=\sizeAR cm
] (AR) at (\xR, 0) {};


\draw[->,very thick]
    (\sizeCL/2 + \xGap/2 - .15, 0) -- (\sizeCL/2 + \xGap/2 + .15, 0);

\draw[->,very thick]
    (\xM + \sizeCM/2 + \xGap/2 - .15, 0) -- (\xM + \sizeCM/2 + \xGap/2 + .15, 0);

\end{tikzpicture}
    \vspace{-.3em}
    \caption{During an execution, the $\accepted_i$ and $\candidates_i$ sets of a correct process $p_i$ monotonically grows and shrinks, respectively.}
    \label{fig:sets-evolution}
\end{minipage}%
\hfil
\begin{minipage}{.5\textwidth}
    \centering
    \begin{tikzpicture}
\definecolor{NavyBlue}{RGB}{20, 20, 180}
\definecolor{ForestGreen}{RGB}{63, 150, 54}

\tikzmath{
\setsWidth = 1.7;
\setsHeight = 2.2;
}


\node[
    ellipse, fill=blue!15,
    minimum width=\setsWidth cm, minimum height=\setsHeight cm
] at (0, .7) {};
\node[
    ellipse, fill=blue!15, 
    minimum width=\setsWidth cm, minimum height=\setsHeight cm, rotate=-60
] at (-.4, 0) {};
\node[
    ellipse, fill=blue!15, 
    minimum width=\setsWidth cm, minimum height=\setsHeight cm, rotate=60
] at (.4, 0) {};

\node[
    ellipse, draw=NavyBlue, thick,
    minimum width=\setsWidth cm, minimum height=\setsHeight cm
] at (0, .7) {};
\node[
    ellipse, , draw=NavyBlue, thick,
    minimum width=\setsWidth cm, minimum height=\setsHeight cm, rotate=-60
] at (-.4, 0) {};
\node[
    ellipse, draw=NavyBlue, thick,
    minimum width=\setsWidth cm, minimum height=\setsHeight cm, rotate=60
] at (.4, 0) {};

\node[text=NavyBlue] at (0, 1.2) {\scriptsize $\candidates_i$};
\node[text=NavyBlue, rotate=-40] at (-.8, -.35) {\scriptsize $\candidates_j$};
\node[text=NavyBlue, rotate=40] at (.8, -.35) {\scriptsize $\candidates_k$};

\node[
    ellipse, draw=ForestGreen, fill=ForestGreen!20, thick,
    minimum width=1.1cm, minimum height=1cm
] at (0, .25) {};
\node[text=ForestGreen] at (0, .25) {\scriptsize $\accepted$};




\end{tikzpicture}
    \caption{After the execution, the $\accepted$ set is the same for all correct processes and is contained in the intersection of their $\candidates$ sets.}
    \label{fig:sets-intersection}
\end{minipage}
\end{figure}

\paragraph{\CAC specification}
Given a correct process $p_i$ and its associated $\candidates_i$ and $\acceptedI{i}$ sets, the following properties define an instance of  \CAC abstraction.%
\footnote{ It can easily be extended to a multi-shot version using execution identifiers such as sequence numbers.}
\begin{itemize}
    \item \CACValidity.
    If  $p_i$ and $p_j$ are correct, $\candidates_i \ne \top$, and  
    $\langle v,j\rangle \in  \candidates_i$, 
    
    then $p_j$ cac-proposed value $v$.
    \item \CACPrediction.
    For any correct process $p_i$ and for any process identity $k$, if, at some point of $p_i$'s execution,
    $\langle v,k\rangle \not\in \candidates_i$, then $p_i$ \emph{never} \cac-accepts $\langle v,k\rangle$ (\ie
    $\langle v,k \rangle \not\in \acceptedI{i}$ holds forever).
    \item \CACNonTriviality.
    For any correct process $p_i$, $\acceptedI{i} \neq \varnothing$ implies $\candidates_i \neq \top$.
    \item \CACLocalTermination. 
    If a correct process $p_i$ invokes $\cacpropose(v)$, its set $\acceptedI{i}$ eventually contains a pair $\langle v',\star\rangle$ (note that $v'$ is not necessarily $v$). 
    \item \CACGlobalTermination.
    If $p_i$ is a correct process and $\langle v,j \rangle \in \acceptedI{i}$,  eventually $\langle v,j\rangle\in \acceptedI{k}$ at every correct process $p_k$.
\end{itemize}
%
The \CACValidity property states that a correct process $p_i$ may include, in its $\candidates_i$ set, and possibly \cac-accept, a pair $\langle v,j \rangle$ from a correct process $p_j$, only if $p_j$ \cac-proposed value $v$, \ie only if there is no identity theft for correct processes. 
The \CACPrediction property states that, \insertft{if, at some point of $p_i$'s execution, some pair $\langle v',k\rangle$ is no longer in $\candidates_i$, then $p_i$ will never accept $\langle v',k\rangle$.}
In other words, $\candidates_i$ provides information about \insertft{which} pairs might be accepted by $p_i$ in the future.
In particular, {if a correct process $p_i$ \cac-accepts a pair $\langle v,j\rangle$, then $\langle v,j \rangle$ was present in $\candidates_i$ from the start of $p_i$'s execution.}
(However, the converse is generally not true; the prediction provided by $\candidates_i$ is, as such, imperfect.)
\ft{I've switched the two descriptions to match the new version of the property.}
This property is at the core of the cooperation provided by a \CAC object.
The \CACNonTriviality property ensures that a trivial implementation that never updates $\candidates_i$ is excluded.
As soon as some process $p_i$ has accepted some pair 
$\langle v,k\rangle$, its $\candidates_i$ set must contain some explicit information about the pairs that might get accepted in the future.\footnote{Ignoring the symbolic value $\top$, $\acceptedI{i}$ and $\candidates_i$ remain finite sets throughout $p_i$'s execution.}

The \CACLocalTermination property states that if a correct process \cac-proposes a value $v$, its $\acceptedI{i}$ set will not remain empty. 
Notice that this does not mean that the pair $\langle v,i\rangle$ will ever be added to $\acceptedI{i}$.
Finally, the \CACGlobalTermination property states that eventually, the $\accepted$ sets of all correct processes are equal.
Let us notice that, in general, no  process $p_i$ can know when no more pairs will be added to 
its set  $\acceptedI{i}$.

\subsection{Termination of the CAC abstraction} \label{sec:termination}
It follows from the definition that, for some correct process $p_i$, if $\candidates_i = \acceptedI{i}$, then $p_i$ will not \cac-accept any new pair $\langle v,j \rangle$.
Using the notations from \Cref{sec:intro:content}, we see that $\ell = |\candidates_i| = |\acceptedI{i}|$.
In this specific case, $p_i$ can detect without ambiguity that the \CAC execution has terminated \insertft{(\ie $p_i$ will not receive any other pair). $p_i$ also knows that all other correct processes will eventually receive exactly the pairs contained in $\acceptedI{i}$.}
We say that $p_i$ knows it terminated.

The most obvious example of ``known termination'' is when only one process \cac-proposes (or is perceived to \cac-propose) a value.
In this case, by \CACValidity, $|\candidates_i|$ eventually equals $1$.

However, in the general case, there might be runs where $|\candidates_i| > |\acceptedI{i}|$ during the whole execution of the abstraction.
In this case, $p_i$ will not be able to know if it has terminated or if new pairs might be added to the $\acceptedI{i}$ set.
This is an inherent feature of the \CAC abstraction, but, as we will see in \Cref{sec:CAC-in-action}, this does not prevent the abstraction from being appropriate to solve complex coordination problems.   

Another side effect of the abstraction is that, it is possible for a correct process $p_j$ to know it terminated because $\candidates_i = \acceptedI{i}$, while some other correct process $p_j$ might never detect its own termination, because $|\candidates_j| > |\acceptedI{j}|$ during the whole run, even though $p_j$ will not \cac-accept any additional value.

\subsection{\CAC with proofs of acceptance} \label{sec:proof:delivery}
The properties of the \CAC abstraction imply that processes \cac-accept pairs asynchronously and in different orders.
In some applications, 
{correct}{} processes must prove to others that they have 
legitimately \cac-accepted some pair $\langle v,j\rangle$.
To support such use cases, the \CAC definition can be enriched with transferable \emph{proofs of acceptance} that a process can use to convince others that the underlying algorithm has been respected.

When using proofs of acceptance, the elements in the $\accepted_i$ sets become triplets $\langle v, j, \pi_v\rangle$, where $\pi_v$ is a cryptographic construct that serves as proof that 
$\langle v,j \rangle$ was added to $\accepted_i$ while following the prescribed algorithm.
We say that $\pi_v$ is valid if there exists a function $\Verify$ such that, for any value $v$ and any proof of acceptance $\pi_v$ pertaining to $v$, the following property holds:
\vspace{-0.5cm}

\begin{equation}\label{eq:validity:acceptance}
   \Verify(v,\pi_v)
   =\ttrue \iff \Exists p_i \text{ correct such that, eventually, }\langle v, \star, \pi_v \rangle\in \acceptedI{i}.
\end{equation}

When $\Verify(v,\pi_v)=\ttrue$, we say that $\pi_v$ is \emph{valid}, and by extension that the triplet $\langle v, j, \pi_v \rangle$ is valid.
When using proofs of acceptance, all properties of the \CAC abstraction are modified to use $\langle v, j, \pi_v \rangle$ triplets.
In this case, the \accepted sets contain triplets
(the \cacpropose operation and the \candidates sets remain unchanged).

\section{CAC: a simple implementation}
\label{sec:simple-algorithm-1}
This section presents an implementation of the \CAC abstraction for $n>4t$. 
Our goal is to show that the \CAC abstraction can be implemented in an easy-to-understand manner.\footnote{A \CAC algorithm with $n>3t$ Byzantine resilience is presented in \Cref{sec:opti-algorithm}. This second algorithm also fulfills the proof of acceptance property.}

\Cref{alg:sb-cdb} works in two phases (the \emph{witness} phase and the \emph{ready} phase), each of them using a specific type of signature (\witsig and \readysig).
During the witness phase, $p_i$ disseminates $\witsig(p_i, \langle v,j\rangle)$ to acknowledge that a value $v$ was \cac-proposed by process $p_j$. As it is signed by $p_i$, it cannot be forged. 
During the ready phase, processes exchange \readysig signatures to collect information about potential competing values that have been \cac-proposed simultaneously, to ensure the \CACPrediction property.
A $\readysig(p_i, M_i)$ signature by $p_i$ embeds a set $M_i$ containing a critical mass of \witsig signatures.
Correct processes need to gather enough $\readysig$ signatures to construct their \candidates and \accepted sets.

{
\begin{algorithm}[ht]
\Init{$\sigs_i \gets \varnothing$; $\candidates_i \gets \top$; $\acceptedI{i} \gets \varnothing$; $M_i\gets\varnothing$.\DontPrintSemicolon}
\smallskip

\Operation{$\cacpropose(v)$}{
    \If{there are no signatures by $p_i$ in $\sigs_i$}{ \label{line:no:broadcast:verif}
        $\sigs_i \gets \sigs_i \cup \{\witsig(p_i, \langle v,i\rangle)\}$\label{line:ope-sig}\Comment*{$p_i$ signs $\langle v,i\rangle$ using a \witsig signature}
        \broadcast $\bundlem(\sigs_i)$. \label{line:ope-bcast}
    }
}
\smallskip

\WhenReceived{$\bundlem(\sigs)$}{
    $\valid_i \gets \{\text{all valid signatures in $\sigs$}\}$\;
    \lIf{$(\Exists p_k,k: \witsig(\star, \langle v_k,k\rangle) \in \valid_i \wedge \witsig(p_k, \langle v_k,k\rangle) \not\in \valid_i)$ 
    }
    {\return}\label{line:new:check:on:valid:i:witsig:initiator}
    $\sigs_i \gets \sigs_i \cup \valid_i$\; \label{line:sigs:i:updated}
    %
    \If{%
    $\Exists p_j:\witsig(p_j, \langle v,j\rangle) \in \sigs_i$ $\land$ $\witsig(p_i, \langle \star,\star\rangle) \notin \sigs_i$
    }{ \label{line:hdl:witness:condition}
        $\sigs_i \gets \sigs_i \cup \{\witsig(p_i, \choice(\{\langle v',k\rangle \mid \witsig(p_k, \langle v',k\rangle) \in \sigs_i\}))\}$ \label{line:hdl-sig-witness} \;
        \Comment{\choice chooses one of the elements in the set given as argument.}
        
        \broadcast $\bundlem(\sigs_i)$\; \label{line:hdl-witness-bcast}
    }
    \If{$|\{j \mid \witsig(p_j, \langle\star,\star\rangle)\in \sigs_i\}| \ge n-t$ $\land$ $\readym(p_i, \star) \notin \sigs_i$}{ \label{line:cond-all-witness}
        $M_i \gets \{ \witsig(\star, \langle \star, \star \rangle) \in \sigs_i\}$\; \label{line:hdl-compute-Mi}
        $\sigs_i \gets \sigs_i \cup \{\readysig(p_i, M_i)\}$\label{line:hdl-sig-ready}\Comment*{$p_i$ signs $M_i$ using a \readysig signature} 
        \broadcast $\bundlem(\sigs_i)$\; \label{line:hdl-ready-bcast} 
    }
    \If{$|\{j\mid\readysig(p_j, \star) \in \sigs_i\}|\ge n-t$}{ \label{line:cond-all-ready}
        \broadcast $\bundlem(\sigs_i)$\; \label{line:hdl-dlv-bcast}
        \If(\Comment*[f]{first time a value is accepted}){$\candidates_i=\top$}{ \label{line:if:first:time}
            $\candidates_i\gets\big\{\langle v,k\rangle\mid \Exists M: \readysig(\star, M)\in \sigs_i\wedge \witsig(\star, \langle v,k\rangle) \in M\big\}$; \label{line:update_candidates}\hspace{-5em}}
        $\acceptedI{i}{}\gets{\hspace{-0.4em}}\left\{\langle v,k \rangle \in \candidates_i%
        \,\left|\,
        \begin{array}{@{}l@{}}
        \text{$2t+1$ \emph{distinct} processes $p_s$ have signed $\readysig(\star, M_s)$}\\
        \text{in $\sigs_i$ such that $\witsig(\star, \langle v,k\rangle)\in M_s$}\\
        \end{array}{\hspace{-0.2em}}\right.\right\}{\hspace{-0.1em}}$\;\label{line:dlv}%
        \lFor{all pairs $\langle v,k \rangle$ that have just been added to $\acceptedI{i}{}$}{$\cacaccept(v,k)$.\DontPrintSemicolon}%
    }
}
\caption{One-shot sig-based \CAC implementation assuming $n>4t$ (code for $p_i$)%
}
\label{alg:sb-cdb}
\end{algorithm}
}

The algorithm works as follows.
When a process $p_i$ 
invokes $\cacpropose(v)$, it
first verifies that it has not already \cac-proposed a value, or that it did not already \broadcast any \witsig (\cref{line:no:broadcast:verif}).
If this verification passes, $p_i$ produces a \witsig for the pair $\langle v,i\rangle$, and \be-broadcasts it in a \bundlem message.
This type of message can simultaneously carry \witsig and \readysig.
As a result, each correct process disseminates its 
complete current knowledge whenever it \be-broadcasts a \bundlem message.
Eventually, this \witsig will be received by all the correct processes.

Let us consider a correct process $p_j$ that receives the \bundlem message, which contains the signature $\witsig(p_i, \langle v,i\rangle)$.
Firstly, $p_j$ checks if the initiator's signature is in the bundle (line~\ref{line:new:check:on:valid:i:witsig:initiator}) and, if so, it saves all the valid signatures into the $\sigsI{j}$ variable; otherwise, it stops processing this message as the sender is Byzantine.

Secondly, if $p_j$ did not already sign (and \be-broadcast) a \witsig, it produces a \witsig for the pair $\langle v,i\rangle$ and \be-broadcasts it (lines~\ref{line:hdl:witness:condition}-\ref{line:hdl-witness-bcast}).
If there are multiple signatures on $\langle v, \star \rangle$ in $\sigs_j$, \cref{line:hdl-sig-witness} imposes that the $p_j$ chooses and signs only one of those pairs.
Thirdly, $p_j$ checks whether it can sign and send a \readysig.
When it receives \witsig signatures from at least $n-t$ processes, $p_j$ produces a \readysig on a set of messages $M_j$ and disseminates it (lines~\ref{line:cond-all-witness}-\ref{line:hdl-ready-bcast}).
$M_j$ contains all the \witsig received by $p_j$.
This \readysig is added to the $\sigsI{j}$ set and \be-broadcast in a \bundlem message.
Hence, the information about the \witsigs known by $p_j$ will be received by every correct process along with the \readysig, thus ensuring the \CACGlobalTermination property.

Finally, $p_i$ verifies if it can \cac-accept a value.  
To this end, it waits for \readysig signatures from $n-t$ processes, then it \cac-accepts all values that are present in at least $2t+1$ sets $M$ (lines \ref{line:cond-all-ready}-\ref{line:dlv}).
The $2t+1$ bound and the assumption that $n>4t$ ensure (\Cref{lemma:pre:extended:prediction}) that, if $p_i$ \cac-accepts a value later on, then it has already been added to the $\candidates_i$ set, thus ensuring the \CACPrediction property. 
A \cacaccept callback is triggered at this point, it is used by algorithms that build upon \CAC to know when new values are added to the \accepted set.  For space reasons, the proof of correctness of Algorithm~\ref{alg:sb-cdb} is provided in \Cref{sec:proof-algo-cac-impl}.

\section{CAC in Action: Solving Low Contention Problems}
\label{sec:CAC-in-action}

The \CAC abstraction can solve cooperation problems by combining the optimistic conflict avoidance of the abstraction with a back-off strategy when conflicts occur. This section thoroughly explores two of these applications. The first one is a solution to a new naming problem, called \emph{short naming}. The naming problem makes it possible for processes to claim new names and to associate them with a public key. 
Short naming  is a variant of the naming problem where the new
names attributed to processes have low entropy. The \CAC abstraction
\insertft{naturally lends itself to such a problem, as it allows runs in which a single proposer puts forward a value and directly obtains agreement on it, thus capturing the case where a process successfully claims a unique short-name for itself.\footnote{By contrast, weak agreement primitives such as crusader agreement~\cite{D82,AW23} require all correct processes to propose a value in every execution, leading to a mismatch with the short-naming problem. Using crusader agreement in \Cref{alg:snaming} would, for instance, require a convoluted strategy in which a correct process $p_i$ first advertises its claim to some name $v$, so that other processes can support this claim by proposing the pair $(i,v)$.}}
\ft{I agree with the reviewer that this claim is exagerated.}

The second application studied is the well known consensus problem. We explore a \CAC-based solution to this problem denoted \emph{Cascading Consensus}, a new optimistically terminating consensus algorithm that ensures an early decision in favorable circumstances. Moreover, this algorithm uses information provided by the \CAC abstraction to reduce synchronization and communication complexity in case of contention. More precisely, this algorithm uses the \CAC abstraction to disseminate values. If contention occurs, \ie if termination is not guaranteed, then only the processes that proposed a value participate in the conflict resolution. This behavior is made possible thanks to the information given by the \candidates set. In that sense, this algorithm goes beyond similar existing solutions~\cite{MR00,MA06,PS03,KKSD23,KTZ21}, and, to the best of our knowledge, the \CAC abstraction is the only existing abstraction that makes it possible to implement an algorithm with such a behavior.
These examples could be extended to many other distributed applications, \eg shared account asset-transfer protocols, access control, naming services, \etc

The goal of this section is to present new ways of solving distributed problems using the \CAC abstraction. We would like to remind the reader that the main contributions of this paper are the definition, the formalisation and the implementation of the \CAC abstraction, not its applications. Furthermore, as far as we know, the behavior of this abstraction is fundamentally different from what has been proposed before. Hence, comparisons with existing work would require extended experimental analysis, which is out of the scope of this paper.

\subsection{The fault-tolerant asynchronous \snaming problem}
\label{sec:sname}
Many distributed applications, including cryptocurrency~\cite{BTC, ETH,TPPKP23}, decentralized identity management~\cite{G24, sovrin, survey-ssi}, or distributed storage~\cite{Trautwein_2022}, involve numerous participating devices that are typically identified by their public keys. For practical purposes, however, applications often choose shorter, more human-manageable names for devices. To formalize this, we define \emph{\snaming} as the problem of choosing such short human-manageable names. A \snaming object provides each process $p_i$ with one operation $n_i \leftarrow \snamingclaim(\pk,\pi)$ that allows it to claim a name $n_i$, starting from its public key, \pk, and its proof of knowledge of the associated secret key, $\pi$. 
The object also provides $p_i$ with an (initially empty) set $\Names_i$, which associates names with public keys. A $\Names_i$ set is composed of triples $\langle n, \pk, \pi\rangle$ where $n$ is the attributed name, $\pk$ is the associated public key, and $\pi$ and the proof of knowledge of the corresponding secret key. The object provides the following properties.

\begin{itemize}
    \item \SNUnicity.
    Given a correct process $p_i$, $\Forall \langle n_j, \pk_j, \pi_j \rangle, \langle n_k, \pk_k, \pi_k \rangle \in \Names_i$, either $n_j \ne n_k$ or $j=k$.
   
    \item \SNShortNames.\footnote{The function $\mcprefix()$ outputs the longest common prefix between two string, \\ \eg $\mcprefix(\mathtt{``abcdefg"}, \mathtt{``abcfed"}) = \mathtt{``abc"}$.}
    If all processes are correct, and given one correct process $p_i$, eventually we have $\Forall \langle n_j,\pk_j,\star \rangle, \langle n_k,\pk_k,\star \rangle\in \Names_i$: \\
        If $|\mcprefix(\pk_j, \pk_k)| \ge |\mcprefix(\pk_j,\pk_\ell)|, \Forall \langle \star,\pk_\ell,\star \rangle \in \Names_i$
        then $|\mcprefix(\pk_j,\pk_k)| + 1 \ge |n_j|$.
   
    \item \SNAgreement.
    Let $p_i$ and $p_j$ be two correct processes. If $\langle n,\pk,\pi \rangle \in \Names_i$ and if the process that invoked $\snamingclaim(\pk,\pi)$ is correct, then eventually $\langle n,\pk,\pi \rangle \in \Names_j$.

    \item \SNTermination.
    If a correct process $p_i$ invokes $\snamingclaim(\pk,\pi)$, then eventually $\langle \star,\pk,\star \rangle \in \Names_i$.
\end{itemize}
The \SNShortNames property captures the fact that the names given to the processes are as short as possible, thereby being easy to remember for humans. 
If there are no Byzantine processes, each name should be the smallest possible when comparing it to other attributed names.
The property only considers this difference eventually, \ie while a process might have successfully claimed a name, it may take a long time for the process it was concurring with to get its own name.

\paragraph{Existing \snaming approaches}
Existing systems follow either of two approaches, which, however, do not solve exactly the \snaming problem. 
\begin{itemize}
    \item The first approach ignores the input public keys and relies on consensus. Each process chooses its name, independently of its public key, and submits it to the consensus algorithm. In case of contention, the consensus algorithm decides which process wins in a first-come, first-served manner.
    The problem with this method is that it leverages consensus---hence requiring additional computability power (\eg partial synchrony or randomization), even if the probability of contention is low. Examples of this solution are NameCoin \cite{KCMPBN15}, Ethereum Name Service \cite{ENS}, and DNSSec \cite{DNSSec}.

    \item The second method directly uses the input public keys as the name and does not require consensus.  If the underlying cryptography is perfectly secure and secret keys are only known to their legitimate users, then the associated public keys are assumed unique, and no conflict can occur because no two processes can claim the same name.
    The problem with this method is that it does not satisfy the \SNShortNames property as public keys consist of long chains of random characters, \emph{which are hard for humans to remember}.
   Some systems circumvent the problem by using functions to map a random string to something that humans can remember: petname systems~\cite{SAKR09}, tripphrases~\cite{V08}, or Proquint IDs~\cite{W09}.
    However, these techniques do not reduce the entropy of the identifier, and they are mainly used to prevent identity theft (\eg phishing). 
\end{itemize}

\paragraph{Solving \snaming with \CAC}
Assuming perfect public/private keys, the \CAC primitive makes it possible to satisfy the \SNShortNames property of \snaming without requiring consensus.
The idea is to let processes claim sub-strings of their public keys.
For example, let a process $p$ have a public key $\mathtt{``abcdefghij"}$.
It will first claim the name $\mathtt{``a"}$ using one instance of the \CAC primitive.
If there is no conflict, \ie if the size of the \candidates set is $1$ after the first acceptance, then the name $\mathtt{``a"}$ belongs to $p$ and is associated with its public key.
On the other hand, if there is a conflict, \ie another process claimed the name $\mathtt{``a"}$ and the size of the candidate set is strictly greater than $1$ after the first acceptance, then $p$ claims the name $\mathtt{``ab"}$.
This procedure ensures that a process can always obtain a name.
Indeed, because we assume perfect cryptographic primitives, only one process knows the secret key associated with its public key.
Therefore, if $p$ conflicts with all its claims on the subsets of its public key, it will eventually claim the name $\mathtt{``abcdefghij"}$.
No other process can claim the same name and prove it knows the associated secret key. We formally describe this algorithm and prove that it solves the \snaming problem in \Cref{sec:sname:appendix}.


\subsection{A ``synchronize only when needed'' \CAC-based consensus algorithm: Cascading Consensus}
\label{sec:cas-cons-tot}

\paragraph{Consensus definition}
Consensus is a cooperation abstraction that allows a set of processes to agree on one of the values proposed by one of them. 
Consensus offers one operation \emph{propose} and one callback \emph{decide} and is defined by the following four properties.
\begin{itemize}
    \item \CValidity.
    If all processes are correct and a process decides a value $v$, then $v$ was proposed by some process.
    
    \item \CAgreement.
    No two correct processes decide different values.

    \item \CIntegrity.
    A correct process decides at most one value.
    
    \item \CTermination.
    If a correct process proposes a value $v$, then all correct processes eventually decide some value (not necessarily $v$).
\end{itemize}

Note that this definition differs slightly from the usual theoretical definition of consensus in that not all processes need to participate in all executions. This reflects what happens in practical consensus-based systems~\cite{DRZ18}, including cryptocurrencies~\cite{MXCSS16}, denylists~\cite{FGR23}, or auditable registers~\cite{ADMPA23}. In particular, the \CTermination property only guarantees termination when some correct process proposes a value.

\paragraph{A novel \CAC-based cascading consensus algorithm}
\df{I've revised the following, to account for DISC reviews PLEASE CHECK YOU AGREE}%
Building upon the \CAC abstraction, we present a new consensus algorithm, called {\it \CCons} (CC), that adopts an \emph{optimistic contention-aware} strategy to offer several fast paths for a varying set of favorable circumstances, including non-unanimous settings. This contrasts with existing optimistic consensus algorithms, which can typically only exploit one (usually unanimity). Specifically, if $n\geq 5t+1$ and there is unanimity, CC decides in 2 rounds without synchrony---as one \cac instance suffices---thus matching the best existing algorithms for this case~\cite{SR08,SVW22,P06} (\Cref{tab:cac:parameters}). If $n\geq 3t+1$ and there is unanimity, CC decides in 3 rounds without synchrony, surpassing existing optimistic algorithms that either terminate in more than 3 rounds~\cite{KADCW07,CMLRS06} and/or require synchrony to exploit their fast path~\cite{KADCW07,RC06,L06-2,CL02,GHMVZ17,P06} (\Cref{tab:cac:parameters}).
When there is no unanimity, CC further offers a progressive degradation of its fast-path, in contrast with existing unanimity-based algorithms~\cite{L06-2,CL02,GHMVZ17,AGGRW05,CMLRS06,SR08,SVW22,P06}, which fall back to a ``slow-path'' as soon as several conflicting values are proposed. Specifically, CC leverages the fact that not all processes need to propose a value to restrict conflict resolution to the set of processes that actually issued proposals. This yields two advantages. (i) These few processes are more likely to experience synchronous network phases \cite{BCRCVB24} (which are required to guarantee that a deterministic consensus algorithm terminates~\cite{DDS87,DLS88,BMR15}), and (ii) these ``restricted'' synchronous phases tend to exhibit shorter network delays, leading to overall heightened efficiency. This local approach is more efficient than full-scale consensus as validated by a recent experimental study \cite{BCRCVB24}.

%


In a nutshell, CC (\Cref{alg:ccons}, \Cref{sec:cascading-consensus}) disseminates messages over the entire system using the \CAC implementation of \Cref{sec:opti-algorithm} extended with proofs of acceptance (cf. \Cref{sec:proof:delivery}). \insertft{When the first \CAC dissemination fails to deliver a (fast) decision, CC} exploits a \emph{Restrained Consensus algorithm} (solving a relaxed consensus variant defined in \Cref{sec:restr-cons-rc}). This algorithm makes it possible for a subset $\Pi'$ of processes to agree on a set of values, and to prove to the rest of the system that all the processes in $\Pi'$ did agree on this value. Hence, it makes it possible to solve the consensus problem locally, and to inform other processes about the result of this consensus.
Both the \CAC  and  Cascading Consensus algorithms are fully asynchronous, \ie they do not require any (partial) synchrony assumptions.

Restrained Consensus, on the other hand, may fail to terminate if there are Byzantine processes or if the network behaves asynchronously (exhibiting long delays). For this reason, Cascading Consensus combines Restrained Consensus with timers \insertft{as a first fallback strategy} to resolve conflicts rapidly among a small subset of processes during favorable synchronous phases.
When circumstances are unfavorable (e.g. when network delays exceed timeouts, or under Byzantine failures), the timer expires, and CC falls back to a slow-path mode, which guarantees 
safety properties in all cases, and terminates (albeit more slowly). Note that the use of timers does not prevent CC from working under asynchrony. Timers are local, and when they expire, CC can fall back to a probabilistic asynchronous or to a partially synchronous algorithm. 

CC leverages four sub-algorithms (two \CAC instances, an instance of Restrained Consensus, and an instance of a standard consensus, which is used as fallback). It works in four steps, each step being associated with a termination condition that is more likely to be met than the previous one (as shown experimentally in \cite{BCRCVB24}).

In the first step, processes propose values using the first \CAC instance. Let $p_i$ be a process that proposed a value. If there is no contention (contention-awareness) \ie the $\candidates_i$ set of the first \CAC instance has size $1$, $p_i$ can terminate: the value proposed by $p_i$ becomes the decided value.
Otherwise, if the size of the $\candidates_i$ set is greater than $1$ after \cac-accepting the value proposed by $p_i$, $p_i$ must resolve the conflict with the other processes that proposed a value, whose pairs are in $\candidates_i$ (contention-awareness).
In this case, conflicting processes proceed to a second step, which involves an instance of the \emph{Restrained Consensus} (\RC) algorithm presented in \Cref{sec:restr-cons-rc}.
If the conflicting processes are correct and benefit from stable network delays, the \RC algorithm is guaranteed to succeed. In this case, the concerned processes disseminate the result of this step to the whole system using the second \CAC instance (third step).
If, on the other hand, some of the processes participating in the RC algorithm are Byzantine, or if messages from correct processes are delayed too much, the \RC algorithm fails.
This failure is detected by the second \CAC instance (third step), which, in this case, returns
\candidates sets with more than $1$ pair (contention-awareness).
In this case, the \CCons algorithm proceeds to its final fourth step, handing the final decision to a \emph{Global Consensus} (\GC), \ie any consensus algorithm based on additional assumptions such as partial synchrony \cite{BMR15,DDS87,DLS88}, randomization \cite{B83,MMR15}, or 
information on failures \cite{CHT96}. 
The implementation of \GC can be chosen without any constraint.
For example, if an asynchronous probabilistic consensus algorithm is chosen to instantiate \GC, then CC implements consensus under fully asynchronous assumptions.

Note that, in a single execution, not all processes necessarily perform the same number of steps. For example, some processes may accept a single value after 2 rounds, reaching a decision in the first \CAC~instance. Other processes may, instead, require additional steps to reach the same decision because their candidate set contains additional values. The \CACPrediction property guarantees that the latter processes can only accept the value accepted by the former. 

\begin{table}[tb]
\caption{Notations for the different abstractions used in this section.}
\label{tab:consensus-notations}
\vspace{-1em}
\fontsize{8}{12}\selectfont
\resizebox{\textwidth}{!}{
\begin{tabular}{|c|c|c|c|}
    \hline
    \textbf{Abstraction} & \textbf{Operations} & \textbf{Communication} & \textbf{\# participants} \\
    \hline \hline
    \makecell{Contention-Aware \\ Cooperation (\CAC)} & \makecell{$\cacpropose(v)$ \\ $\cacaccept(v,i)$} & Asynchronous & $n$ \\
    \hline
    \makecell{Cascading \\ Consensus (CC)} & \makecell{$\cconspropose(v)$ \\ $\cconsdecide(v)$} & \makecell{Async. for whole system \\ Sync. for RC} & $n$ \\
    \hline
    \makecell{Restrained \\ Consensus (\RC)} & \makecell{$\rconspropose(v)$ \\ $\rconsdecide(E,S_e,S_r)$ \\ $\rconserror()$} & Synchronous & \makecell{$\ell$ \\ (where \insertft{typically} $\ell \ll n$)} \\
    \hline
    \makecell{Global \\ Consensus (\GC)} & \makecell{$\tconspropose(v)$ \\ $\tconsdecide(v)$} & Any & $n$ \\
    \hline
\end{tabular}
}
\end{table}

\label{sec:tab:conditions}
\begin{table}[tb]
\caption{Summary of the ``progressively degrading'' conditions of Cascading Consensus (instantiated with the \CAC algorithm of \Cref{sec:opti-algorithm}), and their associated round complexity.}
\vspace{-1em}
\label{tab:conditions}
\fontsize{8}{12}\selectfont
\begin{tabular}{|c|c|c|c|c|}
    \hline
    \makecell{\textbf{Condition}} &  \makecell{\textbf{Assumpt.} \\ \textbf{needed}} & \makecell{\textbf{Execution path}} & \makecell{\textbf{Nb of system-} \\ \textbf{wide rounds}} & \makecell{\textbf{Nb of} \\ \textbf{\RC rounds}} \\
    \hline \hline
    Unanimity (fast path) & $n>5t$ & $\CAC_1$ (fast path) & 2 & N/A \\
    \hline
    \makecell{Unanimity (slow path)} & $n>3t$  & $\CAC_1$ (slow path) & 3 & N/A \\
    \hline
    \makecell{All procs of \RC \\ correct and sync.} & $n>5t$ & \makecell{$\CAC_1$; $\RC$; \\ $\CAC_2$ (fast path)} & 4 & 1 \\
    \hline
    \makecell{All procs of \RC \\ correct and sync.} & $n>3t$ & \makecell{$\CAC_1$; $\RC$; \\ $\CAC_2$ (slow path)} & 5 & 1 \\
    \hline
    \makecell{$\ge 1$ Byzantine proc. in \\ \RC or async. period} & $n>3t$ & \makecell{$\CAC_1$; $\RC$; \\ $\CAC_2$; $\GC$} & \makecell{5 + $\GC$ \\ rounds} & 1 \\
    \hline
\end{tabular}
\end{table}

\Cref{tab:conditions} summarizes the termination conditions of the CC algorithm and their associated round complexity.
The table considers two types of rounds: the fourth column counts \emph{system-wide rounds}%
---\ie for one asynchronous round, each process has to send $n$ messages.
The final column counts the asynchronous rounds executed by 
\RC---\ie for \RC round, the $\ell$ processes that execute \RC have to send a message to all other $\ell-1$ involved processes.
With fewer processes involved, the asynchronous rounds of \RC will typically be faster (measured in wall-clock time) than those of the whole network \cite{BCRCVB24}.
The ``execution path'' column details where in the algorithm a process terminates by listing the sub-algorithm instances (noted $\CAC_1$, $\RC$, $\CAC_2$, and $\GC$) that intervene in a process execution.
For instance, the first row describes the most favorable scenario in which a correct process terminates after the first \insertft{two rounds} of the first \CAC instance.\footnote{\insertft{In the table, \emph{Unanimity} refers to the unanimity of the proposers (since not all processes are required to propose in our algorithm). We have omitted an even more favorable case, in which \emph{all} correct processes propose the same value (\textit{i.e.}, there is a pre-agreement between correct processes), and Byzantine processes remain silent. In this limit case, CC saves one further round under the conditions presented in the first two rows: it decides in one round when $n>5t$ and two rounds when $n>3t$.}}

We detail the workings of the Restrained Consensus algorithm (\RC) (\Cref{sec:restr-cons-rc}), and the operations of \CCons in \Cref{sec:cascading-consensus}.

\section{Conclusion}
\label{sec:conclusion}
The paper has introduced a new cooperation abstraction denoted ``Contention-Aware Cooperation'' (CAC).
This  abstraction allows an arbitrary set of processes to propose values while multiple value acceptances are triggered.
Furthermore, each acceptance comes with information about other acceptances that can possibly occur.
This paper is the first to formalize such a cooperation abstraction.
Two implementations of \CAC have been presented.
The first one is a simple algorithm that works in asynchronous networks when $n>4t$.
The second uses fine-tuned thresholds to improve efficiency and Byzantine resilience, and to reduce the probability of contention.
This second implementation works in three asynchronous rounds if $n>3t$ and in two asynchronous rounds in favorable cases when $n>5t$.

This new cooperation abstraction can be used in low-contention distributed applications to improve efficiency or remove the need for synchronization.
The paper proposed two such examples, where the \CAC abstraction can be used to build distributed algorithms.
The first is an optimistically terminating consensus algorithm denoted \CCons. This algorithm (as some other consensus algorithms, \eg \cite{BGMR01,SVW22}), can optimistically terminate when there is no contention
or when the inputs satisfy specific patterns.
However, differently from other algorithms that do not use the \CAC abstraction, \CCons is the first to use information about contention to restrain synchronization to the processes that actually proposed a value. Furthermore, unlike other optimistically terminating consensus algorithms, cascading consensus terminates optimistically even if multiple processes propose different values.
The second example
is a short-naming algorithm, which works deterministically in fully asynchronous networks. 
It allows processes to claim shorter names based on their public keys.
However, contrary to other asynchronous naming algorithms, the claimed name is a sub-string of the public key, thus reducing the size of the name space, making it easier for humans to handle. 
This paper is the first to introduce this new solution for the naming problem, where the entropy of names can be reduced in fully asynchronous networks, where processes can be faulty.

%
More generally, the \CAC abstraction can be used to optimistically or deterministically solve other distributed cooperation problems where contention is low, \eg shared account asset-transfer protocols \cite{AFRT20,CGK20,GKMPS19}, or distributed access control mechanism~\cite{FGR23,G24}.
These applications will be explored in future work. 

Finally, another interesting direction for future work would be to design an algorithm that implements \CAC without cryptographic signatures, or to prove that such an algorithm does not exist.


\printbibliography

@article{AFRT20,
	title        = {Money Transfer Made Simple: a Specification, a Generic Algorithm, and its Proof},
	author       = {Alex Auvolat and Davide Frey and Michel Raynal and Fran{\c c}ois Ta{\"i}ani},
	year         = 2020,
	journal      = {Bulletin of EATCS},
	volume       = 132,
	pages        = {22--43}
}

@article{AGGRW05,
	title        = {Fault-Scalable {B}yzantine Fault-Tolerant Services},
	author       = {Michael Abd-El-Malek and Gregory R. Ganger and Garth R. Goodson and Michael K. Reiter and Jay J. Wylie},
	year         = 2005,
	journal      = {Proc. 20th {ACM} Symposium on Operating Systems Principles (SOSP'05)},
	volume       = 39,
	pages        = {59--74}
}

@inproceedings{AW23,
  author       = {Hagit Attiya and
                  Jennifer L. Welch},
  title        = {Multi-Valued Connected Consensus: {A} New Perspective on Crusader
                  Agreement and Adopt-Commit},
  booktitle    = {{OPODIS}},
  series       = {LIPIcs},
  volume       = {286},
  pages        = {6:1--6:23},
  publisher    = {Schloss Dagstuhl - Leibniz-Zentrum f{\"{u}}r Informatik},
  year         = {2023}
}

@inproceedings{B83,
	title        = {Another Advantage of Free Choice (Extended Abstract): Completely Asynchronous Agreement Protocols},
	author       = {Michael Ben-Or},
	year         = 1983,
	booktitle    = {Proc. 2nd ACM Symposium on Principles of Distributed Computing (PODC'83)},
	volume       = {},
	pages        = {27--30}
}

@inproceedings{BCRCVB24,
author = {Berger, Christian and Rodrigues, L\'{\i}vio and Reiser, Hans P. and Cogo, Vinicius and Bessani, Alysson},
title = {Chasing Lightspeed Consensus: Fast Wide-Area Byzantine Replication with Mercury},
year = {2024},
publisher = {Association for Computing Machinery},
doi = {10.1145/3652892.3700756},
pages = {158–171},
numpages = {14},
booktitle = {Middleware '24}
}

@inproceedings{BGMR01,
	title        = {Consensus in One Communication Step},
	author       = {Francisco Vilar Brasileiro and Fab{\'{\i}}ola Gr\'{e}ve and Achour Most{\'{e}}faoui and Michel Raynal},
	year         = 2001,
	booktitle    = {Proc. 6th International Conf. on Parallel Computing Technologies (PaCT'01)},
	publisher    = {Springer},
	series       = {LNCS 2127},
%	volume       = 2127,
	pages        = {42--50}
}

@inproceedings{BMR15,
	title        = {Minimal Synchrony for {B}yzantine Consensus},
	author       = {Z. Bouzid and A. Most{\'{e}}faoui and M. Raynal},
	year         = 2015,
	booktitle    = {Proc. 34th {ACM} Symposium on Principles of Distributed Computing (PODC'15)},
	publisher    = {{ACM}},
	pages        = {461--470}
}

@article{BT85,
	title        = {Asynchronous Consensus and Broadcast Protocols},
	author       = {Gabriel Bracha and Sam Toueg},
	year         = 1985,
	journal      = {J. {ACM}},
	volume       = 32,
	number       = 4,
	pages        = {824--840}
}

@article{CHT96,
	title        = {The Weakest Failure Detector for Solving Consensus},
	author       = {Tushar Deepak Chandra and Vassos Hadzilacos and Sam Toueg},
	year         = 1996,
	journal      = {J. {ACM}},
	volume       = 43,
	number       = 4,
	pages        = {685--722}
}

@article{CL02,
	title        = {Practical Byzantine fault tolerance and proactive recovery},
	author       = {Miguel Castro and Barbara Liskov},
	year         = 2002,
	journal      = {{ACM} Trans. Comput. Syst.},
	volume       = 20,
	number       = 4,
	pages        = {398--461}
}

@inproceedings{CMLRS06,
	title        = {{HQ} Replication: A Hybrid Quorum Protocol for {B}yzantine Fault Tolerance},
	author       = {James Cowling and Daniel Myers and Barbara Liskov and Rodrigo Rodrigues and Liuba Shrira},
	year         = 2006,
	booktitle    = {Proc. 7th Symposium on Operating Systems Design and Implementation (OSDI'06)},
	volume       = {},
	number       = {},
	pages        = {177--190}
}

@article{D82,
	title        = {The {B}yzantine Generals Strike Again},
	author       = {Danny Dolev},
	year         = 1982,
	journal      = {J. Algorithms},
	volume       = 3,
	number       = 1,
	pages        = {14--30}
}

@article{DDS87,
  author       = {Danny Dolev and
                  Cynthia Dwork and
                  Larry J. Stockmeyer},
  title        = {On the minimal synchronism needed for distributed consensus},
  journal      = {J. {ACM}},
  volume       = {34},
  number       = {1},
  pages        = {77--97},
  year         = {1987},
}

@article{DLS88,
  author       = {Cynthia Dwork and
                  Nancy A. Lynch and
                  Larry J. Stockmeyer},
  title        = {Consensus in the presence of partial synchrony},
  journal      = {J. {ACM}},
  volume       = {35},
  number       = {2},
  pages        = {288--323},
  year         = {1988},
  url          = {https://doi.org/10.1145/42282.42283},
  doi          = {10.1145/42282.42283},
  bibsource    = {dblp computer science bibliography, https://dblp.org}
}

@inproceedings{DRZ18,
	title        = {{BEAT:} Asynchronous {BFT} Made Practical},
	author       = {Sisi Duan and Michael K. Reiter and Haibin Zhang},
	year         = 2018,
	booktitle    = {Proc. 25th {ACM} {SIGSAC} Conference on Computer and Communications Security (CCS'18)},
	publisher    = {{ACM}},
	pages        = {2028--2041}
}

@inproceedings{FGR23,
	title        = {The Synchronization Power (Consensus Number) of Access-Control Objects: the Case of AllowList and DenyList},
	author       = {Davide Frey and Matthieu Gestin and Michel Raynal},
	year         = 2023,
	booktitle    = {Proc. 37th Int'l Symposium on Distributed Computing (DISC'23)},
	series       = {LIPICs},
	volume       = 281,
	pages        = {21:1--21:23}
}

@article{FLP85,
	title        = {Impossibility of Distributed Consensus with One Faulty Process},
	author       = {Michael J. Fischer and Nancy A. Lynch and Michael S. Paterson},
	year         = 1985,
	journal      = {J. ACM},
	volume       = 32,
	pages        = {374--382}
}

@article{FM97,
	title        = {An Optimal Probabilistic Protocol for Synchronous {B}yzantine Agreement},
	author       = {Pesech Feldman and Silvio Micali},
	year         = 1997,
	journal      = {{SIAM} J. Comput.},
	volume       = 26,
	number       = 4,
	pages        = {873--933}
}

@inproceedings{FM88,
  author       = {Paul Feldman and
                  Silvio Micali},
  title        = {Optimal Algorithms for Byzantine Agreement},
  booktitle    = {{STOC}},
  pages        = {148--161},
  publisher    = {{ACM}},
  year         = {1988}
}

@inproceedings{G98,
	title        = {Round-by-Round Fault Detectors: Unifying Synchrony and Asynchrony (Extended Abstract)},
	author       = {Eli Gafni},
	year         = 1998,
	booktitle    = {Proc. 17th {ACM} Symposium on Principles of Distributed Computing (PODC'98)},
	publisher    = {{ACM}},
	pages        = {143--152}
}

@inproceedings{YNG98,
  author       = {Jiong Yang and
                  Gil Neiger and
                  Eli Gafni},
  title        = {Structured Derivations of Consensus Algorithms for Failure Detectors},
  booktitle    = {{PODC}},
  pages        = {297--306},
  publisher    = {{ACM}},
  year         = {1998}
}

@phdthesis{G24,
  title={Privacy Preserving and fully-Distributed Identity Management Systems},
  author={Gestin, Mathieu},
  year={2024},
  school={Universite de rennes 1}
}

@article{GKMPS19,
	title        = {The Consensus Number of a Cryptocurrency},
	author       = {Rachid Guerraoui and Petr Kuznetsov and Matteo Monti and Matej Pavlovi{\v c} and Dragos-Adrian Seredinschi},
	year         = 2022,
	journal      = {Distributed Computing},
	volume       = 35,
	pages        = {1--15}
}

@article{KKSD23,
	title        = {Cuttlefish: Expressive Fast Path Blockchains with FastUnlock},
	author       = {Lefteris Kokoris{-}Kogias and Alberto Sonnino and George Danezis},
	year         = 2023,
	journal      = {CoRR},
	volume       = {abs/2309.12715},
	doi          = {10.48550/ARXIV.2309.12715},
	url          = {https://doi.org/10.48550/arXiv.2309.12715},
	eprinttype   = {arXiv},
	eprint       = {2309.12715}
}

@inproceedings{KTZ21,
	title        = {Revisiting optimal resilience of fast {B}yzantine consensus},
	author       = {Petr Kuznetsov and Andrei Tonkin and Yan Zang},
	year         = 2021,
	booktitle    = {Proc. 40th  ACM Symposium on Principles of Distributed Computing (PODC'21)},
	pages        = {343--353}
}

@article{L06,
	title        = {Fast {P}axos},
	author       = {Leslie Lamport},
	year         = 2006,
	journal      = {Distributed Computing},
	volume       = 19,
	pages        = {79--103}
}

@article{L06-2,
	title        = {Lower bounds for asynchronous consensus},
	author       = {Leslie Lamport},
	year         = 2006,
	journal      = {Distributed Computing},
	volume       = 19,
	pages        = {104--125}
}

@article{L98,
	title        = {The Part-Time Parliament},
	author       = {Leslie Lamport},
	year         = 1998,
	journal      = {{ACM} Trans. Comput. Syst.},
	volume       = 16,
	number       = 2,
	pages        = {133--169}
}

@article{LSP82,
	title        = {The {B}yzantine Generals Problem},
	author       = {Leslie Lamport and Robert E. Shostak and Marshall C. Pease},
	year         = 1982,
	journal      = {{ACM} Trans. Program. Lang. Syst.},
	volume       = 4,
	number       = 3,
	pages        = {382--401}
}

@article{MA06,
	title        = {Fast {B}yzantine Consensus},
	author       = {Jean{-}Philippe Martin and Lorenzo Alvisi},
	year         = 2006,
	journal      = {{IEEE} Trans. Dependable Secur. Comput.},
	volume       = 3,
	number       = 3,
	pages        = {202--215}
}

@article{MMR15,
	title        = {Signature-Free Asynchronous Binary {B}yzantine Consensus with t>n/3, O(N2) Messages, and O(1) Expected Time},
	author       = {Achour Most\'{e}faoui and Hamouma Moumen and Michel Raynal},
	year         = 2015,
	journal      = {J. ACM},
	numpages     = 21
}

@inproceedings{MR00,
	title        = {Low cost consensus-based Atomic Broadcast},
	author       = {Achour Most{\'{e}}faoui and Michel Raynal},
	year         = 2000,
	booktitle    = {Proc. 2000 Pacific Rim International Symposium on Dependable Computing (PRDC'00)},
	publisher    = {{IEEE} Computer Society},
	pages        = {45--52}
}

@inproceedings{MXCSS16,
	title        = {The Honey Badger of {BFT} Protocols},
	author       = {Andrew Miller and Yu Xia and Kyle Croman and Elaine Shi and Dawn Song},
	year         = 2016,
	booktitle    = {Proc. 23rd ACM Conference on Computer and Communications Security (CCS'16)},
	volume       = {},
	pages        = {31--42}
}

@article{PS03,
	title        = {Optimistic atomic broadcast: a pragmatic viewpoint},
	author       = {Fernando Pedone and Andr{\'{e}} Schiper},
	year         = 2003,
	journal      = {Theor. Comput. Sci.},
	volume       = 291,
	number       = 1,
	pages        = {79--101}
}

@article{PSL80,
	title        = {Reaching Agreement in the Presence of Faults},
	author       = {Marshall Pease and Robert Shostak and Leslie Lamport},
	year         = 1980,
	journal      = {J. ACM},
	volume       = 27,
	pages        = {228--234}
}

@book{R13,
	title        = {Concurrent programming: Algorithms, principles and foundations},
	author       = {Michel Raynal},
	year         = 2013,
	publisher    = {Springer},
	pages        = {1--515}
}

@book{R18,
	title        = {Fault-tolerant message-passing distributed systems: an algorithmic approach},
	author       = {Michel Raynal},
	year         = 2018,
	publisher    = {Springer},
	pages        = {1--459}


}

@inproceedings{SAKR09,
	title        = {Security Usability of Petname Systems},
	author       = {Md. Sadek Ferdous and Audun J{\o}sang and Kuldeep Singh and Ravishankar Borgaonkar},
	year         = 2009,
	booktitle    = {Identity and Privacy in the Internet Age},
	pages        = {44--59}
}

@inproceedings{SR08,
	title        = {Bosco: One-Step {B}yzantine Asynchronous Consensus},
	author       = {Yee Jiun Song and Robbert van Renesse},
	year         = 2008,
	booktitle    = {22nd Int'l Symposium on Distributed Computing},
	publisher    = {Springer},
	series       = {LNCS },
	volume       = 5218,
	pages        = {438--450}
}

@inproceedings{TPPKP23,
author = {Tonkikh, Andrei and Ponomarev, Pavel and Kuznetsov, Petr and Pignolet, Yvonne-Anne},
title = {CryptoConcurrency: (Almost) Consensusless Asset Transfer with Shared Accounts},
year = {2023},
publisher = {Association for Computing Machinery},
doi = {10.1145/3576915.3616587},
booktitle = {Proceedings of the 2023 ACM SIGSAC Conference on Computer and Communications Security},
pages = {1556–1570},
numpages = {15},
series = {CCS '23}
}

@inproceedings{SVW22,
	title        = {Consensus on Demand},
	author       = {Jakub Sliwinski and Yann Vonlanthen and Roger Wattenhofer},
	year         = 2022,
	booktitle    = {Proc. 24th Int'l Symposium on Stabilization, Safety, and Security of Distributed                           Systems},
	series       = {LNCS },
	publisher    = {Springer},
	volume       = 13751,
	pages        = {299--313}
}

@book{T06,
	title        = {Synchronization algorithms and concurrent programming},
	author       = {Gadi Taubenfeld},
	year         = 2006,
	publisher    = {Pearson Education/Prentice Hall}
}

@misc{V08,
	title        = {Tripphrases},
	author       = {Bret Victor},
	year         = 2008,
	note         = {Accessed: 2023-12-11},
	howpublished = {\url{http://worrydream.com/tripphrase/}}
}

@article{W09,
	title        = {A Proposal for {P}roquints: Identifiers that are Readable, Spellable, and Pronounceable},
	author       = {Daniel Shawcross Wilkerson},
	year         = 2009,
	journal      = {CoRR},
	volume       = {abs/0901.4016},
	url          = {http://arxiv.org/abs/09011.4016},
	eprinttype   = {arXiv},
	eprint       = {0901.4016}
}

@inproceedings{Z06,
  author       = {Piotr Zielinski},
  title        = {Optimistically Terminating Consensus: All Asynchronous Consensus Protocols
                  in One Framework},
  booktitle    = {Proc. 5th Int'l Symposium on Parallel and Distributed Computing (ISPDC'06)},
  pages        = {24--33},
  publisher    = {{IEEE} Computer Society},
  year         = {2006}
}

@inproceedings{CGK20,
  author       = {Daniel Collins and
                Rachid Guerraoui and        
                  Matteo Monti and
                  Athanasios Xygkis and
                  Matej Pavlovic and
                  Petr Kuznetsov and
                  Yvonne{-}Anne Pignolet and
                  Dragos{-}Adrian Seredinschi and  Andrei Tonkikh},
  title        = {Online Payments by Merely Broadcasting Messages}, 
  year         = {2020},
   booktitle    = {Proc. 50th Int'l Conference on Dependable Systems and Network (DSN20)},
   publisher  ={IEEE},
	pages        = {1--13}
}

@inproceedings{KCMPBN15,
  title={An Empirical Study of Namecoin and Lessons for Decentralized Namespace Design.},
  author={Kalodner, Harry A and Carlsten, Miles and Ellenbogen, Paul M and Bonneau, Joseph and Narayanan, Arvind},
  booktitle={WEIS},
  volume={1},
  pages={1--23},
  year={2015}
}

@misc{ENS,
  title={Ethereum name service},
  author={Johnson, N and Griffith, V},
  year={2018},
  publisher={Ethereum Name Service},
  comment = {online, \url{https://docs.ens.domains/}}
}

@misc{DNSSec,
    title = {DNS Security Extensions (DNSSEC)},
    author= {P. Hoffman},
    Publisher = {IETF},
    year = {2023}
}

@InProceedings{PSNR18,
author="Pass, Rafael
and Shi, Elaine",
editor="Nielsen, Jesper Buus
and Rijmen, Vincent",
title="Thunderella: Blockchains with Optimistic Instant Confirmation",
booktitle="Advances in Cryptology -- EUROCRYPT 2018 ",
year="2018",
pages="3--33"
}

@inproceedings{KADCW07,
author = {Kotla, Ramakrishna and Alvisi, Lorenzo and Dahlin, Mike and Clement, Allen and Wong, Edmund},
title = {Zyzzyva: speculative byzantine fault tolerance},
year = {2007},
doi = {10.1145/1294261.1294267},
pages = {45–58},
numpages = {14},
series = {SOSP '07}
}

@INPROCEEDINGS{P06,

  author={Zielinski, Piotr},

  booktitle={Fifth International Symposium on Parallel and Distributed Computing}, 

  title={Optimistically Terminating Consensus: All Asynchronous Consensus Protocols in One Framework}, 

  year={2006},
  pages={24-33},
  doi={10.1109/ISPDC.2006.37}
}

@InProceedings{RC06,
author="Ramasamy, HariGovind V.
and Cachin, Christian",
title="Parsimonious Asynchronous Byzantine-Fault-Tolerant Atomic Broadcast",
booktitle="Principles of Distributed Systems",
year="2006",
pages="88--102",
}

@inproceedings{GHMVZ17,
author = {Gilad, Yossi and Hemo, Rotem and Micali, Silvio and Vlachos, Georgios and Zeldovich, Nickolai},
title = {Algorand: Scaling Byzantine Agreements for Cryptocurrencies},
year = {2017},
address = {New York, NY, USA},
doi = {10.1145/3132747.3132757},
pages = {51–68},
numpages = {18},
series = {SOSP '17}
}

@techreport{sovrin,
     title = {Sovrin: A Protocol and Token for Self-Sovereign Identity and Decentralized Trust},
     institution = {Sovrin Foundation},
     author = {Sovrin Foundation},
     year = {2018},
}

@article{survey-ssi,
author = {Mühle, Alexander and Grüner, Andreas and Gayvoronskaya, Tatiana and Meinel, Christoph},
year = {2018},
month = {11},
pages = {80-86},
title = {A survey on essential components of a self-sovereign identity},
volume = {30},
journal = {Computer Science Review},
doi = {10.1016/j.cosrev.2018.10.002}
}

@inproceedings{Trautwein_2022, 
   series={SIGCOMM ’22},
   title={Design and evaluation of IPFS: a storage layer for the decentralized web},
   url={http://dx.doi.org/10.1145/3544216.3544232},
   DOI={10.1145/3544216.3544232},
   booktitle={Proceedings of the ACM SIGCOMM 2022 Conference},
   publisher={ACM},
   author={Trautwein, Dennis and Raman, Aravindh and Tyson, Gareth and Castro, Ignacio and Scott, Will and Schubotz, Moritz and Gipp, Bela and Psaras, Yiannis},
   year={2022},
   month=aug, pages={739–752},
   collection={SIGCOMM ’22} }

@techreport{BTC,
	title = {Bitcoin: A Peer-toPeer Electronic Cash System},
	author = {Satoshi Nakamoto},
	url = {https://bitcoin.org/bitcoin.pdf},
	month = {march},
	year = {2009}
}

@MISC{ETH,
author="Buterin, Vitalik",
year="2014",
title="Ethereum: A Next-Generation Smart Contract and Decentralized Application Platform."
}

@InProceedings{ADMPA23,
  author =	{Attiya, Hagit and Del Pozzo, Antonella and Milani, Alessia and Pavloff, Ulysse and Rapetti, Alexandre},
  title =	{{The Synchronization Power of Auditable Registers}},
  booktitle =	{27th International Conference on Principles of Distributed Systems (OPODIS 2023)},
  pages =	{4:1--4:23},
  series =	{Leibniz International Proceedings in Informatics (LIPIcs)},
  ISBN =	{978-3-95977-308-9},
  ISSN =	{1868-8969},
  year =	{2024},
  volume =	{286},
  editor =	{Bessani, Alysson and D\'{e}fago, Xavier and Nakamura, Junya and Wada, Koichi and Yamauchi, Yukiko},
  publisher =	{Schloss Dagstuhl -- Leibniz-Zentrum f{\"u}r Informatik},
  address =	{Dagstuhl, Germany},
  URL =		{https://drops.dagstuhl.de/entities/document/10.4230/LIPIcs.OPODIS.2023.4},
  URN =		{urn:nbn:de:0030-drops-194940},
  doi =		{10.4230/LIPIcs.OPODIS.2023.4}
}
\appendix
\section{Byzantine Resiliency Bound}
We now prove that the \CAC abstraction can only be implemented in asynchronous Byzantine-prone systems, if the number of processes, $n$, is such that $n \geq 3t+1$, $t$ being the maximum number of Byzantine processes. The proof of this result, given in \Cref{thm:3tbound}, hinges on the following lemma.

\begin{lemma}
\label{lem:acceptpropose}
Let $p_x$ and $p_y$ be two correct processes, then $p_x$ can only accept a value $v$ from $p_y$ if $p_y$ proposed this value. 
\end{lemma}
\begin{proof}
Let us assume $p_x$ accepts a value $v'$ from $p_y$ without $p_y$ having proposed $v'$.
By \CACNonTriviality we have that $\candidates_x\neq \top$. Thus, the \CACValidity property implies that if $p_y$ did not propose $v'$, then the pair $\langle  v',y \rangle \not \in \candidates_x$. But then by the \CACPrediction property, we have that $p_x$ never \cac-accepts $\langle v,k\rangle$, contradicting the initial assumption. 

\end{proof}

\begin{theorem}
\label{thm:3tbound}
There is no algorithm implementing the \cac abstraction for $n\le 3t$. 
\end{theorem}

\begin{proof}

Let us assume an algorithm $A$ implementing \CAC exists for $n\le 3t$. Let us partition the set of processes into three sets $P_1$, $P_2$, and $P_3$ with $|P_i|\le t\, \forall i \in \{1,2,3\}$. For any $i \in \{1,2,3\}$, there are executions in which all the processes in $P_i$ are Byzantine.

Now, let $E_1$ be an execution in which (i) all processes are correct, (ii)
a process $p_x\in P_3$ \cac-proposes value $v_1$, (iii) no other process
\cac-proposes any value, and (iv) all messages from $P_2$ \df{We do not
  need to and from, only from, right?} are delayed until time
$\tau_1$, while all other messages are delivered promptly. Observe
that $A$ cannot wait for protocol messages from more than $n-t$
processes. Therefore, because $n\le 3t$, $A$ cannot wait for
protocol messages from more than $2t$ processes. So by \CACValidity,
\CACPrediction, \CACLocalTermination and \CACGlobalTermination, processes in $P_1$ will
\cac-accept value $v_1$ without needing the messages from $P_2$. We can
then assume $\tau_1$ to be some time after the processes in $P_1$ have
\cac-accepted $v_1$. Moreover, by \Cref{lem:acceptpropose}, processes in $P_1$ will \cac-accept only value $v_1$ because all processes are correct and no other process \cac-proposed any other value.

Let us now consider a similar execution $E_2$ in which (i) all
processes are correct, (ii) a process $p_x\in P_3$ \cac-proposes value
$v_2$, (iii) no other process \cac-proposes any value, and (iv) all
messages from $P_1$\df{We do not need to and from, only from, right?}
are delayed until time $\tau_2$, while all other messages are
delivered promptly. Analogously to what happens in $E_1$, processes in
$P_2$ will \cac-accept value $v_2$, and only value $v_2$, without needing
the messages from $P_1$ and thus $\tau_2$ can be some time after the
processes in $P_2$ have \cac-accepted $v_2$.

Let us now consider an execution $E_{12}$ in which (i) all processes
in $P_3$ are Byzantine, (ii) no correct process \cac-proposes any value,
(iii) a process $p_x\in P_3$ acts as if it was proposing value $v_1$
to the processes in $P_1$ and as if it was proposing $v_2$ to the
processes in $P_2$. Further, let us assume that all messages from $P_2$
to $P_1$ and all messages from $P_1$ to $P_2$ are delayed by
asynchrony until time $\tau=\max(\tau_1,\tau_2)$.

Execution $E_{12}$ is indistinguishable from $E_1$ to the processes in
$P_1$ until time $\tau_1$, while it is indistinguishable from $E_2$ to the processes in
$P_2$ until time $\tau_2$. So processes in $P_1$ should \cac-accept only $v_1$ while processes
in $P_2$ should \cac-accept only $v_2$. But this contradicts Global
Termination. Hence, we cannot have $n \le 3t$.

\end{proof}

\section{Proofs of Algorithm \ref{alg:sb-cdb}~~(Non optimal CAC Algorithm)}
\label{sec:local:termination:proof}
\label{sec:proof-algo-cac-impl}

The proof that \Cref{alg:sb-cdb} is a signature-based implementation of the \CAC abstraction under the assumption $n>4t$ follows from the following lemmas. 
In the following, 
$\mathit{var}_x^{\tau}$ denotes the value of variable $\mathit{var}$ at process $p_x$ at time point $\tau$.
    
    


\begin{lemma}[\CACValidity]
If $p_i$ and $p_j$ are correct processes, $\candidates_i \ne \top$ and $\langle v,j \rangle \in \candidates_i$, then $p_j$ cac-proposed value $v$.
\end{lemma}

\begin{proof}
Let $p_i$ and $p_j$ be two correct processes.
If $\candidates_i \ne \top$, it implies $p_i$ executed the line \ref{line:update_candidates}.
Furthermore, if a tuple $\langle v,j \rangle$ is still in the $\candidates_i$ set after the execution of line \ref{line:update_candidates} by $p_i$, it means that there exists a $\witsig(p_j, \langle v, j \rangle)$ in $\sigs_i$ thanks to line \ref{line:new:check:on:valid:i:witsig:initiator}. We assume $p_j$ is correct.
Hence, due to the unforgeability assumption of cryptographic signatures, the only process able to produce such a signature is $p_j$ itself.
The only place in the algorithm where a correct process can produce such a signature is during a $\cacpropose(v)$ invocation.
\end{proof}

\begin{lemma} \label{lemma:pre:extended:prediction}
For any two correct processes $p_i$ and $p_j$, a process $p_k$ (possibly Byzantine), and a value $v_k$, if $\sigs_i$ contains $\readysig(\star, M_s)$ signatures, $s\in \{1, \ldots, 2t+1\}$, from $2t+1$ distinct processes with $\witsig(\star, \langle v_k,k\rangle)\in M_s$ for all $s$,
then $\langle v_k,k\rangle \in \candidates_j$ from the start of $p_j$'s execution.
\end{lemma}

\begin{proof}
Let $p_i$ and $p_j$ be two correct processes, $p_k$ a process (possibly Byzantine), and $v_k$ a value.
Assume that at some point $\tau$ of $p_i$'s execution $\sigs_i$ contains $\readysig(\star, M_s)$ signatures from $2t+1$ distinct processes such that $\witsig(\star, \langle v,k\rangle)\in M_s$ for all $s$, i.e.
\begin{equation} \label{pre:extended:prediction:lemma:assumption}
    \big|\big\{ p_s \,\big|\, \readysig(p_s, M_s)\in\sigs^\tau_i : \witsig(\star, \langle v_k,k\rangle)\in M_s\big\}\big|\ge 2t+1.
\end{equation}
Let $\langle v_\ell,\ell \rangle$ be the first value \cac-accepted by $p_j$ at \cref{line:dlv}.
The proof considers two cases according to two time periods: the period before $p_j$ accepts $\langle v_\ell,\ell \rangle$ (Case 1), and the period after (Case 2).

\begin{itemize}
    \item Case 1. 
    In this case, before $p_j$ executes \cref{line:update_candidates,line:dlv} for $\langle v_\ell, \ell\rangle$, $\candidates_j$ retains its initial value, namely $\top$, and by definition of $\top$, $\langle v_k,k \rangle \in \candidates_j$, the lemma holds.
    
    \item 
    Case 2. 
    Let's now turn to the value of $\candidates_j$ after $p_j$ has accepted $\langle v_\ell, \ell\rangle$. Let  $\tau_\ell$ be the time when $p_j$ adds $\langle v_\ell,\ell \rangle$ to $\acceptedI{j}$.
    We show that $\langle v_k,k\rangle \in \candidates_j^{\tau_\ell}$ when $p_j$ accepts $\langle v_\ell, \ell\rangle$.
    The proof considers two sets of processes, denoted $A$ and $B$.
\begin{itemize}
    \item $A$ is the set of processes whose $\readysig$ signatures are known to $p_j$ at time point $\tau_\ell$.
    Because of the condition at \cref{line:cond-all-ready}, to add $\langle v_\ell,\ell \rangle$ to $\accepted_j^{\tau_\ell}$, $A$ must contain at least $n-t$ processes.
    \item $B$ is the set of processes $p_s$ that have signed a $\readysig(p_s,M_s)$ signature with $\witsig(\star, \langle v_k,k \rangle)\in M_s$, so that this \readysig signature is known to $p_i$ at time point $\tau$.
    From \cref{pre:extended:prediction:lemma:assumption}, $B$ contains at least $2t+1$ distinct processes.
\end{itemize}
We have $|A \cap B| = |A|+|B|-|A \cup B| \geq (n-t)+(2t+1)-n=t+1$ processes, which means that there is at least one correct process $p_r \in A \cap B$.
Because $p_r\in A$, $p_r$ has signed $\readysig(M_r,r)$ which was received by $p_j$ by time $\tau_\ell$, hence 
$\readysig(M_r,r)\in \sigs_j^{\tau_\ell}$.
Furthermore, because $p_j$ is correct, and because $\langle v_\ell,\ell\rangle$ was the first value \cac-accepted by $p_j$, $\candidates_j$ was updated at time $\tau_\ell$ at \cref{line:update_candidates}, from which point onward the following holds:
\begin{equation} \label{eq:msg:M_r:in:candidates:j}
   \big\{ \langle v, s \rangle \;\big|\; \witsig(\star, \langle v,s\rangle)\in M_r\big\} \subseteq \candidates_j.
\end{equation}

Because $p_r\in B$, $p_r$ has signed $\readysig(p_r, M'_r)$ which was received by $p_i$ by time $\tau$ and  where $\witsig(\star,\langle v_k, k \rangle) \in M'_r$.
Because $p_r$ is correct, due to the condition at \cref{line:cond-all-witness}, it only produces at most one $\readysig(p_r,\star)$ signature, therefore $M_r=M'_r$, and $\witsig(\star,\langle v_k, k \rangle) \in M_r$.
By \cref{eq:msg:M_r:in:candidates:j}, $\langle v_k, k \rangle\in \candidates_j^{\tau_\ell}$.
Due to the condition at \cref{line:if:first:time}, $\candidates_j$ is only updated once, when $\langle v_\ell, \ell\rangle$ is accepted by $p_j$, as a result $\langle v_k, k \rangle\in \candidates_j$ after $p_j$ accepts $\langle v_\ell, \ell\rangle$, which concludes the lemma.\qedhere
\end{itemize}
\end{proof}

\begin{lemma}[Extended Prediction] \label{lemma:extended:prediction}
For any two correct processes $p_i$ and $p_j$, if $\langle v,k \rangle \in \acceptedI{i}$ then 
$\langle v,k\rangle \in \candidates_j$ from the start of $p_j$'s execution.
\end{lemma}
\begin{proof}
Let $p_i$ and $p_j$ be two correct processes.
Assume that $\langle v_k,k \rangle \in \acceptedI{i}$.
Consider the set of processes $S=\{p_s\}$ that have signed a $\readysig(\star,M_s)$ signature with $\witsig(\star,\langle v_k,k \rangle)\in M_s$, so that this \readysig signature is known to $p_i$ when it accepts $\langle v_k,k \rangle$.
By construction of $\acceptedI{i}$ at \cref{line:dlv}, $S$ contains at least $2t+1$ distinct processes.
\cref{lemma:pre:extended:prediction} applies, concluding the proof.\qedhere
\end{proof}

\begin{corollary}[\CACPrediction]
For any correct process $p_i$ and for any process identity $k$, if, at some point of its execution, $\langle v,k \rangle \notin \candidates_i$, then $p_i$ \emph{never} cac-accepts $\langle v,k\rangle$ (\ie $\langle v,k \rangle \not\in \acceptedI{i}$ holds forever).
\end{corollary}

\begin{proof}
The corollary follows from the contrapositive of \Cref{lemma:extended:prediction} when $p_j=p_i$.
\end{proof}

\begin{lemma}[\CACNonTriviality]
If process $p_i$ is   correct, $\accepted_i\neq \emptyset$ implies
$\candidates_i\neq \top$.
\end{lemma}

\begin{proof}
This is an immediate consequence of lines~\ref{line:update_candidates}-\ref{line:dlv} where, if  $\candidates_i=\top$, it is set to a non-$\top$ value before that $\accepted_i$ is updated.
\end{proof}

\begin{lemma}\label{lemma:pre:local:termination}
If a correct process $p_i$ \cac-proposes a value $v$, then each correct process $p_j$
\begin{itemize}
    \item broadcasts its own $\witsig(p_j,\langle \star,\star \rangle)$ signature in a $\bundlem$ message at \emph{\cref{line:ope-bcast}} or \emph{\ref{line:hdl-witness-bcast}};
    \item broadcasts its own $\readysig(p_j,M_j)$ signature in a $\bundlem$ message, with $|M_j|\ge n-t$, at \emph{\cref{line:hdl-ready-bcast}};
    \item eventually receives \readysig signatures from at least $n-t$ distinct processes.
\end{itemize}
\end{lemma}

\begin{proof}
Suppose $p_i$ has \cac-proposed a value $v$.
In that case, it has necessarily broadcast a $\bundlem(\sigsI{i})$ message, where $\sigsI{i}$ contains a signature $\witsig(p_i,\langle \star,\star \rangle)$, either during {the} invocation of $\cacpropose(v)$ at \cref{line:ope-bcast} (if it has not done it previously) or during the handling of a received $\bundlem$ message at \cref{line:hdl-witness-bcast}. 
All correct processes will therefore broadcast a $\bundlem$ message containing their own $\witsig(\star,\langle \star,\star \rangle)$ signature, either because they have received $p_i$'s $\bundlem(\sigsI{i})$ message, because they have received the $\bundlem$ message of another process, or because they invoked the \cacpropose operation themselves before receiving any valid \bundlem message.
As all $c \ge n-t$ correct processes broadcast their own $\witsig(\star,\langle \star,\star \rangle)$ signature in a $\bundlem$ message, all correct processes eventually receive these messages (thanks to the best effort broadcast properties and since the network is reliable) and pass the condition at \cref{line:cond-all-witness}.
This implies that all correct processes sign and broadcast their own $\readysig(\star,M)$ signature in a $\bundlem$ message 
(at \cref{line:hdl-sig-ready,line:hdl-ready-bcast}). $M$ contains the whole list of \witsig received so far, due to condition at \cref{line:cond-all-witness}, $|M|\ge n-t$.
As with \witsig signatures, these messages are eventually received by all correct processes, which eventually receive \readysig signatures from at least $n-t$ distinct processes and pass the condition at \cref{line:cond-all-ready}.
\end{proof}

\newcommand{\setC}{\mathsf{C}}
\newcommand{\setS}{\mathsf{S}}
\newcommand{\indic}{\mathds{1}}
\begin{lemma}\label{lemma:presence}
    Let $\setC$ be a set such that $|\setC| \le c$ (with $c>0$), where $c\ge 3t+1$. Let $\mathcal{S} = \{\setS_1, \ldots, \setS_c\}$ be a set of $c$ subsets of $\setC$ that each contain at least $c-t$ elements, i.e. $\forall i \in \{1, \ldots, c\}, |\setS_i| \ge c-t$. Then, there is at least one element $e\in \setC$ that appears in at least $2t+1$ sets $\setS_i$, i.e.
    \begin{equation*}
        \Exists e\in \setC: |\{\setS_i\mid e\in \setS_i\}|\ge 2t+1.
    \end{equation*}
\end{lemma}
\begin{proof}
    We prove \Cref{lemma:presence} by contradiction. Let us assume there are no element $e\in \setC$ that appears in at least $2t+1$ sets $\setS_i$. This implies that, in the best case, each element of $\setC$ appears at most in $2t$ of the sets in $\mathcal{S} = \{\setS_1, \ldots, \setS_c\}$, {\it i.e.}
    \begin{align}
        \Forall e\in \setC: &|\{\setS_i\mid e\in \setS_i\}|\le 2t,\nonumber\\
        \Forall e\in \setC: &\sum_{\setS_i\in \mathcal{S}} \indic_{\setS_i}(e)\le 2t,\label{eq:bound:freq:e:in:Si}
    \end{align}
    where $\indic_{\setS_i}$ is the indicator function for the set $S_i$, i.e.
    \begin{equation}
        \indic_{\setS_i}(e) = \left\{\begin{array}{@{}l}
            1\text{ if }e\in \setS_i,\\
            0\text{ otherwise.}
        \end{array}\right.
    \end{equation}
    For each $\setS_i \in\mathcal{S}$, we further have $\setS_i \subseteq C$ and therefore
    \begin{equation}\label{eq:card:Si:indic:func}
        |\setS_i|=\sum_{e\in C}\indic_{\setS_i}(e).
    \end{equation}
    Combining \cref{eq:card:Si:indic:func,eq:bound:freq:e:in:Si} yields
    \begin{align}
        \sum_{\setS_i\in \mathcal{S}} |\setS_i| = \sum_{\setS_i\in \mathcal{S}} \sum_{e\in C} \indic_{\setS_i}(e)
        = & \sum_{e\in C} \sum_{\setS_i\in \mathcal{S}} \indic_{\setS_i}(e)\tag{by inverting the sums}\\
        \le & \sum_{e\in C} 2t\tag{using \cref{eq:bound:freq:e:in:Si}}\\
        \le &\, c\times 2t\tag{as $|\setC|\le c$ by assumption.}
    \end{align}
    However, by lemma assumption $c\ge 3t+1$ and $\forall \setS_i\in\mathcal{S}, |\setS_i|\ge c-t$. As a result, $$\sum_{\setS_i\in \mathcal{S}} |\setS_i| \ge c(c-t) \ge c\times(2t+1).$$
    As $c>0$, the two last inequalities contradict each other, proving \Cref{lemma:presence}. 
\end{proof}

\begin{lemma}[\CACLocalTermination] \label{lemma:local:termination}
If a correct process $p_i$ \cac-propose a value $v$, then its set $\accepted_i$ eventually contains a pair $\langle v',\star\rangle$ (Note that $v'$ can be different from $v$).
\end{lemma}

\begin{proof}
Consider a correct process $p_i$ that \cac-proposes a value $v$.
\cref{lemma:pre:local:termination} applies.
As each correct process signs and sends a \readysig signature using a \bundlem message, and as \bundlem messages are disseminated using best effort broadcast, $p_i$ eventually receives the \readysig of each correct process, and by extension, it receives each of their $M_j$ sets. Without loss of generality, we assume there are $c$ correct processes, with $n\ge c \ge n-t$, and their identifiers goes from $1$ to $c$, \ie $p_1, \ldots, p_c$ are correct processes. 

By conditions at line \ref{line:cond-all-witness} and \ref{line:hdl-compute-Mi}, each $M_j$ set sent by a correct process contains at least $n-t$ \witsig, and out of those $n-t$ \witsig, at least $c-t\ge n-2t$ are \witsig signed by correct processes. Let $\setS_j$ be the set of \witsig in $M_j$ signed by correct processes, \ie $\setS_j = \{\witsig(p_k, \langle \star, \star\rangle)\mid \forall p_k \text{ correct }, \witsig(p_k, \langle \star, \star\rangle)\in M_j\}$, therefore, $|\setS_j|\ge c-t, \forall j \in \{1, \ldots,c\}$. We note $\mathcal{S} = \{\setS_1, \ldots, \setS_c\}$ the set of $\setS_j$ sets sent by correct processes. Finaly, we note $\setC=\bigcup_{k=1}^{c}\setS_k$ the set of \witsig signed by correct processes and sent in the $M_j$ sets by correct processes. As each correct process only produces one \witsig signature during an execution, $|\setC|=|\bigcup_{k=1}^{c}\setS_k|\le c$. Hence, \Cref{lemma:presence} can be applied.

Therefore, among the $c$ sets $\setS_j$ that $p_i$ eventually receives, at least one \witsig signature is present in $2t+1$ of those sets.
Therefore, the pair associated to this \witsig will eventually verify condition at line \ref{line:dlv} at $p_i$. Thus, $p_i$ will add this pair to its $\accepted_i$ set.
\end{proof}

\begin{lemma}[\CACGlobalTermination]\label{lemma:global:termination}
If, for a correct process $p_i$, $\langle v,j \rangle \in \accepted_i$, then eventually $\langle v,j \rangle \in \accepted_k$ at each correct process $p_k$.
\end{lemma}
\begin{proof}  
Consider two correct processes $p_i$ and $p_k$.
Assume $p_i$ adds $\langle v,j \rangle$ to $\accepted_i$.
By construction of \cref{line:dlv}, $p_i$ has saved the \readysig signatures of $2t+1$ processes 
\begin{equation*}
    \{\readysig(p_{i_1}, M_1),\readysig(p_{i_2}, M_2),...,\readysig(p_{i_{2t+1}}, M_{2t+1})\}
\end{equation*}
where $\witsig(p_{\ell_k}, \langle v,\ell_k \rangle) \in M_{i_k}$, for some $\{\ell_1,\ell_2,...\ell_{2t+1}\} \subseteq [1..n]$. $p_i$ \broadcastText{}s all these signatures at \cref{line:hdl-dlv-bcast}.
Let us note $R_i^{v_j}$ this set of \readysig signatures.
\begin{subobservation} \label{subobs:pk:2t+1}
$p_k$ will eventually receive the $2t+1$ signatures in $R_i^{v_j}$. 
\end{subobservation}

\begin{proof}
This trivially follows from the best effort broadcast properties and network's reliability.
\end{proof}

\begin{subobservation}\label{subobs:pk:n-t}
$p_k$ will eventually receive at least $n-t$ \readysig signatures.
\end{subobservation}

\begin{proof}
As $R_i^{v_j}$ contains the signatures of $2t+1$ distinct processes (\cref{line:dlv}), $t+1$ of these processes must be correct.
W.l.o.g, assume $p_{i_1}$ is correct.
$p_{i_1}$ has signed $\readysig(p_{i_1}, M_1)$ at \cref{line:hdl-sig-ready} with $\witsig(p_{\ell_1}, \langle v,\ell_1 \rangle) \in M_{i_1}$.
As a result, at \cref{line:hdl-compute-Mi}, 
\begin{equation}
\witsig(p_{\ell_1}, \langle v,\ell_1\rangle) \in \sigs_{i_1},
\end{equation}
which implies that $p_{i_1}$ \broadcastText{}s $\witsig(p_{\ell_1}, \langle v,\ell_1\rangle)$ at \cref{line:hdl-ready-bcast}.
As the network is reliable, all correct processes will eventually receive $\witsig(p_{\ell_1}, \langle v,\ell_1\rangle)$, and if they have not done so already will produce a \witsig signature at \cref{line:hdl-sig-witness} and \broadcastText{} it at \cref{line:hdl-witness-bcast}.
As there are at least $n-t$ correct processes, all correct processes will eventually receive $n-t$ \witsig signatures, rendering the first part of \cref{line:cond-all-witness} true.
If they have not done so already, all correct processes will therefore produce a \readysig signature at \cref{line:hdl-sig-ready}, and \broadcastText it at \cref{line:hdl-ready-bcast}.
As a result $p_k$ will eventually receive at least $n-t$ \readysig signatures
\end{proof}

\begin{subobservation} \label{subobs:cond:candidates:k:fulfilled}
There exists some $\ell\in[1...n]$ and some set $M_\ell$ of \witsig signatures such that 
\begin{itemize}
    \item eventually, $\readysig(p_\ell, M_\ell)\in\sigs_k$;
    \item and $\witsig(p_j, \langle v,j\rangle) \in M_\ell$.
\end{itemize}
\end{subobservation}

\begin{proof}
When $p_i$ accepts $\langle v,j \rangle$, \cref{line:dlv} implies that $\langle v,j \rangle\in \candidates_i$, and therefore because of \cref{line:update_candidates} that
$
    \Exists p_\ell,M_\ell: \readysig(p_\ell, M_\ell)\in \sigs_i\wedge \witsig(p_j, \langle v,j\rangle) \in M_\ell.
$
Because $p_i$ \be-broadcasts $\sigs_i$ at \cref{line:hdl-dlv-bcast}, $p_k$ will eventually receive the signatures contained in $\sigs_i$, and add then to its own $\sigs_k$ at \cref{line:sigs:i:updated}, including $\readysig(p_\ell, M_\ell)$.
\end{proof}

\begin{subobservation}
Eventually, $\langle v,j\rangle\in\accepted_k$.
\end{subobservation}

\begin{proof}
The previous observations have shown the following:
\begin{itemize}
    \item By \cref{subobs:pk:n-t}, the condition of \cref{line:cond-all-ready} eventually becomes true at $p_k$. 
    \item By \cref{subobs:cond:candidates:k:fulfilled}, $\sigs_k$ eventually contains $\readysig(p_\ell, M_\ell)\in$ for some $\ell\in[1...n]$ such that $\witsig(p_j, \langle v,j\rangle) \in M_\ell$. 
    \item By \cref{subobs:pk:2t+1}, $p_k$ eventually receives the $2t+1$ signatures in $R_i^{v_j}$. 
\end{itemize}
When the last of these three events occurs, $p_k$ passes through the condition at \cref{line:cond-all-ready}, then \cref{line:update_candidates} leads to 
\begin{equation} \label{eq:v:j:candidate:k}
    \langle v,j\rangle\in \candidates_k,
\end{equation}
and by \Cref{subobs:pk:2t+1}, the selection criteria at \cref{line:dlv} is true for $v$, which, with \Cref{eq:v:j:candidate:k}, implies that $\langle v,j \rangle \in \accepted_k$, concluding the proof of the Lemma.
\end{proof}
\end{proof} 
\section{Contention-Aware Cooperation: An Optimal Implementation}

\label{sec:opti-algorithm}

\begin{table}[ht]
\begin{center}
\begin{tabular}{|c|l|}
\hline {\bf Variable} & \multicolumn{1}{c|}{\bf Meaning}\\
\hline $\sigs_i$      & set of valid signatures known by $p_i$ \\
\hline $\sigcount_i$  & sequence number of the signatures generated by $p_i$ \\
\hline
\end{tabular}
\end{center}
\caption{\CAC algorithm parameters and variables}
\label{table:parameters}
\end{table}
\def\optimversion{}
\ifdefined\optimversion
  \renewcommand{\mylinelabel}[1]{\label{line:#1}}
  \renewcommand{\cdbalgoTitle}{One-shot optimal signature-based CAC implementation (code for $p_i$)\label{alg:cdb_impl_opti}}
  \renewcommand{\OptionalElse}[1]{{\color{red}\Else{\normalcolor{}#1}}}
\else
  \renewcommand{\mylinelabel}[1]{\label{line-no:#1}}
  \renewcommand{\cdbalgoTitle}{One-shot signature-based CAC implementation (code for $p_i$)\label{alg:cdb_impl}}
  \renewcommand{\OptionalElse}[1]{#1}
\fi

\ifdefined\optimversion
\begin{algorithm}
\caption{One-shot signature-based \CAC implementation (code for $p_i$) (Part I)\df{Double check usage of wit_count to make sure it is consistent}}
\label{alg:cdb_impl:part1}
\Init{$\accepted_i \gets \varnothing$; $\candidates_i \gets \top$; $\sigs_i \gets \varnothing$; $\blacklist_i \gets \varnothing$; $\sigcount_i \gets 0$. \DontPrintSemicolon}
\medskip

\Function{$\witcount(\langle v,j\rangle, \sigs)$}{
  $S \gets \big\{k:\witsig(p_k,\langle v,j\rangle,\star)\in\sigs \big\}$%
  \Comment*{$p_k$ has backed $\langle v, j\rangle$.}
  \return $|S|$.
}
\medskip
\Operation{$\cacpropose(v)$}{
    \If{no $\witnessm(\star)$ or $\readym(\star)$ already \be-broadcast by $p_i$}{ \label{line:verif-init}
        $\sigs_i \gets \sigs_i \cup \big\{
        \witsig(p_i,\langle v,i\rangle,\sigcount_i)
        \big\}$; $\sigcount_i\texttt{++}$\; \label{line:new:witisg:when:init:bcast}
        \broadcast $\witnessm(\sigs_i)$\label{line:verif-end}.
    }
}
\rememberlines{firstPartOfCDBAlgo}
\label{alg:cdb:part:1}
\end{algorithm}
\fi

\begin{algorithm}
\caption{\cdbalgoTitle{} (Part II).\label{alg:cdb:part:2} \df{Double check usage of wit_count to make sure it is consistent}}
\resumenumbering{firstPartOfCDBAlgo}
\WhenReceived(\Comment*[f]{invalid messages are ignored}){$\witnessm(\sigs)$\label{line:wit:received}}{
    $\sigs_i \gets \sigs_i \cup sigs$\label{line:accum:sig:when:witMsg}\;
    \If{$p_i$ has not signed any \witsig or \readysig statement yet\hspace{-0.2em}~\label{line:cond:witisg:when:new:witnessm}}{
        \mylinelabel{init-choice} $\langle v,j\rangle \gets \choice(\sigs_i)$%
        \Comment*{\choice chooses one of the elements in $\sigs_i$}
        $\sigs_i \gets \sigs_i \cup \{\witsig(p_i,\langle v,j\rangle,\sigcount_i)\}$; $\sigcount_i\texttt{++}$\; \mylinelabel{new:witisg:when:new:witnessm}
        \broadcast $\witnessm(\sigs_i)$\; \label{line:init:brd} 
    }
    \If{there are \witsig from at least $\lfloor\frac{n+t}{2}\rfloor+1$ processes in $\sigs_i$\mylinelabel{intersec-condit}}{
        \ForAll{$\langle v,j\rangle$ \SuchThat $\witcount(\langle v,j\rangle, \sigs_i)\ge 2t+k$}{ \label{line:qw-witness}
            \If{$\readysig(p_i,\langle v, j\rangle,\star) \notin \sigs_i$}{ \mylinelabel{check:no:readysig:yet:Byz:quorum}
                $\sigs_i \gets \sigs_i \cup \{\readysig(p_i,\langle v,j\rangle,\sigcount_i)\}$; $\sigcount_i\texttt{++}$\; \mylinelabel{new:readysig:Byz:quorum}
                \broadcast $\readym(\sigs_i)$\; \label{line:ready-time-one}
            }
        }
        \If{$\Exists \langle v,j\rangle: \witcount(\langle v,j\rangle,\sigs_i) \ge n-t$ \cAnd $\Forall \langle v',j'\rangle \ne \langle v,j\rangle, \witcount(\langle v',j'\rangle, sigs_i)=0$ \cAnd $n>5t$ }{ \label{line:fast-path-condition}
              \If{$\langle v,j\rangle$ has not been accepted yet}{
              $\candidates_i \gets \langle v,j\rangle$\; \label{line:candidates:witness}
              $\accepted_i \gets \{\langle v, j, \sigs_i\rangle\}$\;\label{line:fast:path:send}%
              $\cacaccept(v,j)$ \Comment*{Fast-path, no other pair $\langle v', j' \rangle \ne \langle v,j\rangle$ will be accepted.}
              }%
        }
    }
    $P \gets \big\{j \mid$ $\witsig(p_j,\langle \star,\star\rangle,\star) \in \sigs_i\big\}$\;
    \If{$|P| \geq n-t$ \cAnd $\readym(\star)$ not already broadcast by $p_i$}{ \label{line:condition:lock:begin} \label{line:lock-begin}
        \uIf{$n>5t$ \cAnd $\Exists \langle v,j\rangle :\witcount(\langle v,j\rangle, \sigs_i)\ge |P|-2t$}{ \label{line:first:unlocking:mechanism}
            \If{$\witsig(p_i,\langle v,j \rangle,\star)\notin\sigs_i$}{ \label{line:fast-path-resend:start}
                $\sigs_i \gets \sigs_i \cup \big\{\witsig(p_i, \langle v,j\rangle,\sigcount_i)\big\}$\; 
                $\sigcount_i\texttt{++}$; \label{line:fast-path-resend:new:witsig}
                \broadcast $\witnessm(\sigs_i)$\; \label{line:fast-path-resend:end}
            }
        }
    \Else{            
        $M \gets \big\{\langle v,j \rangle \mid \witsig(\star,\langle v,j \rangle,\star) \in \sigs_i\big\}$\; \mylinelabel{slow-path-resend:start}
        $T \gets \big\{\langle v,j\rangle \mid \witcount(\langle v,j\rangle, \sigs_i) \ge \maxfn(n-(|M| + 1)t,1)\}$\; 
        \ForAll{$\langle v,j\rangle\in T$ \textbf{such that} $\witsig\big(p_i,\langle v,j\rangle,\star\big)\notin \sigs_i$\mylinelabel{cond:witsig:unlocking:slow:path}}{ \mylinelabel{unlock-classique}
                $\sigs_i \gets \sigs_i \cup \{\witsig(p_i,\langle v,j\rangle,\sigcount_i)\}$\; \mylinelabel{new:witsig:unlocking:slow:path}
                $\sigcount_i\texttt{++}$;
                \broadcast $\witnessm(\sigs_i)$. \label{line:slow-path-resend:end}
        }}
    
        
    }
}
\medskip

\WhenReceived(\Comment*[f]{invalid messages are ignored}){$\readym(\sigs)$}{ \label{line:ready:received}
    \If{$\Exists \langle v',j'\rangle $ \textbf{such that} $\witcount(\langle v',j\rangle,\sigs)\ge 2t+k$}{
    \mylinelabel{check:readyMsg:valid}
    $\sigs_i \gets \sigs_i \cup \{\sigs\}$\; \label{line:accum:sig:when:readyMsg}
    \ForAll{$\langle v,j\rangle$ \SuchThat $\witcount(\langle v,j\rangle,\sigs_i)\ge 2t+k$}{ \label{line:qw-ready}
        \If{$\readysig(p_i,\langle v,j\rangle,\star)\notin\sigs_i$}{\mylinelabel{check:no:readysig:yet:receive:readymsg}
            $\sigs_i \gets \sigs_i \cup \{\readysig(p_i,\langle v,j\rangle,\sigcount_i)\}$; \label{line:new:readysig:receive:readymsg}
            $\sigcount_i\texttt{++}$\;
            \broadcast $\readym(\sigs_i)$\; \label{line:ready-time-two} 
        }
    }
    $\candidates_i \gets \candidates_i \cap \big\{\langle v,j \rangle : \witcount(\langle v,j\rangle, \sigs_i)\ge k \big\}$\; \label{line:C:fill}
    \ForAll{$\langle v,j\rangle \in \candidates_i$ \SuchThat $\big|\big\{j:\readysig(p_j,\langle v,j\rangle,\star)\in \sigs_i\big\}\big|\ge n-t$\label{line:qr-ready}}{
         $\accepted_i \gets \accepted_i \cup \{\langle v, j, \sigs_i\rangle\}$\; \label{line:cbdeliver:when:readymsg}
         $\cacaccept(v,j)$.
        }
    }
}


\end{algorithm}

\begin{figure}
    \centering
    \includegraphics[width=1\linewidth]{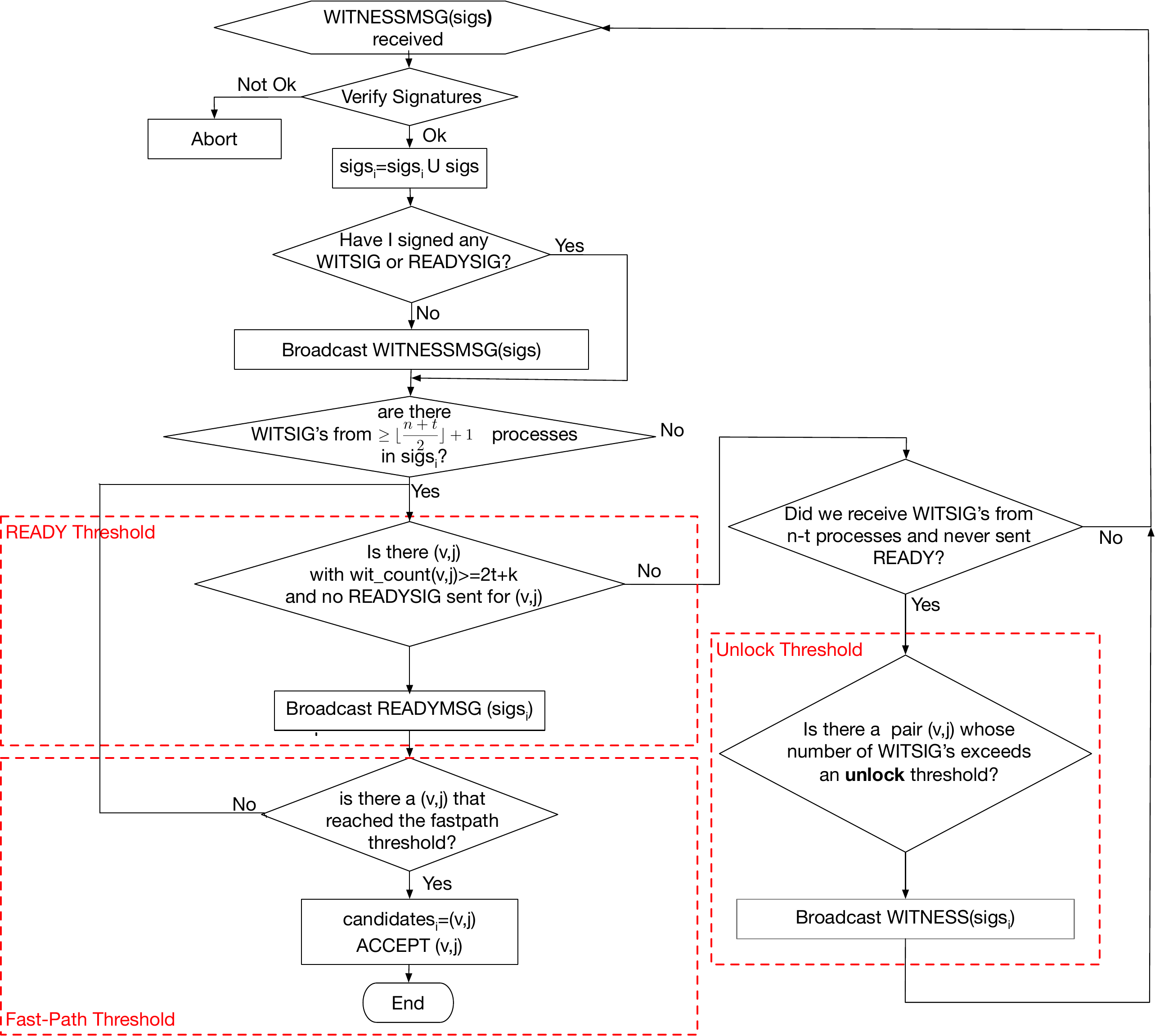}
    \caption{Flowchart of the \CAC implementation; workflow to process \witnessm messages for process $p_i$.}
    \label{fig:witness}
\end{figure}
\begin{figure}
    \centering
    \includegraphics[height=0.55\paperheight]{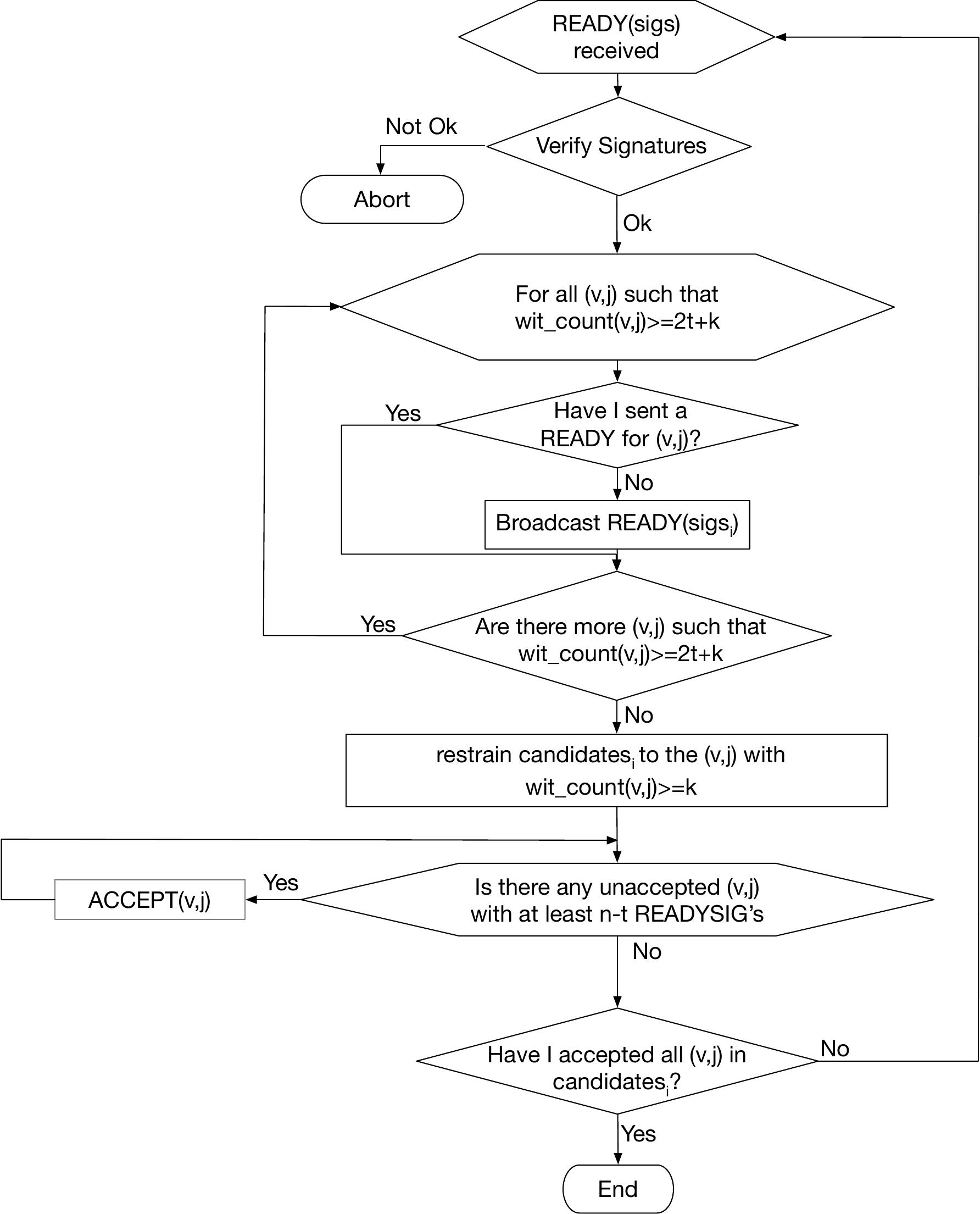}
    \caption{Flowchart of the \CAC implementation; workflow to process \readym messages for process $p_i$.}
    \label{fig:ready}
\end{figure}

\subsection{An optimal implementation of 
the \CAC abstraction}
\Cref{alg:cdb_impl:part1} and \Cref{alg:cdb_impl_opti} are the two parts of a signature-based algorithm that implements the \CAC abstraction with optimal Byzantine resilience.
Furthermore, the implementation has a good case latency of $2$ asynchronous rounds when $n>5t$ and $3$ asynchronous rounds when $n>3t$.
Those best-case latencies are optimal as we analyze in \Cref{sec:best-case-latency-proof}.
The algorithm also respects the proof of acceptance as proven by \Cref{th:proof-of-delivery}.
This optional property comes without additional cost in our implementation.
\Cref{table:parameters} summarizes the parameters and variables of the implementation and \Cref{fig:witness} and \Cref{fig:ready} is a flow-chart visually describing the algorithm. 

In the first part of the description of the algorithm, we omit the ``fast-path'' mechanism (lines \ref{line:fast-path-condition} to \ref{line:fast:path:send} and lines \ref{line:first:unlocking:mechanism} to \ref{line:fast-path-resend:end}).
Those are totally optional, for example, if $n<5t$ they are never executed. 

The algorithm works in two phases: \witness and \ready. 
During each of these phases, correct processes sign and propagate two types of ``statements'' (\witsig statements during the \witness phase, and \readysig statements during the \ready phase), using two types of messages (\witnessm messages and \readym messages).
Those statements are signatures of pairs $\langle v,j\rangle$, where $v$ is a value and $j$ is the identifier of the process that initially \cac-propose $v$ (if $p_j$ is correct).

The signed statements produced by a node $p_i$ are uniquely identified through a local sequence number $\sigcount_i$, which is incremented every time $p_i$ signs a new statement (at lines~\ref{line:new:witisg:when:init:bcast}, \ref{line:new:witisg:when:new:witnessm}, \ref{line:new:readysig:Byz:quorum}, \ref{line:new:witsig:unlocking:slow:path}, and \ref{line:new:readysig:receive:readymsg}).
When communicating with other processes, a correct process always propagates all the signed statements it has observed or produced so far.
(These statements are stored in the variable $\sigs_i$.)
To limit the power of Byzantine nodes, correct nodes only accept messages that present no ``holes'' in the sequence of statements they contain, \ie if a message \textsc{xxMsg} (\ie \witnessm or \readym) contains a statement signed by $p_j$ with sequence number $k$, then \textsc{xxMsg} must contain one statement by $p_j$ for all earlier sequence numbers $k'\in \{0,\cdots,k-1\}$ to be considered valid.
Furthermore, a valid \textsc{xxMsg} contains a signature of the pair $\langle v,j\rangle$ by $p_j$, the process that \cac-proposed the value.
Similarly, a valid message can only contain valid signatures. Invalid messages are silently dropped by correct processes (not shown in the pseudo-code for clarity).

In the first phase, processes exchange \witnessms to accumulate votes on potential pairs to accept.
A vote for a pair takes the form of a cryptographic signature on the message, which we refer to as \witsig. Each \witm can thus contain one or more \witsigs.
In the second phase, processes use \readyms to propagate cryptographic proofs that certain pairs have received
enough support/votes.
We refer to one such proof as \readysig.
Receiving a sufficient number of \readysigs triggers the \cac-acceptance.
In \Cref{alg:cdb_impl_opti}, the notation $\witsig\big(p_i,\langle v,j \rangle,s_i\big)$ stands for a \witness statement signed by the process $p_i$ with sequence number $s_i$ of value $v$ proposed by the process $p_j$.
Similarly, the notation $\readysig\big(p_i,\langle v,j\rangle,s_i\big)$ denotes a \ready statement signed by the process $p_i$ with sequence number $s_i$ of value $v$ initiated by the process $p_j$.

The algorithm relies on a parameter, $k$, which determines which pairs should enter the $\candidates_i$ set.
Specifically, a process adds a pair $\langle v,j\rangle$ to $\candidates_i$ only if it has received at least $k$ \witsigs in favor of $\langle v,j\rangle$ from $k$ different processes.
The value of $k$ strikes a balance between utility and fault tolerance.
In particular, for $k=1$, any two distinct pairs generated during an execution have a chance of being \cac-accepted and thus enter the $\candidates_i$ set, thus decreasing the probability of ``known'' termination for $p_i$, see \Cref{sec:termination}.
But in general, only pairs that $k$ distinct processes have witnessed can enter the $\candidates_i$ set.
In either case, the algorithm works for $n\geq 3t+k$.
Therefore, $k$ must be chosen by the algorithm's implementer to balance Byzantine resilience and known termination probability.

In the following, we begin by describing each of the two phases of the algorithm without the fast-path, while referencing the pseudocode in~\Cref{alg:cdb_impl_opti}.
Then, we describe the specificity of the fast path.


\subsection{{\normalfont\witness} phase}
Let us consider a correct process $p_i$ that \cac-proposes value $v$.
If $p_i$ has not yet witnessed any earlier value, it signs $\langle v,i \rangle$ and propagates the resulting \witsig to all the participants in a \witnessm (lines~\ref{line:verif-init}-\ref{line:verif-end}).
We refer to process $p_i$ as the initiator of value $v$.

When $p_i$ receives a \witnessm, it accumulates the \witsigs the message contains into its local signature set $\sigs_i$ (\cref{line:accum:sig:when:witMsg}). 
The process then checks whether it has already 
witnessed an earlier pair (\cref{line:cond:witisg:when:new:witnessm}).
If it has not, it selects one of the \witsigs is in its local signature set $\sigs_i$, and signs a new \witsig for the corresponding pair.
It then broadcasts a new \witnessm containing all \witsigs it has observed or produced so far (lines~\ref{line:init-choice} to \ref{line:init:brd}).
Because channels are reliable, this behavior ensures that all correct processes eventually witness some pair, which they propagate to the rest.

Once a process has received \witsigs from a majority of correct processes (\cref{line:intersec-condit})---the majority is ensured by the threshold $\lfloor\frac{n+t}{2}\rfloor+1$---it enters the \ready phase of the algorithm.
More precisely, it sends---if it has not done so already---a \readym for each of the pairs that have collected a quorum of $2t+k$ \witsigs in their favor (lines~\ref{line:qw-witness}-\ref{line:ready-time-one}).
The \readym contains a \readysig for the considered pair, and each of the \witsigs received so far. 
Intuitively, this \ready phase ensures that correct processes discover all the pairs that can potentially be \accepted before accepting their first pair.
(We discuss this phase in detail just below.)

However, receiving \witnessms from a majority of processes does not guarantee the presence of a pair with $2t+k$ \witsigs.
Indeed, up to this point, each correct process was only allowed to vote once.
For example, each correct process can vote for its own value it \cac-proposes.
Hence, a correct process may even receive $n-t$ \witnessm without reaching the quorum of $2t+k$ \witnessm for any pair.
When this happens, we say that the algorithm has reached a \emph{locked} state, which can be resolved using an \emph{unlocking mechanism} (lines~\ref{line:lock-begin} to~\ref{line:slow-path-resend:end}).

The first unlocking mechanism (from lines \ref{line:first:unlocking:mechanism} to \ref{line:fast-path-resend:end}) is used when the fast path may have been used and will be described in \Cref{sec:fast:path:describe}.
The second unlocking mechanism ensures that at least one pair reaches the $2t+k$ threshold at line \ref{line:qw-witness} or \ref{line:qw-ready}.
Once a process enters the unlocking mechanism, it sends a \witnessm for each pair that received at least $\maxfn\big(n- (|M|+1)t,1\big)$ \witsigs in their favor.
This threshold ensures that at least one pair reaches $2t+k$ \witsigs.
Thanks to this mechanism, all correct processes eventually broadcast at least one \readym.
\subsection{{\normalfont\ready} phase}
The \ready phase starts by sending a \readym at
\cref{line:ready-time-one}.
When a correct process, $p_i$, receives a \readym, it first checks that it indeed contains at least $2t+k$ valid signatures for a given pair (\cref{line:check:readyMsg:valid}).
If not, the message was sent by a malicious process and is thus ignored.
After this verification step, $p_i$ signs and broadcasts a \readym for all the pairs with at least $2t+k$ \witsigs.
This ensures that all correct processes eventually share the same knowledge about potentially acceptable values.

Then, process $p_i$ computes its current $\candidates_i$ set by only keeping the values that are backed by at least $k$ \witsigs.
Then, process $p_i$ \cac-accepts all the pairs in the $\candidates_i$ set that have received at least $n-t$ \witsigs.

\subsection{Fast-path} \label{sec:fast:path:describe}
We now detail the optimization of the \CAC algorithm that introduces an optimal latency path that can be followed by a process $p_i$ when $n\ge 5t+1$ and all the \witsigs that $p_i$ receives are in favor of a unique value.
This fast-path can be seen from lines~\ref{line:fast-path-condition} to \ref{line:fast:path:send} in \Cref{alg:cdb_impl_opti}. 

The optimization requires an additional condition to the algorithm.
If a process uses the fast-path for a pair $\langle v,j\rangle$, its $\candidates_i$ set only contains $\langle v,j\rangle$ (\cref{line:candidates:witness}.
Hence, no pair different from $\langle v,j \rangle$ can be accepted by any correct process to satisfy the \CACGlobalTermination and \CACPrediction properties.
Thereby, the unlocking mechanism of the algorithm is also modified.
Namely, a condition to send new \witnessms is added to the algorithm to ensure that, if a process could have taken the fast-path, then all the correct processes only send \witsigs in favor of this pair.

This mechanism (the condition at \cref{line:first:unlocking:mechanism}) is used if a process may have taken the fast-path, whereas the original mechanism at \cref{line:unlock-classique} is used in all other cases.
If a process $p_i\ne p_k$ uses the fast-path, then $n\ge 5t+1$ and it received $n-t$ signatures in favor of one pair, for example, $\langle v,j\rangle$, and no signatures in favor of $v'$.
Therefore, $p_k$ receives a minimum of $n-2t$ messages from the same processes as $p_i$, among which $t$ can have been sent by Byzantine processes.
Hence, $p_k$ receives at least $n-3t$ messages in favor of $v$, and $t$ in favor of $v'$ among the first $n-t$ \witnessms it receives.
Furthermore, if $p_k$ received $t$ messages from Byzantine processes, it means that it can still receive messages from $t$ correct processes.
If $p_i$ did use the fast-path, those new messages will back $v$.
Therefore, if at least $n-3t\le |P|-2t\le n-2t$---where $|P|$ is the number of unique processes from which $p_i$ received \witsigs---\witnessm received by $p_k$ are in favor of a unique pair $\langle v,j\rangle$, then another correct process $p_i$ may have taken the fast path.
Furthermore, when a correct process does use the fast-path for a pair $\langle v,j\rangle$, it accepts it along with a \candidates set containing only the pair $\langle v, j \rangle$.
In other words, if a process did use the fast-path, then no other pair should be \accepted.
Therefore, if a correct process is in a locked state, and if it detects that a process might have taken the fast-path for a pair $\langle v,j\rangle$, it should only send new \witsigs in favor of $\langle v,j\rangle$.
If every correct process detects that a process might have taken the fast path, then every process that did not vote in favor of $\langle v,j\rangle$ will do so.
Therefore, each correct process will receive at least $2t+k$ \witsigs in favor of $v$ and will send a \readym in its favor, and no other pair will reach the $2t+k$ threshold.

\subsection{Proof of the algorithm}

We now prove that \Cref{alg:cdb_impl_opti} is a valid implementation of the \CAC abstraction.
The algorithm is proven for $n \ge 3t+k \ge 3t+1$.

The different lemmas used to prove \Cref{alg:cdb_impl_opti} use the following notations:
Let $\sigs_i^{\tau}$ be the set $\sigs_i$ of the process $p_i$ at time $\tau$.  
Let $\accepted_i^{\tau}$ be the state of the set $\accepted_i$ at time ${\tau}$.
Let $\candidates^{\tau}$ be the state of the set $\candidates_i$ at time ${\tau}$.
Let $\witsig(\sigs_i^{{\tau}}, v)$ be the \witsigs relative to $v$ in $\sigs_i^{{\tau}}$, and let $\readysig(\sigs_i^{{\tau}}, v)$ be the \readysigs relative to $v$ in $\sigs_i^{{\tau}}$.
Let $\mathsf{max}(\sigs_i^{{\tau}}, p_j)$ be the \readysig or \witsig with the greatest $\sigcount$ from process $p_j$ in $\sigs_i^{{\tau}}$.

\begin{lemma}[\CACValidity] \label{lemma:brd-validity}
If $p_i$ and $p_j$ are correct, $\candidates_i \ne \top$ and $\langle v,j\rangle \in  \candidates_i$, then $p_j$ cac-proposed value $v$.
\end{lemma}

\begin{proof}
Let $p_i$ and $p_j$ be two correct processes $p_i$ and let $\langle v_j, j \rangle \in \candidates_i^\tau$ for some time $\tau$.
Furthermore, let us assume $\candidates_i \ne \top$. Hence, $\candidates_i$ has been modified by $p_i$.
There are only two lines in \Cref{alg:cdb_impl_opti} where $p_i$ can modify $\candidates_i$.
Either it did it at \cref{line:C:fill} and it received $k$ \witsigs backing $\langle v_j, j \rangle$, or it modified it at \cref{line:candidates:witness}, and $n > 5t$ and $p_i$ received at least $n-t$ \witsigs backing $\langle v_j, j \rangle$ and no \witsig backing another pair (\cref{line:fast-path-condition}). 
    
In both cases, $p_i$ considers the pair $\langle v_j, j \rangle$ to be valid only if $\witsig(p_j, \langle v_j, j \rangle, \star)\in \sigs_i$\footnote{This condition is an implicit condition stated in the description of the algorithm and assumed by the comment at lines \ref{line:wit:received} and \ref{line:ready:received}.}, \ie there exist a \witsig in $\sigs_i$ from the proposer of the value.
In both cases, there is a signature of $\langle v_j, j \rangle$ by $p_j$ in $\sigs_i$.
Hence, at time $t$, $p_i$ received a \witsig from $p_j$. Furthermore, we assume that $p_j$ is correct and that cryptographic signatures cannot be impersonated.
Therefore, the only process able to sign a value using $p_j$'s secret key is $p_j$ itself.
Hence, $p_j$ did \cac-propose value $v_j$ in both cases.
\end{proof}

\begin{lemma} \label{lemma:conf-set-validity}
For any two correct processes $p_i$ and $p_j$, if $\langle v, f, \star \rangle \in \acceptedI{i}$ and $\langle v',o, \star \rangle \in \accepted_j$, then $\langle v,k\rangle, \langle v',\ell\rangle \in \big(\candidates_i \cap \candidates_j\big)$.
\end{lemma}

\begin{proof}
Let $p_i$ and $p_j$ be two correct processes, and let $\langle v, f, \star \rangle \in \accepted_i$ and $\langle v', o, \star \rangle \in \accepted_j$.
We note $\tau_v$ the time $\langle v, f, \star \rangle$ is added to $\accepted_i$ and we note $\accepted_i^{\tau_v}, \candidates_i^{\tau_v}$ the state of the sets $\accepted_i$ and $\candidates_i$ at this time.
Using the \CACGlobalTermination property, we know that $p_i$ will eventually \cac-accept $v'$.

With this setup, $p_i$ cannot \cac-accept one of these tuples using the fast path---if $p_i$ uses the fast path for $\langle v, f, \star \rangle$, there can only be a maximum of $2t$ \witsigs in favor of $\langle v', o, \star \rangle$, no correct process will send a \readysig in favor of $\langle v', o, \star \rangle$.
Hence, both $\langle v, f, \star \rangle$ and $\langle v', o, \star \rangle$ are \cac-accepted at \cref{line:cbdeliver:when:readymsg}.

The following uses a proof by contradiction, we assume $\langle v, f, \star \rangle$ is \cac-accepted first and $\langle v', o, \star \rangle \notin \candidates_i^{\tau_v}$.

Furthermore, we use the following notations:
Let $\witfunc(\sigs_i^{t}, \langle v,f\rangle)$ be the \witsig signatures for the pair $\langle v,f\rangle$ in $\sigs_i^{\tau}$, and let $\readyfunc(\sigs_i^{\tau}, \langle v,f\rangle)$ be the \readysigs relative to the pair $\langle v,f\rangle$ in $\sigs_i^{t}$.
Let $\mathsf{max}(\sigs_i^{t}, p_j)$ be the \readysig or \witsig with the greatest $\sigcount$ from process $p_j$ in $\sigs_i^{t}$.

At $\tau_v$, due to the condition at \cref{line:qr-ready}, $|\readyfunc(\sigs_i^{\tau_v}, \langle v,f \rangle)|\ge n-t$.
However, $\langle v',o\rangle \notin \candidates_i^{\tau_v}$ by assumption.
Hence, $|\witfunc(\sigs_i^{t_v}, \langle v',o\rangle)| < k$ (\cref{line:C:fill,line:qr-ready}).
In the following, we use two characteristics of the algorithm:
\begin{enumerate}
    \item A correct process does not send a \witnessm if it already sent a \readym (\cref{line:cond:witisg:when:new:witnessm,line:ready:received}); and    
    \item A correct process only accepts complete sequences of messages, \ie signature received from correct processes can be assumed FIFO.\footnote{This condition is  implicitly stated in the description of the algorithm and assumed by the comment at \cref{line:wit:received,line:ready:received}.}
\end{enumerate}
Among the \readysigs in $\readyfunc(\sigs_i^{\tau_v}, \langle v,f \rangle)$, at least $n-2t$ are sent by correct processes. Using the second point of the previous remark, we know that if $p_l$ is correct and given $k=\maxfn(\sigs_i^{\tau}, p_l)$, we received all messages from $p_l$ with $\sigcount$ lesser than $k$.
Furthermore, if there exists a \readysig from $p_l$ in $\sigs_i^{t_v}$ and if $p_l$ is correct, using the first point of the previous remark, we know that there are no \witsigs from $p_l$ in $\sigs_i^{\tau}$ that are not in $\sigs_i^{\tau_v}$, for all $\tau\ge \tau_v$. 

Hence, the only \witsigs in $\witfunc(\sigs_i^{\tau}, \langle v',o\rangle)$ that are not in $\witfunc(\sigs_i^{\tau_v}, \langle v',o\rangle)$ for $\tau>\tau_v$ are the one sent by correct processes whose \readysigs weren't in $\readyfunc(\sigs_i^{\tau_v}, \langle v,f\rangle)$---we call them the set of \emph{missed processes}---or the one sent by Byzantine processes.
Because $\readyfunc(\sigs_i^{\tau_v}, \langle v,f\rangle)$ contains the signature from at least $n-t$ processes, we know that the set of missed processes is lesser or equal to $t$.
Hence, a maximum of $2t$ additional \witsigs can be received by $p_i$ after $\tau_v$.
(Up to $t$ from correct processes whose \readysigs weren't in $\readyfunc(\sigs_i^{\tau_v}, \langle v,f\rangle)$, and up to $t$ from Byzantine processes that do not respect this constraint.)
Therefore, $p_i$ can receive up to $|\witfunc(\sigs_i^{\tau_v}, \langle v',o\rangle)| + 2t$ \witsig in favour of $\langle v',o\rangle$ during the whole execution of the algorithm.
However, we said that $|\witfunc(\sigs_i^{\tau_v}, \langle v',o\rangle)| < k$. Hence, $|\witfunc(\sigs_i^{\tau_v}, \langle v',o\rangle)| + 2t < 2t +k$

Therefore, $\langle v',o\rangle$ will never reach the $2t+k$ threshold (\cref{line:qw-witness} or \ref{line:qr-ready}) and $\langle v',o, \star\rangle$ cannot be \cac-accepted by a correct process, hence contradicting the assumption.
Therefore, $\langle v, \star, \star \rangle, \langle v', \star, \star \rangle \in \candidates_i\cap \candidates_j$.
\end{proof}

\begin{corollary}[\CACPrediction] \label{corr:prediction-opti}
For any correct process $p_i$ and for any process identity $k$, if, at some point of $p_i$'s execution, $\langle v,k\rangle \not\in \candidates_i$, then $p_i$ \emph{never} \cac-accepts $\langle v,k\rangle$ (\ie $\langle v,k \rangle \not\in \acceptedI{i}$ holds forever).
\end{corollary}

\begin{proof}
The corollary follows from the contrapositive of \cref{lemma:conf-set-validity} when $p_j=p_i$.
\end{proof}

\begin{lemma}[\CACNonTriviality] \label{lemma:non-triv-opti}
For any correct process $p_i$, $\acceptedI{i}\neq\emptyset$ $\Rightarrow$  $\candidates_i\neq\top$.
\end{lemma}

\begin{proof}
This property is directly verified. When a process \cac-accepts a value, it first intersects its $\candidates_i$ set with a finite set.
Hence, at this point in time, $\candidates_i \ne \top$.
\end{proof} 

\begin{lemma} \label{lemma:nmt}
If a correct process broadcasts a \witnessm at \emph{\cref{line:verif-end}} or \emph{\ref{line:init:brd}}, then eventually we have $|\{j:\witsig(p_j,\langle\star,\star\rangle,\star)\in \sigs_i\}| \ge n-t$.
\end{lemma}

\begin{proof}
If a correct process broadcasts a \witnessm at \cref{line:verif-end} or \ref{line:init:brd}, it is sure all the correct processes will eventually receive this \witnessm (thanks to the best effort broadcast properties).
Hence, each correct process $p_j$ will answer with a \witnessm containing a $\witsig(p_j, \langle\star, \star\rangle, \star)$ if they did not already do so (lines \ref{line:cond:witisg:when:new:witnessm} to \ref{line:init:brd}).
Therefore, if $p_i$ broadcasts a \witnessm, it is sure that eventually, $|\{j:\witsig(p_j,\langle\star,\star\rangle,\star)\in \sigs_i\}| \ge n-t$.
\end{proof}

\begin{lemma}[\CACLocalTermination]\label{lemma:local-delivery}
If a correct process $p_i$ invokes $\cacpropose(v)$, its set $\acceptedI{i}$ eventually contains a pair $\langle v',\star\rangle$ (note that $v'$ is not necessarily $v$).
\end{lemma}

\begin{proof}
Let a correct process $p_i$ \cac-proposes a value $v$. 
To prove the \CACLocalTermination property, three different cases must be explored.
\begin{itemize}
    \item In the first case, $p_i$ signs and \be-broadcasts a \witsig in favour of $\langle v,i\rangle$.
    It eventually receives $n-t$ \witsigs (\Cref{lemma:nmt}) among which at least $2t+k$ \witsigs are in favour of a unique pair $\langle v',j\rangle$ (either $\langle v',j\rangle=\langle v,i\rangle$ or $\langle v',j\rangle\ne \langle v,i\rangle$), \ie $\exists \tau$ such that $|\{l: \witsig(p_l, \langle v',j\rangle, \star) \in \sigs_i^{\tau}\}| \ge 2t+k$.
    Hence, $\langle v',j\rangle$ satisfies the condition \cref{line:qw-witness} or \ref{line:qw-ready} and $p_i$ will broadcast a \readym along with $\sigs_i^{\tau}$.
    Thanks to the best effort broadcast properties and because $p_i$ is correct, the $n-t$ correct processes will eventually receive $\sigs_i^{\tau}$.
    
    Let $p_\kappa$ be a correct process that receives the \readym from $p_i$ and $\sigs_i^{\tau}$ at time $\tau'$.
    It will add all the signatures from $\sigs_i^{\tau}$ to $\sigs_\kappa^{\tau'}$ (\cref{line:accum:sig:when:readyMsg}).
    Therefore, $|\{l: \witsig(p_l, \langle v',j\rangle, \star) \in \sigs_\kappa^{\tau'}\}| \ge 2t+k$ and $v'$ satisfies the condition at \cref{line:qw-ready}.
    Process $p_\kappa$ will eventually send a \readym at line \ref{line:ready-time-two} along with its set $\sigs_\kappa$ where $\readysig(p_\kappa, \langle v',j\rangle, \star) \in \sigs_\kappa$ (\cref{line:new:readysig:receive:readymsg,line:ready-time-two}).
    Each correct process will eventually send such \readym.
    Hence, eventually, $p_i$ will receive $\readysig(\star, \langle v',j \rangle, \star)$ from the $n-t$ correct processes.
    Hence, the condition at \cref{line:qr-ready} will eventually be verified, and $p_i$ will \cac-accept $\langle v',j \rangle$.
     
    \item In the second case, $p_i$ signs and \be-broadcasts a \witnessm in favour of $\langle v,i \rangle$, and among the $n-t$ responses it receives (\Cref{lemma:nmt}), there are less than $2t+k$ \witsigs messages in favour of $\langle v,i \rangle$ or any other $\langle v', j \rangle$, \ie $|\{j \mid \witsig(p_j,\langle \star,\star\rangle,\star)\in \sigs_i^{\tau}\}| \ge n-t$ and $\Nexists \langle v', j \rangle,$ such that $|\{\witsig(\star,\langle v', j \rangle,\star)\in \sigs_i^{\tau}\}|\ge 2t+k$.
    In this case, $p_i$ is stuck.
    It cannot send a \readym or \cac-accept a value, but it cannot wait for new \witnessm either, because all the Byzantine processes could act as if they crashed.\footnote{Let us recall that, except for the unlocking mechanisms from \cref{line:condition:lock:begin} to \ref{line:slow-path-resend:end}, a correct process can only produce one \witsig during the execution of the algorithm.}
    It must use one of the unlocking mechanisms implemented from lines~\ref{line:lock-begin} to~\ref{line:slow-path-resend:end}. 
    
    We analyze the two possible unlocking mechanisms:
    \begin{itemize}
        \item The first unlocking mechanism (from \cref{line:first:unlocking:mechanism} to \ref{line:fast-path-resend:end}) is used by $p_i$ if a correct process $p_j$ might have used the fast-path.
        If $p_j$ might have used the fast-path at time $\tau$, $|\{\witsig(\star, \langle v_f,f\rangle, \star) \in \sigs_j^{\tau'}\}|\ge n-t$ and $|\{\witsig(\star, \langle v_o,o\rangle, \star) \in \sigs_j^{\tau'}\}|= 0, \forall \langle v_f,f \rangle \ne \langle v_o,o \rangle$.
        Let $|P|$ be the number of processes from which $p_i$ received \witsigs, \ie $P = \big\{j \mid$ $\witsig(p_j,\langle\star,\star\rangle,\star) \in \sigs_i\big\}$, and $|P| \ge n-t$.
        We consider the worst case scenario where $T_j^{\tau'}= \{\witsig_{t}, \cdots, \witsig_{2t} \cdots, \witsig_{n}\}$ is the set of \witsigs received by $p_j$ at time $\tau'$, where $\{\witsig_{t}, \cdots, \witsig_{2t}\}$ are messages sent by Byzantine processes and $T_i^{\tau}= \{\witsig_1, \cdots, \witsig_{t-1}, \witsig_{t'}, \cdots, \witsig_{2t'}, \witsig_{2t+1}, \cdots, \witsig_{|P|}\}$ is the set of \witsigs received by $p_i$ at time $\tau$ where $\{\witsig_{t'}, \cdots, \witsig_{2t'}\}$ are messages sent by Byzantine processes, $\witsig_{i} \ne \witsig_{i'}, \forall i \in \{t, \cdots, 2t\}$.
        We have $|T_i^{\tau} \cap T_j^{\tau'}|\ge |P|-2t \ge n-3t$. 
        
        Therefore, if $\exists \langle v_f, f\rangle$ such that $|{\witsig(\star, \langle v_f, f\rangle, \star) \in \sigs_i^{\tau}}|\ge |P|-2t$, $p_j$ might have used the fast-path (this condition is verified by  \cref{line:first:unlocking:mechanism}).
        In this case, processes send a new \witnessm only in favor of $\langle v_f, f \rangle$ (if they did not already do so). 
        Eventually, $p_i$ will receive all the \witnessm sent by the correct processes.
        Therefore, if $n-2t$ correct processes sent a \witnessm message in favor of $\langle v_f, f \rangle$, then the $t$ correct processes that did not vote for this pair in the first place will send a new \witnessm in its favor.
        Therefore, the correct processes will eventually receive $n-t\ge 2t+k$ \witnessm messages in favor of $\langle v_f, f\rangle$, and they will send a \readym in favor of this pair.
        Hence, each correct process will receive $n-t$ \readysig in favor of $\langle v_f, f\rangle$, and will \cac-accept it (\cref{line:cbdeliver:when:readymsg}).
        Otherwise, if there are less than $n-2t$ correct processes that sent \witsigs in favor of $\langle v_f, f\rangle$ in the first place, $p_i$ will eventually receive the messages from the $t$ correct processes that it missed, the condition at \cref{line:first:unlocking:mechanism} will no longer be true and $p_i$ will resume to the second unlocking mechanism.  
        
        \item With the second unlocking mechanism, a correct process sends a new \witnessm only if it received at least $\maxfn\big(n-(|M|+1)t,1\big)$ \witsigs (\cref{line:unlock-classique}) where $M = \{\langle v',\kappa \rangle:$ $\witsig\big(\star,\langle v',\kappa \rangle,\star\big) \in \sigs_i \}$.
        First, let us prove that either a correct process $p_j$ sends a \readym message after receiving the first messages of the $n-t$ correct processes, or a pair $\langle v',\kappa \rangle$ eventually satisfies the following condition at all correct processes: $\{\witsig(\star, \langle v',\kappa \rangle, \star) \in \sigs_l\}\ge \maxfn\big(n-(|M|+1)t,1\big), \forall p_l$, a correct process.
        
        Let us assume that the previous assumption is wrong, \ie no correct process sends a \readym message after receiving the first \witnessm from the $n-t$ correct processes, and there is a process $p_l$ such that $\{\witsig(\star, \langle v',\kappa \rangle, \star) \in \sigs_l\}< \maxfn\big(n-(|M| + 1)t,1\big)$.
        The first part of the assumption implies that $\Forall \langle v',\kappa \rangle$, $\{\witsig(\star, \langle v',\kappa \rangle, \star) \in \sigs_l\}< 2t+k$, for all $p_l$ correct.
        We know (thanks to the best effort broadcast properties) that each correct process will eventually receive the first $n-t$ \witsig sent by correct processes, let $\sigs_{tot}$ be this set.
        Furthermore, the worst case scenario is when each pair in $\sigs_{tot}$ is backed by a minimal number of \witsigs, \ie the scenario where each pair in $\sigs_{tot}$ is backed by $\frac{n-t}{|M|}$ \witsigs---otherwise, by the pigeonhole argument, we have one pair that is backed by more signatures and which is more likely to reach the $\maxfn\big(n-(|M|+1)t,1\big)$ threshold.
        Hence, $\forall \langle v',\kappa \rangle$ such that $\lfloor\frac{n-t}{|M|}\rfloor| +1 \ge |\{\witsig(\star, \langle v',\kappa \rangle,\star) \in \sigs_{tot}\}|\ge\lfloor\frac{n-t}{|M|}\rfloor$. Furthermore, we see that $\lfloor\frac{n-t}{|M|}\rfloor \ge \maxfn(n- (|M| + 1) t ,1)$.
        Hence, the hypothesis is contradicted.
        We know that either a correct process $p_j$ sends a \readym while receiving the first value of the $n-t$ correct processes, or a value $v'$ eventually satisfies the following condition at all correct processes: $\{\witsig(\star, \langle v',\kappa \rangle, \star) \in \sigs_l\}\ge \maxfn\big(n-(|M|+1)t,1\big), \forall p_l$ correct processes.
    \end{itemize}
    In both cases, each correct process will eventually send a \readym along with the $2t+k$ \witsigs they received in favor of a unique pair, hence falling back to the first case. 
    
    \item The third case occurs when no \witnessm in support of $\langle v,i\rangle$ is sent by $p_i$ to the other processes---another \witnessm was already broadcast (\cref{line:verif-init})---\ie $\witsig(p_i, \langle v,i\rangle, \star) \notin \sigs_i^{\tau}$ for any time $\tau$ of the execution.
    However, even if $p_i$ does not broadcast a \witnessm in favor of $\langle v,i\rangle$ (due to the condition at \cref{line:verif-init}), it has already sent a \witnessm in favor of some pair $\langle v',j\rangle$ (again, because of the condition at \cref{line:verif-init}), thus falling back to the first or the second case.
\end{itemize}
Therefore, if a correct process $p_i$ \cac-proposes a value, it will always \cac-accept at least one pair.
\end{proof}

\begin{lemma}[\CACGlobalTermination] \label{lemma:global-delivery}
If $p_i$ is a correct process and $\langle v,j \rangle \in \acceptedI{i}$,  eventually $\langle v,j\rangle\in \acceptedI{k}$ at every correct process $p_k$.
\end{lemma}

\begin{proof}
Let $p_i$ and $p_j$ be two correct processes. Let $p_i$ \cac-accept a tuple $\langle v,j, \star \rangle$, but $p_j$ does not.
Two cases can be highlighted: 
\begin{itemize}
    \item In the first case, $p_i$ received $n-t$ \readysig in favour of $\langle v,j \rangle$ (\cref{line:qr-ready}). The \readym that is used to send those signatures contains $2t+k$ \witsig in favor of $\langle v,j\rangle$ (thanks to the verification at \cref{line:check:readyMsg:valid}). Furthermore, it did not satisfy the condition \cref{line:fast-path-condition}. 
    
    \item In the second case, $n > 5t$, and $p_i$ received more than $n-t$ \witsig in favour of $\langle v,j\rangle$ and no signatures in favour of another pair (\cref{line:fast-path-condition}). 
\end{itemize}
In both cases, $p_i$ broadcasts a \readym in favour of $\langle v,j\rangle$ (\cref{line:ready-time-one} or \ref{line:ready-time-two}).
Each \readym in favor of $\langle v,j\rangle$ sent by a correct process contains at least $2t+k$ valid \witsig in favor of $\langle v,j\rangle$ (\cref{line:qw-witness,line:qw-ready}). 
Process $p_i$ is correct, therefore each correct process will eventually receive at least one \readym associated with the proof that $2t+k$ \witsig in favour of $\langle v,j\rangle$ exists.
Hence, $\langle v,j \rangle$ will eventually reach the $2t+k$ threshold at each correct process.
When a correct process receives $2t+k$ \witnessm in favor of a pair, it sends a \readym in its favor (\cref{line:ready-time-one} or \ref{line:ready-time-two}).
Therefore, each correct process will send a \readym relative to $\langle v,j \rangle$ at \cref{line:ready-time-two}.
Because there are $n-t$ correct processes, each correct process will receive $n-t$ \readym in favor of $\langle v,j\rangle$, and $p_j$ will eventually \cac-accept $\langle v,j\rangle$ (\cref{line:cbdeliver:when:readymsg}). 
\end{proof}
 
\begin{theorem}\label{th:fullproof}
If $n \ge 3t+k \ge 3t+1$, then \Cref{alg:cdb_impl_opti} implements the {\em \CAC} abstraction.
\end{theorem}

\begin{proof}
Using \Cref{lemma:brd-validity}, \Cref{corr:prediction-opti}, \Cref{lemma:non-triv-opti}, \Cref{lemma:local-delivery}, and \Cref{lemma:global-delivery}, \Cref{alg:cdb_impl_opti} implements the \CAC abstraction.
\end{proof}

\begin{corollary}
    \Cref{alg:cdb_impl_opti} can implement the \CAC abstraction with $n \ge 3t+1$ (\Cref{th:fullproof}), which is optimal in term of Byzantine resilience as proven in \Cref{thm:3tbound}. 
\end{corollary}

\begin{lemma}[Proof of acceptance] \label{th:proof-of-delivery}
There exists a function $\Verify$ such that, for any proof of acceptance $\pi_v$, the following property holds

\begin{equation*}
   \Verify(v,\pi_v)
   =\ttrue \iff \Exists p_i \text{ correct such that, eventually, }\langle v, \star, \pi_v \rangle\in \acceptedI{i}.
\end{equation*}
\end{lemma}

\begin{proof}
Let $\pi_v$ be a candidate proof of \cac-acceptance. Let the function $\Verify(\pi_v,v)$ return true if and only if $\pi_v$ contains valid \readysigs on the value $v$ from at least $n-t$ processes in $\Pi$.
Hence, at least $n-2t\ge t+k$ correct processes signed \readysigs in favor of $v$.
Furthermore, correct processes only propagate their signatures via a \readym, which is a best-effort broadcast.
Therefore, all the correct processes eventually receive those $n-2t$ \readysigs. 
Furthermore, a \readym from a correct process contains at least $2t+k$ \witsigs (\cref{line:qw-witness} or \ref{line:qw-ready}).
Hence, all the correct processes will receive $2t+k$ \witsig backing $v$, and all the correct processes will send a \readym backing $v$, and they will eventually \cac-accept $v$.
In other words, $\Verify(\pi_v,v) = \ttrue \Rightarrow \exists p_i$ correct such that $p_i$ \cac-accepts $\langle v, \star, \pi_v \rangle$.

Let a correct process $p$ \cac-accept a tuple $\langle v,\star,\pi_v=\sigs_i \rangle$.
Then $\sigs_i$ contains all the signatures $p$ sent and received before the \cac-acceptance.
To \cac-accept a value $v$, $\sigs_i$ must contain at least $n-t$ \readysigs in its favor (\cref{line:qr-ready}). Hence $\Verify(\sigs_i,v)=\ttrue$.
Therefore, $\Verify(\pi_v) = \ttrue \Leftrightarrow \Exists p_i$ correct such that $p_i$ \cac-accepts $\langle v, \star, \pi_v \rangle$.
\end{proof}

\subsection{Optimality of the best case latency} \label{sec:best-case-latency-proof}
This section explores the theoretical best-case latency of the abstraction.
More precisely, it proves that the optimized \Cref{alg:cdb_impl_opti} reaches the lower bound with respect to Byzantine resilience when fast-path is enabled.

\begin{theorem}\label{th:fast:pth:optimality}
If a {\em \CAC} algorithm allows processes to \cac-accept after all the correct processes have only broadcast one message, then $n \ge 5t+1$.
\end{theorem}

\begin{proof}
\insertmg{As a preliminary argument, we prove that, if a process \cac-proposed a value, then other correct processes have to send a message once they received it. First, processes are symmetrical, they run the same algorithm. Thus, once they received a proposition, they cannot wait for other correct processes to send a message if they do not. Furthermore, they cannot \cac-accept the proposition as is, otherwise the \CACPrediction property could be trivially violated. Finally, they don't know if another value will be \cac-proposed in the future. Hence, once correct processes are \cac-proposed a value, they have to broadcast a message.} 

We can now prove that the bound $n\ge 5t+1$ is optimal. This proof is done by contradiction.
Let $T_1,T_2,T_3,T_4$ be partitions of $\Pi$.
Let $|T_1| = |T_2| = |T_3| = t$.
Let $n \le 5t$.
We consider $p_1 \in T_1$ and $p_2\in T_2$ two processes.
Two values $v$ and $v'$ are \cac-proposed by two correct processes $p_v$ and $p_{v'}$ respectively.
The assumption is that $p_1$ \cac-accepts a value $v$ after all correct processes broadcast one message.
In the best case, $p_1$ received messages from processes that only received the broadcast from $p_v$. We build three executions $E_1, E_2$ and $E_3$

In $E_1$, processes in $T_2$ are Byzantine, and act as if they crashed.
Processes in $T_1$, $T_3$ and $T_4$ receive the proposition for the value $v$ and then broadcast a message. This broadcast can only contain information about $v$.

In $E_2$, processes in $T_3$ are Byzantine; they can send messages with information about $v$ to the processes in $T_1$ and messages with information about $v'$ to the processes in $T_2$.
Processes in $T_2$ only known about $v'$ and broadcast messages that can only contain information about this value, while processes in $T_4$ only know about $v$ and broadcast messages that can only contain information about $v$. 

In $E_3$, processes in $T_1$ are Byzantine and act as if they crashed.
Processes in $T_2$ and in $T_3$ both only know about $v'$ and broadcast messages that can only contain information about this value, while processes in $T_4$ only know about $v$ and broadcast messages that can only contain information about this value.

Because of the asynchrony of the network, correct processes can only wait for $n-t$ messages. Thus, from $p_1$'s point of view, $E_1$ and $E_2$ are indistinguishable if it receives messages from $T_1$, $T_3$, and $T_4$ first.
In both cases, it must \cac-accept a value.
Thus, in both of them, it \cac-accepts the value $v$ in one round because it only received messages with information about $v$.

Furthermore, from $p_2$'s point of view, $E_2$ and $E_3$ are indistinguishable if it receives messages from $ T_2$, $ T_3$, and $T_4$ first.
In both of them, it sees $2t$ messages that only contain information about $v$ and $2t$ messages that only contain information about $v'$. 
Thus, whether $v$ or $v'$ should be \cac-accepted is undetermined, and processes must send new messages to decide. We further assume that messages from $T_1$ are further delayed and received after those new messages.
A second round of communication is necessary, and both values could eventually be \cac-accepted.

However, the assumption was that $p_1$ \cac-accepts in one round, \ie it can participate in the second round of communication, but the result should not impact the fact that only $v$ is \cac-accepted.
However, $v'$ can also be \cac-accepted, hence contradicting the \CACPrediction and \CACGlobalTermination property.
Therefore, the proportion of Byzantine processes for a best-case latency of one round is at least $n \ge 5t+1$.
\end{proof}

\begin{corollary}
The fast-path proposed by \emph{\Cref{alg:cdb_impl_opti}} makes it possible to terminate after all correct processes broadcast once if $n\ge 5t+1$. Hence, the fast path of this algorithm is optimal with respect to Byzantine resilience as proven in \Cref{th:fast:pth:optimality}.
\end{corollary}

\subsection{Proof of the latency of Algorithm \ref{alg:cdb_impl_opti}}
\label{sec:proof-latency-cb}

This section analyzes the latency properties of our algorithm.

\begin{theorem}
Let $x$ values be \cac-proposed by $x$ processes.
In the worst case, processes that implement \emph{\Cref{alg:cdb_impl_opti}} exchange $2 \times x \times n^2$ messages and \cac-decide after four rounds of best-effort broadcast.  
\end{theorem}

\begin{proof}
Let $x$ values be \cac-proposed by $x$ processes.
Each process will broadcast an initial \witnessm---one best effort broadcast round, and $x \times n$ messages.
After the reception, the $n-x$ processes that did not broadcast answer those broadcasts with new \witnessm---a second best effort broadcast round, and $n(n-x)$ messages.
Because of the conflict, the processes have to use the unlocking mechanism for each of the values they did not already witness ---third best effort broadcast round and $(x-1)n^2$ messages.
Finally, each process sends a \readym in favor of each value and \cac-accepts---fourth best effort broadcast round and $xn^2$ messages.
Therefore, in the worst case, the values are \cac-accepted after four best-effort broadcast rounds, and $xn+n(n-x)+(x-1)n^2+xn^2=2xn^2$ messages are exchanged.
\end{proof}

\begin{theorem}
The best case latency of the \emph{\Cref{alg:cdb_impl_opti}} when $n<5t+1$ is three asynchronous rounds.
\end{theorem}

\begin{proof}
When $n<5t+1$, processes cannot use the fast path.
The best case for the implementation is when there are no conflicts.
In this case, a process $p_i$ broadcasts an initial \witnessm in favor of value $v$---first asynchronous round.
Then, each process broadcasts its own \witnessm in favor of $v$---second asynchronous round.
Finally, each process broadcasts a \readym message, and \cac-decides---third asynchronous round.
The correct processes \cac-accept a value after three asynchronous rounds.
\end{proof}

\begin{theorem}
The best case latency for a correct process in \emph{\Cref{alg:cdb_impl_opti}} is two asynchronous rounds when $n \ge 5t+1$.
\end{theorem}

\begin{proof}
Let a correct process $p_i$ be the unique process to \cac-propose value $v$.
Let $n\ge5t+1$.
First, it broadcasts a \witnessm in favor of $v$---first asynchronous round.
Then, each correct process broadcasts a \witnessm in favor of $v$---second asynchronous round.
When $p_i$ receives the $n-t$ \witnessm of the correct processes, it uses the fast path and accepts $v$.
Thus, the best-case latency of the optimized version of the \CAC implementation is two asynchronous rounds.
\end{proof}

\section{A \CAC-based short-naming algorithm}\label{sec:sname:appendix}

\subsection{Description of the algorithm}

Given a character string $s$, $s[i]$ denotes its prefix of length $i$, e.g. $\mathtt{``abcdefghijk"}[3] = \mathtt{``abc"}$.



\Cref{alg:snaming} implements the short naming abstraction presented in \Cref{sec:sname}. 
This implementation uses two steps: a claiming phase and a commitment phase.
The claiming phase verifies (and proves) that no other process tries to claim the same name.
The commitment phase is used to actually associate a name with a public key, once this association has been successfully claimed.
The claiming phase uses multiple \CAC instances. 
Each instance is associated with a name.   
If the \CAC instance \cac-accepts the value \cac-proposed by a process $p_i$ and $|\candidates_i|=1$, it means that there is no contention on the attribution of the name.
If $p_i$ is the only process claiming this name, then it can commit to this name; otherwise, there is a conflict.
In the latter case, the invoking processes will claim a new name by adding one character from its public key to the old name. 
The \CAC instances for the claiming phase are stored in a dynamic dictionary, $\claimlist$.
This dictionary dynamically associates a \CAC instance to a name.
It is initiated as an empty dictionary, and whenever a process invokes the \cacpropose operation on a specific name---\ie when a process executes $\claimlist[name].\cacpropose(\pk)$---or when the first \CAC value for a specific name is received, the dictionary dynamically allocate a new \CAC object. 

The commitment phase uses a new set of \CAC instances.
Similarly to the claiming phase, \Cref{alg:snaming} uses one \CAC instance per name.
However, unlike the claiming phase, there is one \CAC instance per name and per process.
When a process knows it successfully claimed a name, \ie no contention was detected, it informs the other processes by disseminating its public key using its \CAC instance associated with the claimed name.
Processes can verify that the commitment does not conflict with another process, because they accepted the associated name in the associated \CAC instance.
However, if a Byzantine process $p$ claimed the same name, it could commit to the name even though the system did not accept its claim.
In this case, the commitment would be rejected by correct processes, as the Byzantine process cannot provide a valid proof of acceptance of the claim.
The only case where multiple processes can commit to the same name is if they are all Byzantine, and all their claims are \cac-accepted.
In this case, they could all commit to the same name, which does not violate the specification and the \SNAgreement property.
In other words, Byzantine processes can share the same names if it does not impact correct processes.
Similarly to the claiming phase, \CAC instances used during the commitment phase are stored in a dynamic dictionary. 

We further assume the \CAC instances only accept valid pairs, \ie for a pair $\langle \pk, \pi \rangle$ and the instance $\commitlist[name]$ or $\claimlist[name]$, $name$ is a sub-string of \pk and $\pi$ is a valid proof of knowledge of the secret key associated to \pk.

This algorithm uses a $\mathsf{VerifySig}(\pk, \pi)$ algorithm. This algorithm returns ``\ttrue'' if and only if $\pi$ is a valid cryptographic signature by the public key $\pk$. 

\begin{algorithm}

\Init{$\Names_i$ $\gets \varnothing$; $\claimlist$ $\gets$ dynamic dictionary of \CAC objects;\\
$\commitlist_i$ $\gets$ dynamic dictionary of \CAC objects; $\prop_i \gets \varnothing$.\DontPrintSemicolon}
\medskip

\Operation{$\snamingclaim(pk, \pi)$}{
    \lIf{$\mathsf{VerifySig}(pk,\pi) = \ffalse$}{\return}
    $\choosename(1,\pk,\pi)$.\DontPrintSemicolon\Comment*{Queries an unused name, starting with $pk[1]$.}
}
\medskip

\InternalOperation{$\choosename(\ell,\pk,\pi)$}{
    $\currname \gets \pk[\ell]$\;
    \While(\Comment*[f]{Looks for the first unused name.}){$\langle \currname,\star \rangle \in \Names_i$}{
        $\ell \gets \ell+1$\;
        \lIf{$\ell > |\pk|$}{\return}
        \label{line:sname:return:cond}
        $\currname \gets \pk[\ell]$\;
    }
    $\prop_i \gets \prop_i \cup \langle \currname,\pk,\pi,\ell \rangle$\;
    $\claimlist[\currname].\cacpropose(\langle \pk,\pi \rangle)$. \DontPrintSemicolon\Comment*{Claims \currname.}\label{line:sname:cdb:brdcast}
}
\medskip

\When{$\claimlist[name].\cacaccept(\langle \pk, \pi \rangle,j)$}{ \label{line:sname:cdb:deliver}
    \lIf{$\mathsf{VerifySig}(\pk,\pi) = \ffalse$ \cOr $name$ is a sub-string of \pk}{\return}
    \If(\Comment*[f]{If $name$ was claimed by $p_i$.}){\insertft{$\exists \pk',\pi':$} $\langle name,\pk',\pi', \ell \rangle \in \prop_i$}{ \label{line:sname:cond:in:prop}
        $\prop_i \gets \prop_i \setminus \langle name,\pk',\pi',\ell \rangle$\; 
        \lIf(\Comment*[f]{If no conflict, commit to $name$.}){$|\claimlist[name].\candidates_i| = 1$ \cAnd $\pi' = \pi$}{%
            $\commitlist_i[name].\cacpropose(\langle \pk', \pi' \rangle)$%
        } \label{line:sname:cond:candidate:size} \label{line:sname:brb:broadcast}
        \lElse(\Comment*[f]{$p_i$ claims a name with more digits (back-off strategy).}){%
            $\choosename(\ell+1,\pk,\pi')$.\DontPrintSemicolon%
            \label{alg:sname:choose:rec}%
        }
    }
}
\medskip

\When{$\commitlist_i[name].\cacaccept(\langle \pk, \pi \rangle,j)$}{\ft{I've changed $_j$ into $_i$. Please check it's correct.}
    \lIf{$\mathsf{VerifySig}(pk,\pi)=\ffalse$ \cOr $name$ is a sub-string of \pk}{\return} \label{line:sname:validity:verif}
    $\wait\big(\langle \pk,\pi \rangle \in \claimlist[name].\accepted_i\big)$\; \label{line:sname:wait:cond}
    \lIf{$\langle name,\star,\star \rangle \notin \Names_i$}{
        $\Names_i \gets \Names_i \cup \{\langle name,\pk,\pi \rangle\}$.\hspace{7em}\Comment*[f]{The association between $name$ and $pk$ is committed by $p_i$.}\DontPrintSemicolon
    } \label{line:sname:name:update}
}
\caption{Short naming algorithm implementation (code for $p_i$)}
\label{alg:snaming}
\end{algorithm}


\subsection{Proof of the algorithm}
The proof that \Cref{alg:snaming} implements the Short Naming abstraction defined in \Cref{sec:sname} follows from the subsequent lemmas.

\begin{lemma}[\SNUnicity] \label{lemma:snaming:unicity}
Given a correct process $p_i$, $\Forall \langle \Names_j, \pk_j, \pi_j \rangle, \langle n_k, \pk_k, \pi_k \rangle \in \Names_i$, either $n_j \ne n_k$ or $j=k$.
\end{lemma}

\begin{proof}
Let $p_i$ be a correct process such that $\Exists \langle n_j,\pk_j,\pi_j \rangle, \langle n_k,\pk_k,\pi_k \rangle \in \Names_i$ and $n_j = n_k$, $j \ne k$. 

The only place in the algorithm where $p_i$ updates $\Names_i$ is at \cref{line:sname:name:update}.
To reach this line, $p_i$ must verify the condition $\langle name,\star,\star \rangle \notin \Names_i$ at \cref{line:sname:name:update}.
However, this condition can only be valid once per name. Hence, $p_i$ will only update $\Names_i$ once per name, and two different tuples $\langle n_j,\pk_j,\pi_j \rangle, \langle n_k,\pk_k,\pi_k \rangle$ cannot be present in $\Names_i$ if $j\ne k$.
Hence, either $n_j \ne n_k$, or $j = k$
\end{proof}

\begin{lemma}[\SNAgreement] \label{lemma:snaming:agreement}
Let $p_i$ and $p_j$ be two correct processes. If $\langle n, \pk, \pi\rangle \in \mathsf{Name}_i$ and if the process that invoked $\snamingclaim(\pk, \pi)$ is correct, then eventually $\langle n, \pk, \pi\rangle \in \mathsf{Name}_j$.
\end{lemma}

\begin{proof}
Let $p_i, p_j$ and $p_k$ be three correct processes.
Let $p_k$ invoke $\snamingclaim(\pk,\pi)$.
Let $\langle n,\pk,\pi \rangle \in \Names_i$, where $\langle n,\pk,\pi \rangle \notin \Names_j$ during the whole execution.

If $\langle n,\pk,\pi \rangle \in \Names_i$, then it means that $p_i$ updated $\Names_i$ at \cref{line:sname:name:update}.
This implies that $\commitlist_k[n].\cacaccept(\langle \pk,\pi\rangle, \star)$ was triggered at $p_i$.
Thanks to the \CACGlobalTermination property of the \CAC abstraction, we know that $\commitlist_k[n].\cacaccept(\langle\pk,\pi\rangle, \star)$ will also eventually be triggered at $p_j$.
Because $p_i$ added $\langle n,\pk,\pi \rangle$ to $\Names_i$, we know that the conditions at \cref{line:sname:validity:verif,line:sname:wait:cond} are verified at $p_j$.
However, the condition $\langle name,\star,\star) \notin \Names_i$ at \cref{line:sname:name:update} may not be verified at $p_j$.
However, because $p_k$ is correct, when it invokes $\commitlist_k[n].\cacpropose(\langle \pk,\pi \rangle)$ at \cref{line:sname:brb:broadcast}, it first verified the condition $|\claimlist[n].\candidates_i| = 1$ at \cref{line:sname:cond:candidate:size}.
Hence, only one \cacaccept occurs for the name $n$ at all correct processes, thanks to the \CACPrediction and \CACGlobalTermination properties of the \CAC abstraction.
Therefore, only one tuple can pass the \wait instruction at \cref{line:sname:wait:cond}: $\langle \pk,\pi \rangle$, at the index $n$ of \claimlist.
Hence, all conditions from \cref{line:sname:validity:verif} to \ref{line:sname:name:update} will eventually be verified at $p_j$ and, eventually, $\langle n,\pk,\pi \rangle \in \Names_j$.
This contradicts the hypothesis; thus, the \SNAgreement property is verified.
\end{proof}

\begin{lemma}[\SNShortNames] \label{lemma:snaming:short:naming}
If all processes are correct, and given one correct process $p_i$, eventually we have $\Forall \langle n_j,\pk_j,\star \rangle, \langle n_k,\pk_k,\star \rangle\in \Names_i$: \\
If $|\mcprefix(\pk_j, \pk_k)| \ge |\mcprefix(\pk_j, \pk_\ell)|, \Forall \langle \star,\pk_\ell,\star\rangle \in \Names_i$
then $|\mcprefix(\pk_j,\pk_k)| + 1 \ge |n_j|$.
\end{lemma}

\begin{proof}
We prove \Cref{lemma:snaming:short:naming} by contradiction. Let all the processes be correct, and let $p_i$ be one of them. 
We assume that $\Exists \langle n_j,\pk_j,\star \rangle, \langle n_k,\pk_k,\star \rangle \in \Names_i, \Forall \langle n_l,\pk_l,\star \rangle \in \Names_i$:
\begin{gather*}
    |\mcprefix(\pk_j,\pk_k)| \ge |\mcprefix(\pk_j,\pk_l)|, \text{ and } \\
    |\mcprefix(\pk_j,\pk_k)| + 1 < |n_j|.
\end{gather*}

Let us call $p_j$ the correct process that executed $\snamingclaim(\pk_j,\star)$.
The only place where $\Names_i$ is modified is at \cref{line:sname:name:update}.
To execute this update, $p_i$ verifies with the condition at \cref{line:sname:wait:cond} that the tuple $\langle \pk_j, \star \rangle$ was \cac-accepted at the index $n_j$ of \claimlist.
The validity property of the \CAC abstraction ensures that $p_j$ \cac-proposed $\langle \pk_j,\star \rangle$.
Here, we assume that, because $p_j$ is the only process that knows the secret key associated with $\pk_j$, it is the only process able to execute $\cacpropose(\pk_j,\star)$.
The only place where $p_j$ can \cac-propose at index $n_j$ of \claimlist is at \cref{line:sname:cdb:brdcast}.
Furthermore, correct processes try all the sub-strings of their public keys sequentially, beginning with the first digit of the key.
Hence, to \cac-propose at index $n_j$ of \claimlist, it implies that, either $p_j$ already added the name $n_j[|n_j|-1]$ to $\Names_j$ associated to a public key $\pk_\kappa$, where $\kappa \ne j$, or that a process $p$ \cac-proposed at index $n_j[|n_j|-1]$ of $\claimlist$, and the $\candidates_j$ set of this \CAC instance contained $\langle \pk_\kappa,\star \rangle$, where $\kappa \ne j$.
In the first case, $|\mcprefix(\pk_\kappa,\pk_j)| \ge |n_j|-1$.
By the \SNTermination property of short naming (\Cref{lemma:sname:termination}), we know that, eventually, $\langle n_j[|n_j|-1],\pk_\kappa,\star \rangle \in \Names_i$, which violates the assumption.
Because $p_\kappa$ is correct, in the second case, it will eventually add an name whose size is greater or equal to $|n_j|$ with $\pk_\kappa$ as a public key to $\Names_\kappa$  (\SNTermination).
By the \SNAgreement property of short naming (\Cref{lemma:snaming:agreement}), this name will be added to $\Names_i$.
Hence, eventually, $|\mcprefix(\pk_j,\pk_\kappa)|+ 1 \ge |n_j|$, and $\langle \star,\pk_i,\star \rangle, \langle \star,\pk_\kappa,\star \rangle \in \Names_i$, thus violating the assumption and concluding the proof.
\qedhere

\end{proof}

\begin{lemma} \label{lemma:sname:while:loop}
If a correct process $p_i$ executes $\choosename(j,\pk,\pi)$, $\Forall j \in \{1,\cdots,|\pk|\}$, then either it eventually invokes $\claimlist[\star].\cacpropose(\langle \pk,\pi \rangle)$, or $\langle \star,\pk,\star \rangle \in \Names_i$.
\end{lemma}

\begin{proof}
Assuming $p_i$ is correct 
and $\langle \star,\pk,\star \rangle \notin \Names_i$, let $p_i$ execute $\choosename(j,\pk,\pi)$, $\Forall j \in \{1, \cdots, |\pk|\}$ and $p_i$ does not invoke $\claimlist[\star].\cacpropose(\langle \pk, \pi \rangle)$.
Then, the process must have returned at \cref{line:sname:return:cond}, and the condition at \cref{line:sname:return:cond} must be verified, \ie $i>|\pk|$.
Hence, all the names from $\pk[i]$ to $\pk[|\pk|]$ were already attributed to public keys different from $\pk$.
Hence, there exists a tuple $\langle \pk,\pk',\pi \rangle \in \Names_i$ where $\pk \ne \pk'$.
If $\Names_i$ is updated, then \cref{line:sname:name:update} has necessarily been executed, and the condition at \cref{line:sname:validity:verif} was verified. 
Therefore, using the perfect cryptography assumption, we have $\pk = \pk'$.
\end{proof}

\begin{lemma}[\SNTermination]\label{lemma:sname:termination}
If a correct process $p_i$ invokes $\snamingclaim(\pk,\pi)$, then eventually $\langle \star,\pk,\star \rangle \in \Names_i$.
\end{lemma}

\begin{proof}
\sloppy
Let $p_i$ be a correct process that invokes $\snamingclaim(\pk,\pi)$.
Then, it will execute $\choosename(1,\pk,\pi)$.
Using \Cref{lemma:sname:while:loop}, we know that either $\langle \star,\pk,\star \rangle \in \Names_i$, or $p_i$ invoked $\claimlist[name].\cacpropose(\langle \pk,\pi \rangle)$.
In the first case, \Cref{lemma:sname:termination} is trivially verified.
In the second case, the \CACLocalTermination property of the \CAC primitive ensures that $\claimlist[name].\cacaccept(\langle\pk',\star\rangle, \star)$ will be triggered at \cref{line:sname:cdb:deliver}.
Again, two cases can arise.
In the first case, $p_i$ invokes $\claimlist[name].\cacpropose(\langle \pk',\pi' \rangle)$ at \cref{line:sname:brb:broadcast}.
In the second case, $|\claimlist[name].\candidates_i|>1$ and $\choosename(i+1,\pk)$ is executed.
Let us study the second case first.
Multiple recursions might occur between the \choosename function and the $\claimlist[name].\cacaccept(\langle\pk',\star\rangle, \star)$ callback.
However, either we will end up in the first case and a \cac-propose will be invoked by $p_i$ at \cref{line:sname:brb:broadcast}, or \choosename will be  eventually executed with $i = |\pk|$.
Using the same reasoning as in \Cref{lemma:sname:while:loop}, we know that $p_i$ only receives $\claimlist[name].\cacaccept(\langle\pk',\star\rangle, \star)$ if $name$ is a sub-string of $\pk'$.
Hence, when $\claimlist[\pk].\cacaccept(\langle\pk',\star\rangle, \star)$ is triggered, $\pk=\pk'$.
Therefore, $p_i$ \cac-proposes $\commitlist_i[name].\cacpropose(\langle \pk, \pi \rangle)$ at \cref{line:sname:brb:broadcast}.
More precisely, we know that, if $p_i$ invokes $\snamingclaim(\pk,\pi)$, then $p_i$ will eventually invoke $\commitlist_i[name].\cacpropose(\langle \pk, \pi \rangle)$ or $\langle \star,\pk,\star \rangle \in \Names_i$.

When $\commitlist_i[name].\cacpropose(\langle \pk,\pi \rangle)$ is invoked by $p_i$, and because $p_i$ is correct, we know that $\commitlist_i[name].\cacaccept(\langle\pk,\pi\rangle, \star)$ will be triggered.
Because $p_i$ is correct, $name$ is a sub-string of $\pk$.
Furthermore, the only place where $p_i$ can \cac-propose such value is at \cref{line:sname:brb:broadcast}.
Hence, before this proposition, $p_i$ accepted $\claimlist[name].\cacaccept(\langle\pk,\pi\rangle, \star)$.
Thus, at this time, condition \cref{line:sname:wait:cond} is always verified.
Hence, $\langle name,\pk,\pi \rangle$ is added to $\Names_i$.

Therefore, when a correct process executes $\snamingclaim(\pk,\pi)$, eventually $\langle \star,\pk,\star \rangle$ is added to $\Names_i$.
\end{proof}




\section{Cascading Consensus implementation details} \label{sec:cc:details}

\todo{Check the text correspond to the new operation names \rconsdecide, \rconserror}%
The definition of \CCons has been introduced in Section~{\ref{sec:cas-cons-tot}}. 
It is  a novel consensus algorithm based on the \CAC\ abstraction 
that adapts to the contention by requiring  processes to synchronize "only when needed". 
As said in Section~\ref{sec:cas-cons-tot}, differently from other optimistic 
algorithms it does not force the processes that do not propose a value to synchronize, 
and exploits the set $\candidates$ of the participating processes 
to restrict the full network synchronization.
\ft{We need to explain the link with Restrained consensus, to help the reader connect \CCons with the next subsection.}%

\subsection{Restrained consensus (\RC): definition} \label{sec:restr-cons-rc}
{\em Restrained Consensus} (\RC) is a \insertft{modular} abstraction used as an \insertft{intermediate} building block of \CCons (\CC, presented in \Cref{sec:cascading-consensus}).
\insertft{Following the execution of a first \CAC instance,} \RC provides weak agreement guarantees \insertft{that help conflicting processes progress towards agreement in favorable cases}. \insertft{Crucially, it allows for implementations} in which only a subset $\Pi'$ of the processes in $\Pi$ interact.
\insertft{In particular, processes can only participate in the Restrained Consensus algorithm if they \cac-proposed a value in the first \CAC instance, and if this value was \cac-accepted. This condition is enforced using \emph{proofs of acceptance} obtained from the first \CAC instance (\cref{sec:proof:delivery}). As a result, Byzantine processes can only take part in a \RC instance if their input provides the same \CAC-derived guarantees as that of a correct process (more on this below).}

\insertft{\RC helps conflicting processes \emph{select} the same set of potential values, so that this set can be fed into a second \CAC instance to clinch a definite decision if circumstances align. 
The sets of values returned by \RC to participating processes are guaranteed to be equal only in favorable conditions (in terms of process faults and synchrony). In the presence of asynchrony or Byzantine faults, processes executing \RC may fail to produce a result, or may return diverging outputs, but thanks to the strict conditions under which \RC is executed (input produced by a \CAC instance and proofs of acceptance), all returned values are ensured to be compatible with any other step of the \CCons algorithm.}


This behavior allows the \CCons algorithm we present in \Cref{sec:cascading-consensus} to resolve a conflict efficiently in good cases while falling back to full-fledged consensus when the restrained-consensus algorithm fails.

Formally, \insertft{a process $p_i$ invokes} Restrained Consensus \insertft{through the} operation $\rconspropose(C,\initialproof)$, where $C$ is a set of \insertft{candidate pairs} $\langle v,j \rangle$
\insertft{obtained from the $\candidates_i$ set of a \CAC instance}---with $v$ a value and $j$ the identifier of a process in $\Pi$---and $\initialproof$ is a proof of acceptance \insertft{for a pair $\langle v_i,i \rangle$ proposed by $p_i$ to the same \CAC instance, with $\langle v_i,i \rangle\in C$ (\cref{sec:proof:delivery}).} 
$\initialproof$  proves that the process \insertft{$p_i$} that invokes $\rconspropose()$ is legitimate to do so.

An execution of \RC induces a set $\Pi'\in \Pi$ of processes that contains all processes that either (i) invoke \rconspropose with valid parameters, or (ii) appear in one of the sets $C$ passed as parameter to a \rconspropose invocation.
\ft{We initialy had `A \emph{correct} process $p_i \in \Pi$ is in $\Pi'$ if it invokes the \rconspropose operation with [..]'. This suggested that a Byzantine process cannot  invoke \rconspropose (or more precisely appear to invoke \rconspropose from the point of view of correct processes), but my understanding is that it can. The rewording has its own issues though, as the notion of `local invocation' is ill-defined for Byzantine processes.}%

{\sloppy
RC has two callbacks: $\rconserror()$ and  $\rconsdecide(E,\sigendorse,\sigretract)$, where $E$ is a set of tuples $\langle v,j \rangle$ with $v$ a value and $j$ a process identifier; $\sigendorse$ is a set of signatures on \insertft{the pairs of $E$} by all the processes $p_i$ \insertft{that appear} in $E$; and $\sigretract$ is a set of signatures of the string \texttt{``RETRACT''}.
\ft{Do we use the $\sigretract$ returned by \rconsdecide for anything in \CCons?}%

\insertft{In the following, $\CAC_1$ denotes the \CAC instance that is associated with the input of a Restrained Consensus execution. (We use the same notation when presenting the \CCons algorithm in \Cref{sec:cascading-consensus,sec:ccons:proof}.)} Restrained Consensus is defined by the following properties.
\begin{itemize}
    \item \insertft{\RCValidityExtra. If a correct process $p_i$ executes the callback $\rconsdecide(E,$ $\star,$ $\star)$, then
    \begin{itemize}
        \item $E\neq\emptyset,$ and
        \item $E\subseteq\big\{ \langle v, \star, \pi_v \rangle \mid \Exists p_k \text{ correct such that, eventually, }\langle v, \star, \pi_v \rangle\in \CAC_1.\acceptedI{k}\big\}.$
    \end{itemize}}
    \item \RCValidity.
    If \insertft{a correct process invokes \rconspropose}, all the processes in $\Pi'$ are correct, and the network delays between the processes of $\Pi'$ are less than or equal to $\delta_{RC}$, then at least one \insertft{correct} process $p_i\in \Pi'$ executes the callback $\rconsdecide(E,\sigendorse,\sigretract)$ and \insertft{$\sigendorse$ contains a signature from each process appearing in $E$.}
    \ft{I've removed `$\sigendorse \cup \sigretract$ contains a signature from each process in $\Pi'$', as I don't think it holds. In particular see my remark below on \RCTermination: some (correct) processes might belong to $\Pi'$ formally, yet never participate in the $\RC$ algorithm.}%
    \ft{It looks more like a termination property. Could we rename it into something like RC-Termination-2?}%
    \item \RCAgreement.
    If \insertft{a correct process invokes \rconspropose}, all the processes in $\Pi'$ are correct and if \insertft{two correct processes $p_i$ and $p_j$} execute the callback $\rconsdecide$, respectively with the parameters $\rconsdecide(E,\star,\star)$ and $\rconsdecide(E',\star,\star)$, then $E=E'$.
    \ft{I've removed `and $\sigendorse=\sigendorse'$.', as I am not sure it holds in all corner cases, and we apparently don't use it anywhere else.}%

    \item \RCIntegrity.
    A correct process $p_i$ invokes at most once either $\rconsdecide(\star,\star,\star)$ or $\rconserror()$ (but not both in the same execution).

    \item \RCTermination.
    Any correct process in $\Pi'$ \insertft{that invokes \rconspropose} eventually invokes either $\rconserror()$ or $\rconsdecide()$.
    \ft{PLEASE CHECK: I've added the condition `\emph{that invokes \rconspropose}', as with the definition of $\Pi'$ we have, a correct process $p_i$ could potentially invoke $\rconspropose(C,\star)$ after retracting itself. If some of the processes appearing in $C$ haven't been involved so far (because none of their values was accepted in $\CAC_1$, and they did not appear in any of the $C$ sets earlier), they won't receive any \RC message from anyone, and never invoke $\rconserror()$ or $\rconsdecide()$, although they formally belong to $\Pi'$.}\todo{Update proof accordingly}%
\end{itemize}
\ft{My gut feeling is that we need guarantees on $E$ outside of the favorable case. I've added \RCValidityExtra to provide these guarantees. \todo{Prove \RCValidityExtra, and use it explicitly in the \CCons proofs.}
}

\ft{I've put `a correct process invokes \rconspropose' at the start of \RCValidity and \RCAgreement to make clear the properties of RC only applies if there is at least one invocation to \rconspropose by a correct process. (Otherwise $\Pi'$ is empty, and for instance \RCValidity might not necessarily hold.)}%
\ft{PLEASE CHECK: In \RCAgreement, I have added that $p_i$ and $p_j$ should be correct processes. (`if a \emph{correct} process $p_i$ executes the callback...'). If $p_i$ is Byzantine, it can invoke as many callbacks as its wants.}

\subsection{Restrained Consensus: implementation} \label{sec:rcons:impl}
\label{Restrained consensus: algorithm}
\todo{Check text to match new notation in algorithm (in particular those using $\square_j$ subscripts.)}%
\ft{I'd inject of few references to line numbers in the explanations below to help the reader connect the pseudo code to what is written.}%
\Cref{alg:rcons} implements the Restrained Consensus abstraction \insertft{using signatures}. 
It relies on the \insertft{existence} of a correct proof of acceptance \insertft{$\initialproof$ for a process to invoke the \rconspropose operation (\Cref{line:rconspropose}).}
\ft{PLEASE CHECK: $\initialproof$ was presented as a set, but it's a single proof if I'm correct.}%
\insertft{The goal of a process that participates in the algorithm is to select pairs $\langle v,k\rangle$ in its local set $E_i$ (\cref{line:rconssmess:init:proposed,line:rcons:intersection:proposed,line:rcons:power-C,line:rconsretractmess:update:proposed}) and gather signatures in its set $\sigendorse_i$ (\cref{{line:rconsmess:signatures:union},line:rcons:signatures:init}) so that all the pairs in $E_i$ are signed by all the processes whose identity appears in a least in one of the pairs of $E_i$.}\ft{PLEASE CHECK: We initially had `gather one signature from each process in $\Pi'$', but because some process in $\Pi'$ might retract themselves and refuse to sign, I felt it did not apply.}
The algorithm uses two types of messages, \rconsmess (\cref{line:bcast:rconsmess,line:PiPrime:receive}) and \rconsretractmess (\cref{line:retract:bcast,line:retract:msg:handler}).
The \rconsmess message is used as the primary mechanism to propagate and gather signatures.
For a correct process to send a \rconsmess, it must possess the proof of acceptance $\initialproof$ of one of its own values.
\insertft{On reception, correct processes ignore \rconsmess messages that do not verify this condition (\cref{line:rconss:condition:proof:acc:in:rconss:message}), thus preventing Byzantine processes that did not obtain a proof of acceptance for one of their values during the $\CAC_1$ instance from interfering with the \RC execution.}
\ft{PLEASE CHECKED: I've removed the following sentence, as it seems obsolete (at least I can't see where in the pseudo code this takes place. `When a correct process receives a \rconsmess message, it has to immediately respond with another \rconsmess message if it possesses a proof of acceptance for one of its values.'}%
\insertft{If $\initialproof$ is valid, a process that receives a \rconsmess message conducts further checks (\cref{line:rcons:validity:checks}), and then aggregates both the proofs of acceptance and the signatures contained in the received message (\cref{line:rconsmess:signatures:union}). If it did not invoke \rconspropose earlier, it records the received value pairs, and replies with a \rconsretractmess message to indicate none of its own values where accepted (\crefrange{line:if:no:retract:yet:then:start}{line:if:no:retract:yet:then:end}). If it invoked \rconspropose earlier, it intersects its own set of value pairs $E_i$ with those of the sending processes $E_j$ (\cref{line:rcons:intersection:proposed}). In all cases, the process checks whether it has reached the condition to produce a selection (call to \checkdecision at \cref{line:rcons:call:to:checkdecision}).}
\ft{Do we keep the following sentence? My understanding is that (correct) processes that use \rconsretractmess may still invoke the callback $\rconsdecide$, so saying that they do not participate could be misleading. `The \rconsretractmess message type proves that the process that sends it does not participate in the Restrained Consensus.'}

The algorithm relies on a timer $\timerrc$ (\cref{line:tim-set-1,line:tim-set-2,line:tim-stop}) \insertft{whose timeout} must be chosen to allow correct processes in $\Pi'$ to \insertft{reach a conclusion before the timer ends in favourable cases}. 
\insertft{When all processes in $\Pi'$ are correct and all invoke \rconspropose simultaneously, \Cref{alg:rcons} terminates in one synchronous round (\Cref{tab:conditions} in \Cref{sec:tab:conditions}). In the slightly less favorable case where some correct processes of $\Pi'$ do not invoke \rconspropose (or invoke it too late), \Cref{alg:rcons} terminates in two synchronous round. (The first round is required for the initial \rconsmess broadcasts to reach all participants in $\Pi'$, and the second for process that did not invoke \rconspropose to respond with a \rconsretractmess message to this initial broadcast.)}
Therefore, the duration of $\timerrc$ can be chosen as two times the expected latency of the network $\delta_{RC}$ composed of the processes in $\Pi'$, \ie $\timerrc= 2 \times \delta_{RC}$.

\begin{algorithm}
\Init{$\retract_i \gets \ffalse$; $E_i \gets \varnothing$; $\sigendorse_i \gets \varnothing$; $\sigretract_i \gets \varnothing$; $\proofset_i \gets \varnothing$.\DontPrintSemicolon}
\medskip
\Operation{$\rconspropose(C,\initialproof)$}{\label{line:rconspropose}
    \If{$\retract_i = \ffalse$ \cAnd $p_i$ has not already rcons-proposed \cAnd $\initialproof$ is valid}{ \label{line:condit-retract}
        $E_i \gets C$\label{line:rcons:power-C} ; 
        $\Pi_i' \gets \{ j\ |\ \Forall \langle \star, j \rangle \in C \}$\label{line:PiPrime:init}\;
        $\sigendorse_i \gets \sigendorse_i \cup \{\text{signature by $p_i$ for each element in }E_i\}$\label{line:rcons:signatures:init}\ft{PLEASE CHECK: My gut feeling is that $\sigendorse_i$ is necessarily empty when this line is executed. Simplify accordingly?}\;
        $\proofset_i \gets \{\initialproof\}$\label{line:init:proofset:i}\;
        \broadcast $\rconsmess(\sigendorse_i,E_i, \proofset_i, \Pi_i')$ to processes in $\Pi_i'$\label{line:bcast:rconsmess}\; 
        $\timerrc.\timerstart()$. \label{line:tim-set-1}
    }
}
\medskip

\newcommand{\multilineCondition}[1]{%
  \raisebox{\dimexpr\fontcharht\font`T-\height}{$\left\{\parbox{0.85\linewidth}{#1}\right.$}%
  {\finalhyphendemerits=0 \parfillskip=0pt \par}%
  \prevdepth=\fontchardp\font`y
}

\WhenReceivedFrom{$\rconsmess(\sigendorse_j,E_j,\proofset_j, \Pi_j')$\label{line:PiPrime:receive}}{
   \lIf{\insertft{$\proofset_j$ does not contain a proof of acceptance for some pair $\langle \star, j \rangle\in E_j$}\label{line:rconss:condition:proof:acc:in:rconss:message}%
   \ft{PLEASE CHECK: I've reworded the condition to be clearer. Is it OK?}%
   }{return}
    \If{\insertft{\multilineCondition{$\proofset_j$ contains an invalid proof of acceptance \cOr\\
    one of the signatures in $\sigendorse_j$ is invalid or is not by $p_j$ \cOr\\
    there is no 1-1 mapping between the signatures of $\sigendorse_j$ and the pairs of $E_j$}\label{line:rcons:validity:checks}}
    }{%
        $\rconserror()$; $\timerrc.\timerstop()$
    }
    $\proofset_i \gets \proofset_i \cup \proofset_j$\label{line:proofset:update};
    $\sigendorse_i \gets \sigendorse_i \cup \sigendorse_j$\label{line:rconsmess:signatures:union}\;
    \uIf{$p_i$ has not rcons-proposed before \cAnd $\retract_i = \ffalse$}{ \label{line:has-delivered}
        $\retract_i \gets \ttrue$ ;
        $E_i \gets E_j$\label{line:if:no:retract:yet:then:start}\label{line:rconssmess:init:proposed}\;
        \broadcast $\rconsretractmess$(\insertft{$\langle$}sig. of \texttt{``RETRACT''} by $p_i$ \insertft{$\rangle$}) to processes in $\Pi_j'$\label{line:retract:bcast}\;
        \lIf{$\timerrc$ has not been started\label{line:rcons:timer:after:retract}}{%
            $\timerrc.\timerstart()$.%
            \DontPrintSemicolon\label{line:tim-set-2}
        }\label{line:if:no:retract:yet:then:end}
    }
    \lElse{$E_i \gets E_i \cap E_j$.\DontPrintSemicolon\label{line:rcons:intersection:proposed}\ft{PLEASE CHECK: I added a else here.}\ft{PLEASE CHECK: Shouldn't we update $\sigendorse_i$, and remove signatures about pairs that are not longer in $E_i$?}}
    $\checkdecision()$.\label{line:rcons:call:to:checkdecision}
}
\medskip

\InternalOperation{$\checkdecision()$}{
    \If{\insertft{\multilineCondition{%
    neither \rconsdecide nor \rconserror have already been invoked \cAnd\\
    $\sigendorse_i$ contains the signatures of all the processes \insertft{appearing} in $E_i$
    \label{line:rcons:checkdecision:cond}}}}{%
        \insertft{$\timerrc.\timerstop()$}\;
        \insertft{$E_i\leftarrow\{\langle v,k\rangle\in E_i \mid \proofset_i\text{ contains a valid proof for }\langle v,k\rangle\}$\label{line:pruning:Ei}\ft{Needed to prove \RCValidityExtra.}\;%
        \insertft{\lIf{$E_i = \varnothing$\label{line:cond:Ei:not:empty}}{$\rconserror()$.\DontPrintSemicolon}}}
        $\rconsdecide(E_i,\sigendorse_i,\sigretract_i)$\label{line:rcons:decide}.%
        \DontPrintSemicolon
    }
}
\medskip

\WhenReceived{$\rconsretractmess(\sigretractsingular)$}{\label{line:retract:msg:handler}
    \insertft{\lIf{$\sigretractsingular$ is not a valid signature}{return\ft{I've added this check which I think was implicit}}}
    $E_i \gets E_i \setminus \{\text{value in $E_i$ associated with the process that signed $\sigretractsingular$}\}$\label{line:rconsretractmess:update:proposed}\ft{PLEASE CHECK: Same remark as above, shouldn't we update $\sigendorse_i$?}\;
    $\sigretract_i \gets \sigretract_i \cup \{\sigretractsingular\}$\;
    $\checkdecision()$.
}
\medskip

\lWhen{$\timerrc.\timerend()$}{\label{line:tim-stop}%
    $\rconserror()$.\DontPrintSemicolon
}
\caption{Restrained consensus implementation (code for $p_i$).%
\ft{PLEASE CHECK: I've changed the code to remover power sets, which are not needed, and renamed $C_i$ into $E_i$.}%
\ft{QUESTION: In the test at \cref{{line:condit-retract}}, what does the condition `\emph{a signature in $\sigendorse_j$ has no proof}'? Do you mean the signature is invalid? Something else? (Proof of acceptance do not apply to endorsement signatures AFAIU.)}
}
\label{alg:rcons}
\end{algorithm}

\subsection{Restrained consensus: proof}
The proof that \Cref{alg:rcons} implements the Restrained Consensus abstraction defined in \Cref{sec:restr-cons-rc} follows from the following lemmas.

\insertft{\begin{lemma}[\RCValidityExtra.]
If a correct process $p_i$ executes the callback $\rconsdecide(E,$ $\star,$ $\star)$, then
    \begin{itemize}
        \item $E\neq\emptyset,$ and
        \item $E\subseteq\big\{ \langle v, \star, \pi_v \rangle \mid \Exists p_k \text{ correct such that, eventually, }\langle v, \star, \pi_v \rangle\in \CAC_1.\acceptedI{k}\big\}.$
    \end{itemize}
\end{lemma}
\begin{proof}
Consider a correct process $p_i$ that executes the callback $\rconsdecide(E,$ $\star,$ $\star)$ at \cref{line:rcons:decide}. By the condition at \cref{line:cond:Ei:not:empty}, $E=E_i\neq\emptyset$.
Du to \cref{line:pruning:Ei}, $E=E_i$ further only contains pairs $\langle v_k,k\rangle$ for which $p_i$ received a valid proof of acceptance. As a result, \Cref{eq:validity:acceptance} holds, which concludes the lemma.
\end{proof}}

\begin{lemma}[\RCValidity] \label{lemma:rcons:weak:val}
If \insertft{a correct process invokes \rconspropose}, all the processes in $\Pi'$ are correct, and the network delays between the processes of $\Pi'$ are less than or equal to $\delta_{RC}$, then at least one \insertft{correct} process $p_i\in \Pi'$ executes the callback $\rconsdecide(E,\sigendorse,\sigretract)$ and \insertft{$\sigendorse$ contains a signature from each process appearing in $E$.}
\end{lemma}

\begin{proof}
If \insertft{a correct process invokes \rconspropose}, all the processes in $\Pi'$ are correct, and the network delays between the processes of $\Pi'$ are lesser or equal to $\delta_{RC}$, then \insertft{consider $p_i$, the first process in $\Pi'$ to invoke \rconspropose.}
\insertft{Let us note $C_i$ the first parameter passed by $p_i$ to \rconspropose, i.e. $p_i$ invoked $\rconspropose(C_i,\star)$.}
\ft{I explicitly pick the first process to avoid difficulty in case 2b below when $p_j$'s \rconsretractmess could reach $p_i$ before it has invoked \rconspropose and initialized its set $E_i$, which would cause problem with the current algorithm.}%
$p_i$ verifies the condition at \cref{line:condit-retract} and therefore broadcasts a \rconsmess message to all the processes in \insertft{$\Pi_i'\subseteq\Pi'$ (\crefrange{line:PiPrime:init}{line:bcast:rconsmess})}.
All the processes in $\Pi'_i$ are correct and \insertft{the network delays between the processes of $\Pi_i'$ and $p_i$ are} synchronous.
Therefore, \insertft{a process $p_j$ in $\Pi'_i$} will \insertft{either} 
\begin{itemize}
    \item (Case 1) answer $p_i$'s broadcast with a \rconsretractmess message at \cref{line:bcast:rconsmess} \insertft{(if they have not yet broadcast any message at lines~\ref{line:bcast:rconsmess} or~\ref{line:retract:bcast} when they receive $p_i$'s \rconsmess message)}, or 
    \item \insertft{(Case 2) will have broadcast either a \rconsmess or \rconsretractmess message earlier.}
\end{itemize}
\insertft{In Case 1}, $p_j$'s \rconsretractmess message will reach $p_i$ before the timer $\timerrc$ of $p_i$ runs out.
\insertft{Case 2 gives rise to two sub-cases.
\begin{itemize}
    \item Case 2a: if $p_j$ broadcast a \rconsmess message at \cref{line:bcast:rconsmess} before receiving $p_i$'s own message, it must have invoked $\rconspropose(C_j,\star)$ (since $p_j\in\Pi'_i\subseteq\Pi'$ is correct by lemma assumption). In this case both $p_i$ and $p_j$ must have had one of their proposed values accepted by the $\CAC_1$ instance, because of the conditions on the use of the $\RC$ abstraction. As $C_j$ is $p_j$'s $\CAC_1$ candidate set when it invokes \rconspropose, \CACPrediction applies, and $\langle \star, i\rangle\in C_j$, which implies that $p_i\in \Pi_j'$ is one of the recipients of $p_j$'s \rconsmess message at \cref{line:bcast:rconsmess}.
    \item Case 2b: if $p_j$ broadcast a \rconsretractmess message at \cref{line:retract:bcast} before receiving $p_i$'s \rconsmess message, then $p_j$ must have received earlier some \rconsmess message from some process $p_k$ with a valid proof of acceptance $\initialproof_k$ and a set $\Pi'_k$ of involved processes at \cref{line:PiPrime:receive}. The validity of $\initialproof_k$ (\Cref{sec:proof:delivery}) and \CACPrediction imply that there exists some pair $\langle \star, k\rangle\in C_i$. By definition of $\Pi'$ and lemma assumption, $p_k$ is therefore correct, and invoked $\rconspropose(C_k,\star)$ with $C_k$ equal to its $\CAC_1$ candidate set. By the same argument as above, \CACPrediction implies $p_i\in \Pi'_k$. Therefor $p_i$ must be one of the recipients of $p_j$'s \rconsretractmess message at \cref{line:retract:bcast}.
\end{itemize}
In all cases, $p_i$ therefore} receives a signature from all the processes in \insertft{$\Pi'_i$} before $\timerrc$ runs out.
Either the received signatures are in $\sigendorse_i$ (\ie the processes \insertft{that broadcast} a \rconsmess message) or in $\sigretract_i$ (\ie the processes \insertft{that broadcast} a \rconsretractmess message).
\insertft{\begin{subobservation}\label{subobs:Ei:nonempty}
Once $p_i$ has invoked  \rconspropose, $\langle \star,i\rangle\in E_i\neq\emptyset$ holds for the remaining of $p_i$'s execution.
\end{subobservation}
\begin{proof}
When $p_i$ invokes $\rconspropose(C_i,\initialproof_i)$, because it is correct, one of its values $\langle\star,i\rangle$ must have been accepted by $\CAC_1$, and $\initialproof_i$ is a (valid) proof of acceptance for this value. By \CACPrediction, $\langle\star,i\rangle\in C_i$, and therefore just after \cref{line:rcons:power-C} $\langle\star,i\rangle\in E_i=C\neq\emptyset$. Afterwards, $E_i$ is only modified by $p_i$ at \cref{line:rcons:intersection:proposed,line:rconsretractmess:update:proposed,line:pruning:Ei}. Let us prove that $\langle\star,i\rangle\in E_i$ for the remainder of $p_i$'s execution.
\begin{itemize}
    \item When $p_i$ receives a \rconsmess message at \cref{line:PiPrime:receive} from $p_j$, by using the same argument as for $p_k$ in Case 2b above, we have $p_j\in\Pi'_i\in\Pi'$, and $p_j$ is therefore correct and invoked $\rconspropose(C_j,\star)$. By \CACPrediction, $\langle\star,i\rangle\in C_j=E_j$ at \cref{line:PiPrime:receive}, and therefore by recursion $\langle\star,i\rangle\in E_i \cap E_j$ at \cref{line:rcons:intersection:proposed}.
    \item As $p_i$ never sends a \rconsretractmess message (since it is the first process to invoke \rconspropose and to broadcast a \rconsmess message), and signatures cannot be forged, $p_i$ never receives a \sigretractsingular signature signed by itself at \cref{line:retract:msg:handler}, and never removes $\langle\star,i\rangle\in C_i$ from $E_i$ at \cref{line:rconsretractmess:update:proposed}.
    \item Finally, because of \cref{line:init:proofset:i}, and because $\proofset_i$ only grows during $p_i$'s execution, $\initialproof_i\in\proofset_i$ when $p_i$ reaches \cref{line:pruning:Ei}. As a result, $\langle\star,i\rangle$ is not removed from $E_i$ at \cref{line:pruning:Ei}.
    \qedhere
\end{itemize}
\end{proof}}
From the above reasoning and \Cref{subobs:Ei:nonempty}, we conclude that the condition at \cref{line:rcons:checkdecision:cond} is \insertft{eventually} verified at $p_i$, which then does not meet the condition at \cref{line:cond:Ei:not:empty}, and eventually executes $\rconsdecide(E_i,\sigendorse_i,\sigretract_i)$. \insertft{$\sigendorse$ contains a signature from each process appearing in $E_i$}.
\insertft{Furthermore, one of the values $\langle\star,i\rangle$ accepted by $p_i$ when it invoked \rconspropose belongs to the set $E_i$ produced by the \rconsdecide callback.}
\end{proof}

\begin{lemma}[\RCAgreement] \label{lemma:rcons:weak:agree}
If \insertft{a correct process invokes \rconspropose}, all the processes in $\Pi'$ are correct and if \insertft{two correct processes $p_i$ and $p_j$} execute the callback $\rconsdecide$, respectively with the parameters $\rconsdecide(E,\star,\star)$ and $\rconsdecide(E',\star,\star)$, then $E=E'$. 
\end{lemma}
\insertft{%
\begin{proof}
Assume a correct process $p_{i_0}$ invokes $\rconspropose(C_0,\star)$, and all the processes in $\Pi'$ are correct.
\begin{subobservation}\label{subobs:acc:proof:mean:correct}
If there exists valid a proof of acceptance $\initialproof_\ell$ from $\CAC_1$ for some pair $\langle v_\ell,\ell\rangle$ then $p_\ell\in\Pi'$ and $p_\ell$ is correct.
\end{subobservation}
\begin{proof}
Using \Cref{eq:validity:acceptance} and by \CACPrediction of $\CAC_1$, the existence of $\initialproof_\ell$ implies that $\langle v_\ell,\ell\rangle\in C_0$. By definition of $\Pi'$, this implies that $p_\ell\in\Pi'$, and by lemma assumption, that $p_\ell$ is correct.
\end{proof}
Consider $p_i$ and $p_j$ correct. $p_i$ executes $\rconsdecide(E,\star,\star)$ and $p_j$ executes $\rconsdecide(E',\star,\star)$.
Assume $\langle v_k,k\rangle\in E$. Due to \cref{line:pruning:Ei}, when $p_i$ executes $\rconsdecide(E,\star,\star)$, $\proofset_i$ contains a valid proof of acceptance $\initialproof_k$ for $\langle v_k,k\rangle$. Applying \Cref{subobs:acc:proof:mean:correct} to $\initialproof_k$ yields that $p_k\in\Pi'$ is correct.
\begin{subobservation}\label{subobs:vk:k:in:Ej}
When $p_j$ executes \checkdecision, $\langle v_k,k\rangle\in E_j$.
\end{subobservation}
\begin{proof}
When $p_j$ invokes \checkdecision, $E_j$ has been first initialized at \cref{line:rcons:power-C} or \cref{line:rconssmess:init:proposed} and then possibly updated at \cref{line:rcons:intersection:proposed}.
\begin{itemize}
    \item At \cref{line:rcons:power-C}, because $p_j$ is correct, it invoked $\rconspropose(C_j,\star)$, where $C_j$ is $p_j$'s candidate set $\CAC_1.\candidates_j$ at the time of invocation. Because $\initialproof_k$ is a valid proof of acceptance, using \Cref{eq:validity:acceptance} and by \CACPrediction of $\CAC_1$, $\langle v_k,k\rangle\in C_j$, and therefore just after \cref{line:rcons:power-C}, we have $\langle v_k,k\rangle\in E_j$.
    \item At \cref{line:rconssmess:init:proposed}, $p_j$ has received a message $\rconsmess(\star,E_s,\proofset_s,\star)$ from some process $p_s$. Due to the condition at \cref{line:rconss:condition:proof:acc:in:rconss:message}, $\proofset_s$ contains a valid proof for some pair $\langle \star, s \rangle$. By \Cref{subobs:acc:proof:mean:correct}, $p_s$ is therefore correct, and must have sent its $\rconsmess$ message at \cref{line:bcast:rconsmess}. By a reasoning identical to the previous case, we derive $\langle v_k,k\rangle\in E_s$, and therefore $\langle v_k,k\rangle\in E_j$ after executing \cref{line:rconssmess:init:proposed}.
    \item A \cref{line:rcons:intersection:proposed}, an identical reasoning indicate that  $\langle v_k,k\rangle$ remains in $E_j$ after computing the intersection, which concludes the observation.\qedhere
\end{itemize}
\end{proof}
When $p_j$ executes $\rconsdecide(E'=E_j,\sigendorse_j,\sigretract_j)$ at \cref{line:rcons:decide}, it is within \checkdecision. By \Cref{subobs:vk:k:in:Ej}, jusy after invoking \checkdecision,  $\langle v_k,k\rangle\in E_j$. Due to the condition at \cref{line:rcons:checkdecision:cond}, $\sigendorse_j$ contains a signature from $p_k$. This signature must have been added to $\sigendorse_j$ at \cref{line:rconsmess:signatures:union} following the reception of a message $\rconsmess(\star,\star,\proofset_k,\star)$ from $p_k$ (due to the condition at \cref{line:rcons:validity:checks}). Because $p_k$ is correct, $\initialproof_k\in\proofset_k$, and thanks to \cref{line:proofset:update} and the fact that $\proofset_j$ only grows, $\initialproof_k\in\proofset_j$ afterwards, and in particular when $p_j$ executes \cref{line:pruning:Ei}. We conclude that $\langle v_k,k\rangle\in E_j$ when $p_j$ reaches $\rconsdecide$ at \cref{line:rcons:decide}, and therefore that $\langle v_k,k\rangle\in E'$.

As any pair contained in $E$ is also in $E'$, we have $E\subseteq E'$, and by symmetry $E=E'$.\qedhere
\end{proof}
}%

\begin{lemma}[\RCIntegrity]
A correct process $p_i$ can invoke at most once either $\rconsdecide(\star,\star,\star)$ or \rconserror (but not both in the same execution).
\end{lemma}

\begin{proof}
This lemma is trivially verified by the condition line \ref{line:rcons:checkdecision:cond}. This condition ensures that \rconsdecide callback can only be triggered once, and cannot be triggered if \rconserror has already been triggered. Furthermore, the timer $\timerrc$ is only started once (lines \ref{line:condit-retract} and \ref{line:rcons:timer:after:retract}), hence, the callback \rconserror can only be triggered once.
\end{proof}

\begin{lemma}[\RCTermination]\label{lemma:rcons:termination}
Any correct process in $\Pi'$ eventually executes \rconserror or \rconsdecide.
\end{lemma}

\begin{proof}
A process in $\Pi'$ is a process that executed the \rconspropose operation without receiving any prior message, or that received a \rconsmess message before executing the \rconspropose operation.
In the first case, the process will start $\timerrc$ at \cref{line:tim-set-1}.
In the second case, the process does not meet the condition at \cref{line:has-delivered}.
Therefore, it start $\timerrc$ at \cref{line:tim-set-2}.
These two cases are mutually exclusive.
Once a process starts $\timerrc$, it cannot start it again (lines \ref{line:condit-retract} and \ref{line:rcons:timer:after:retract}).
Finally, when the timer expires, it executes the callback \rconserror.
Therefore, any correct process in $\Pi'$ will terminate.
\end{proof}

\begin{algorithm}[tb]
\Init{$\pi_i \gets \varnothing$. \DontPrintSemicolon}
\smallskip

\lOperation{$\cconspropose(v)$}{%
    $\CAC_1.\cacpropose(v)$. \DontPrintSemicolon
}
\PrintSemicolon
\smallskip

\When{$\CAC_1.\cacaccept(v,j,\pi)$}{
    $\pi_i \gets \pi_i \cup \{\pi\}$\;
    \lIf{($|\CAC_1.\candidates_i|=1$ \cOr all values in $\CAC_1.\candidates_i$ are the same) \cAnd \cconsdecide has not already been triggered}{%
        $\cconsdecide(v)$%
    } \label{line:decide-1}
    \lElseIf{$j=i$}{%
        $\rconspropose(\CAC_1.\candidates_i,\pi)$
    } \label{line:rcons}
    \lElse{%
        $\timercc.\timerstart()$. \DontPrintSemicolon \Comment*[f]{start timer with a duration of $2 \times \delta_{RC} + \delta_{CC}$}%
    }
}
\smallskip

\When{$\RC.\rconsdecide(E,\sigendorse,\sigretract)$}{
    $\CAC_2.\cacpropose(\langle E,\sigendorse,\sigretract, \pi_i \rangle)$.\ft{Where do we use $\sigendorse$ and $\sigretract$?}
}
\smallskip

\When{$\RC.\rconserror()$ is invoked \cOr $\big(\timercc.\timerend()\;\cAnd\;\CAC_1.\accepted_i  \ne \varnothing\big)$}{
    $\CAC_2.\cacpropose(\langle \CAC_1.\accepted_i, \varnothing,\varnothing,\pi_i \rangle)$.
}
\smallskip

\When{$\CAC_2.\cacaccept(\langle E,\star,\star,\star \rangle, j,\pi$)}{
    \lIf{($|\CAC_2.\candidates_i|=1$ \cOr all values in $\CAC_1.\candidates_i$ are the same) \cAnd \cconsdecide has not already been triggered}{%
        $\cconsdecide(\choice(E))$;%
        \DontPrintSemicolon\label{line:decide-2}
    }
    \lElseIf{$p_i$ has not already \ccons-proposed a value}{%
        $\GC.\tconspropose(\langle\CAC_2.\accepted_i, \pi \rangle)$. \DontPrintSemicolon
        \label{line:tcons}
    }
}
\smallskip

\When{$\GC.\tconsdecide(\langle E,\star \rangle)$}{%
    \lIf{\cconsdecide has not already been triggered}{%
    $\cconsdecide(\choice(E))$. \label{line:decide-3}\DontPrintSemicolon 
    }\DontPrintSemicolon 
}
\caption{\CCons implementation (code for $p_i$).
}
\label{alg:ccons}
\end{algorithm}

\subsection{Contention-aware Cascading Consensus: implementation} \label{sec:cascading-consensus}
\Cref{alg:ccons} presents the Cascading Consensus algorithm.
It relies on two instances of the \CAC abstraction, $\CAC_1$ and $\CAC_2$, one instance of Restrained Consensus ($\RC$) and one instance of Global Consensus (\GC) (as all the processes in $\Pi$ participate in it).
The list of all different abstractions is summarized in \Cref{tab:consensus-notations} (where \sigendorse and \sigretract are respectively replaced by $S_e$ and $S_r$).

When a process $p_i$ \cac-accepts a tuple from one of the \CAC instances, it can fall into either of the following two cases.
\begin{enumerate}
    \item $|\{v\mid(v,\star)\in\candidates_i\}|=1$: $p_i$ detects there is no conflict, so it knows that other correct processes cannot \cac-accept any other value, and it can immediately decide the value it received.

    \item $|\{v\mid(v,\star)\in\candidates_i\}|>1$: $p_i$ detects multiple candidate values, so it must continue the algorithm to resolve the conflict.
\end{enumerate}

A conflict in $\CAC_1$ leads to the execution of Restrained Consensus ($\RC$) among the participants involved in the conflict (\cref{line:rcons}).
A conflict in $\CAC_2$ leads to the execution of Global Consensus ($\GC$) among all the system participants (\cref{line:tcons}).

In $\CAC_2$, the set of values \cac-accepted in the prior steps are proposed.
To simplify the presentation of the algorithm, the pseudo-code omits some implementation details.
In particular, $\CAC_2$ verifies the proofs associated with the proposed values.
A correct process $p_i$ considers a set of values $E$ \cac-proposed by a process $p_j$ in $\CAC_2$ only if either one of the following conditions holds:

\begin{itemize}
    \item $p_j$ did not propose one of the values in $E$ during $\CAC_1$---\ie it did not participate in $\RC$---, each value in $E$ is associated with a valid proof of acceptance and $E$ is not empty.
    \item $p_j$ proposed one of the values in $E$ during $\CAC_1$---\ie it participates in $\RC$---and $E$ is signed by all the processes that proposed values in $E$. Furthermore, each process whose value proposed in $\CAC_1$ is eventually accepted signed the string ``$\mathtt{RETRACT}$''.
\end{itemize}
Similarly, the values proposed in the Global Consensus are also associated with a proof of acceptance from the second instance of the \CAC algorithm.
We assume that the Global Consensus implementation cannot decide a value not associated with a valid proof of acceptance.

Note that, if a correct process $p_i$ \cac-accepts a value with $|\candidates_i|=1$, it does not necessarily imply that other correct processes will have the same \candidates set.
The processes that detect a conflict execute one of the consensuses, restrained or global.
However, the algorithm ensures, using acceptance proofs, that only a value that has been \cac-accepted in a previous step can be proposed for the next step.
Hence, if a correct process $p_i$ \cac-accepts a value $v$ with $\candidates_i=\{\langle v, \star\rangle\}$, the other processes will not be able to propose $v'\neq v$ in the following steps of the algorithm---by the prediction property of the \CAC abstraction.
In other words, some correct processes may terminate faster than others, but this early termination does not impact the agreement of the protocol.

Like the \RC algorithm described in \Cref{sec:rcons:impl}, the \CCons algorithm uses a timer, $\timercc$. This timer provides the operation $\timercc.\timerstart()$ to start the timer, and the callback $\timercc.\timerend()$, which is invoked once the time has elapsed.
The duration of $\timercc$ should be long enough to allow the processes participating in Restrained Consensus to terminate if they are in a synchronous period.
Subject to this condition, the algorithm can terminate in 2 synchronous periods for Restrained Consensus plus 1 synchronous period to initiate the second instance of the \CAC abstraction.
Therefore, the duration of $\timercc$ should ideally equal $2 \times \delta_{RC} + \delta_{CC}$, where $\delta_{RC}$ is the likely latency of the sub-network of all participants of Restrained Consensus and $\delta_{CC}$ is the likely latency of the network composed of all the processes in $\Pi$. However, if $\timercc$ is chosen too small, the safety and liveness properties of CC are still ensured. 
}
\subsection{Cascading Consensus: proof}
\label{sec:ccons:proof}

The proof of correctness that the Cascading Consensus algorithm presented in \Cref{alg:ccons} implements consensus follows from the subsequent lemmas.

\begin{lemma}[\CValidity]
    If all processes are correct and a process decides a value $v$, then $v$ was proposed by some process.
\end{lemma}

\begin{proof}
By exhaustion, we explore the three following cases.
\begin{itemize}
    \item If a value is decided at \cref{line:decide-1}, then it is the result of the first \CAC instance. Thanks to the \CACValidity property, we
        know that a process in $\Pi$ proposed this value.
    \item If a value is decided at \cref{line:decide-2}, then it is the result of $\CAC_2$.
        The only values \cac-proposed using $\CAC_2$ are a set of values \cac-accepted from $\CAC_1$, either they are \cac-proposed by a process that participated in \rcons or not. Thanks to the \CACValidity property, we know that a process in $\Pi$ proposed this value.
    \item If a value is decided at \cref{line:decide-3} then the value was decided by the GC instance. However, the values proposed to GC are values accepted by $\CAC_2$. Thanks to the \CValidity of GC and \CACValidity of $\CAC_1$ and $\CAC_2$, we know that a process in $\Pi$ proposed this value.\qedhere
\end{itemize} 
\end{proof}

\begin{lemma}[\CAgreement]
No two correct processes decide different values.
\end{lemma}

\begin{proof}
A correct process that participates in the \CCons can decide at different points of the execution of the algorithm: lines~\ref{line:decide-1}, \ref{line:decide-2} or~\ref{line:decide-3}.
However, if a correct process decides at \cref{line:decide-1} or~\ref{line:decide-2}, not all correct processes will necessarly do so.

Nonetheless, the \CACGlobalTermination property ensure that if a correct process decides before the others, all the correct processes will decide the same value.

Let us assume that a correct process $p_i$ decides a value $v$ at \cref{line:decide-1}.
This implies that $\CAC_1$ outputs a $\CAC_1.\candidates_i$ set of size $1$ or all the values in $\CAC_1.\candidates_i$ are the same for $p_i$ after the first \cac-acceptance.
Using the \CACPrediction and \CACGlobalTermination, we know that $p_i$ will not \cac-accept any value different from $v$ with the $\CAC_1$ instance.
Furthermore, using \CACGlobalTermination, we know that no other correct process can \cac-accept a value $v' \ne v$.
Otherwise, $p_i$ would also \cac-accept it, contradicting the \CACPrediction property.
Therefore, if $p_i$ decides $v$ at \cref{line:decide-1}, all correct processes that do not \ccons-decide at this point will only \cac-accept $v$ with $\CAC_1$.
Furthermore, the values \cac-proposed in $\CAC_2$ are those that were \cac-accepted by $\CAC_1$.
Therefore, $v$ is the only value that is \cac-proposed in $\CAC_2$.
Using the \CACGlobalTermination and \CACPrediction property, we know that all the correct processes will \ccons-decide $v$ at \cref{line:decide-2}.

Similar reasoning can be applied if a correct process decides at \cref{line:decide-2} whereas others do not.
The only values that can be \tcons-proposed are those \cac-accepted in $\CAC_2$.
Therefore, using the \CACGlobalTermination and \CACPrediction properties of \CAC, we know that if a correct process \ccons-decided a value $v$ at \cref{line:decide-2}, then all correct processes that did not \ccons-decide at this point will \ccons-decide $v$ at \cref{line:decide-3}.
    
Finally, if no process decides at \cref{line:decide-1} or~\ref{line:decide-2}, then the \CAgreement property of consensus ensures that all the processes \ccons-decide the same value at \cref{line:decide-3}.
\end{proof}

\begin{lemma}[\CIntegrity]
    A correct process decides at most one value.
\end{lemma}

\begin{proof}
This lemma is trivially verified. All the lines where a process can decide (lines \ref{line:decide-1}, \ref{line:decide-2} and \ref{line:decide-3}) are preceded by a condition that can only be verified if the process did not already triggered \cconsdecide. Hence the \CIntegrity property is verified.
\end{proof}

\begin{lemma}[\CTermination]
If a correct process proposes value $v$, then all correct processes eventually decide some value (not necessarily $v$).
\end{lemma}

\begin{proof}
All the sub-algorithms used in \CCons (\CAC, RC, and GC) terminate.
Furthermore, each algorithm is executed sequentially if the previous one did not decide a value.
The only algorithm that may not be triggered is RC if $\CAC_1$ terminates with a $\candidates_i$ set whose size is greater than $1$ at $p_i$, and if $p_i$ did not \cac-proposed one of the values in $\candidates_i$.
However, we observe that processes not participating in the RC algorithm set a timer $\timercc$ when $\CAC_1$ returns.
Once this timer expires, these processes \cac-propose a value using $\CAC_2$.
Therefore, any correct process that participates in the \CCons terminates.
\end{proof}

\end{document}